\documentclass[11pt]{book}
\usepackage{Wiley-AuthoringTemplate}
\usepackage{chapterbib}
\usepackage[sectionbib,numbers]{natbib}% for numbered citation comment the above line
\usepackage{amsmath}
\usepackage{amssymb}
\usepackage{graphicx}
\usepackage{scalefnt}
\usepackage[font=footnotesize]{caption}
\usepackage[font=footnotesize,labelformat=simple]{subcaption}

\usepackage[nopostdot,nonumberlist,acronym]{glossaries}
\usepackage{xurl}
\usepackage{multirow}
\usepackage{tabularx}
\usepackage{algorithmic}
\usepackage[ruled]{algorithm2e}
\usepackage{longtable}
\usepackage{xcolor}
\usepackage[mediumspace,mediumqspace,Grey,squaren]{SIunits}
\usepackage{hyperref}

\makeindex% for index generation

\newcommand{\quot}[1]{``#1''}
\newcommand*\nestedgls[1]{%
	\protect\ifglsused{#1}{%
		\glsentryshort{#1}%
	}{%
		\glsentrylong{#1}%
	}%
}

\newacronym{NR}{NR}{new radio}
\newacronym{3GPP}{3GPP}{3rd generation partnership project}
\newacronym{eMBB}{eMBB}{enhanced mobile broadband}
\newacronym{uRLLC}{uRLLC}{ultra-reliable and low-latency communications}
\newacronym{mMTC}{mMTC}{massive \gls{MTC}}
\newacronym{VOD}{VOD}{video on demand}
\newacronym{LTI}{LTI}{linear time invariant}
\newacronym{UV}{UV}{ultraviolet}
\newacronym{OCC}{OCC}{optical camera communication}
\newacronym{mmWave}{mmWave}{millimetre wave}
\newacronym{LAA}{LAA}{licence assisted access}
\newacronym{IoT}{IoT}{internet-of-things}
\newacronym{IoE}{IoE}{internet-of-everything}
\newacronym{D2D}{D2D}{device-to-device}
\newacronym{UMTS}{UMTS}{Universal Mobile Telecommunications System}
\newacronym{WLAN}{WLAN}{Wireless Local Area Network}
\newacronym{AP}{AP}{access point}
\newacronym{LiFi}{LiFi}{light-fidelity}
\newacronym{FET}{FET}{field-effect transistor}
\newacronym{PED}{PED}{portable electronic device}
\newacronym{MCRT}{MCRT}{Monte Carlo ray-tracing}
\newacronym{DL}{DL}{downlink}
\newacronym{UL}{UL}{uplink}
\newacronym{ICI}{ICI}{inter-cell interference}
\newacronym{LED}{LED}{light emitting diode}
\newacronym{PD}{PD}{photo-diode}
\newacronym{IFC}{IFC}{in-flight connectivity}
\newacronym{OIFC}{OIFC}{optical \nestedgls{IFC}}
\newacronym{IFE}{IFE}{in-flight entertainment}
\newacronym{IM/DD}{IM/DD}{intensity-modulation and direct-detection}
\newacronym{IR}{IR}{infra-red}
\newacronym{VL}{VL}{visible light}
\newacronym{OW}{OW}{optical wireless}
\newacronym{OWC}{OWC}{optical wireless communications}
\newacronym{CAD}{CAD}{computer-aided-design}
\newacronym{VLC}{VLC}{visible light communications}
\newacronym{SIR}{SIR}{signal-to-interference-ratio}
\newacronym{SNR}{SNR}{signal-to-noise-ratio}
\newacronym{SINR}{SINR}{signal-to-interference-plus-noise-ratio}
\newacronym{QoS}{QoS}{quality-of-service}
\newacronym{RF}{RF}{radio frequency}
\newacronym{EM}{EM}{electromagnetic}
\newacronym{OOK}{OOK}{on-off keying}
\newacronym{PL}{PL}{path loss}
\newacronym{WDM}{WDM}{wavelength-division-multiplexing}
\newacronym{RGB}{RGB}{red-green-blue}
\newacronym{LoS}{LoS}{line-of-sight}
\newacronym{NLoS}{NLoS}{non-\nestedgls{LoS}}
\newacronym{VoM}{VoM}{volume of mobility}
\newacronym{FoV}{FoV}{field-of-view}
\newacronym{SM}{SM}{spatial modulation}
\newacronym{GSM}{GSM}{generalised \nestedgls{SM}}
\newacronym{OSM}{OSM}{optical \nestedgls{SM}}
\newacronym{SISO}{SISO}{single-input-single-output}
\newacronym{SIMO}{SIMO}{single-input-multiple-output}
\newacronym{MISO}{MISO}{multiple-input-single-output}
\newacronym{MIMO}{MIMO}{multiple-input-multiple-output}
\newacronym{OMIMO}{OMIMO}{optical \nestedgls{MIMO}}
\newacronym{VR}{VR}{virtual reality}
\newacronym{AR}{AR}{augmented reality}
\newacronym{MTC}{MTC}{machine type communications}
\newacronym{CAGR}{CAGR}{compound annual growth rate}
\newacronym{InM}{InM}{intensity modulation}
\newacronym{DD}{DD}{direct detection}
\newacronym{TX}{TX}{transmitter}
\newacronym{RX}{RX}{receiver}
\newacronym{PAM}{PAM}{pulse amplitude modulation}
\newacronym{QAM}{QAM}{quadrature amplitude modulation}
\newacronym{MPAM}{$M$-PAM}{$M$-ary pulse amplitude modulation}
\newacronym{PAPR}{PAPR}{peak to average power ratio}
\newacronym{NRZ-OOK}{NRZ-OOK}{non-return-zero \nestedgls{OOK}}
\newacronym{SSK}{SSK}{space shift keying}
\newacronym{A/D}{A/D}{analog-to-digital}
\newacronym{D/A}{D/A}{digital-to-analog}
\newacronym{O/E}{O/E}{optical-to-electrical}
\newacronym{E/O}{E/O}{electrical-to-optical}
\newacronym{DC}{DC}{direct-current}
\newacronym{SER}{SER}{symbol error ratio}
\newacronym{BER}{BER}{bit error ratio}
\newacronym{OFDM}{OFDM}{orthogonal frequency division multiplexing}
\newacronym{OFDMA}{OFDMA}{orthogonal frequency division multiple access}
\newacronym{AWGN}{AWGN}{additive white Gaussian noise}
\newacronym{ML}{ML}{maximum-likelihood}
\newacronym{ME-OSM}{ME-OSM}{minimum error OSM}
\newacronym{CCI}{CCI}{co-channel interference}
\newacronym{CCD}{CCD}{charge-coupled device}
\newacronym{CIR}{CIR}{channel impulse response}
\newacronym{CFR}{CFR}{channel frequency response}
\newacronym{IFOWC}{IFOWC}{in-flight \nestedgls{OWC}}
\newacronym{LD}{LD}{laser diode}
\newacronym{FR}{FR}{frequency reuse}
\newacronym{WR}{WR}{wavelength reuse}
\newacronym{W-LED}{W-LED}{white \nestedgls{LED}}
\newacronym{PB}{PB}{phosphorescence-based}
\newacronym{RR}{RR}{retro-reflector}
\newacronym{LTE}{LTE}{long term evolution}
\newacronym{Wi-Fi}{Wi-Fi}{wireless fidelity}
\newacronym{5G}{5G}{fifth generation}
\newacronym{FAA}{FAA}{Federal Aviation Administration}
\newacronym{FCC}{FCC}{Federal Communications Comission}
\newacronym{ATENAA}{ATENAA}{Advanced Technologies for Networking in Avionic Applications}
\newacronym{RMS}{RMS}{root-mean-squared}
\newacronym{PSU}{PSU}{passenger service unit}
\newacronym{SIC}{SIC}{successive interference cancellation}
\newacronym{SRT}{SRT}{sequential ray tracing}
\newacronym{NSRT}{NSRT}{non-\nestedgls{SRT}}
\newacronym{RDB}{RDB}{ray database}
\newacronym{CCT}{CCT}{correlated colour temperature}
\newacronym{SMD}{SMD}{surface-mounted device}
\newacronym{USGS}{USGS}{United States Geological Survey}
\newacronym{FSO}{FSO}{free space optical communication}
\newacronym{w.r.t.}{w.r.t.}{with respect to}
\newacronym{HPBW}{HPBW}{half power beam width}
\newacronym{PSD}{PSD}{power spectral density}
\newacronym{DSP}{DSP}{digital signal processing}
\newacronym{UE}{UE}{user equipment}
\newacronym{BRDF}{BRDF}{bidirectional reflectance distribution function}
\newacronym{BSDF}{BSDF}{bi-directional scatter distribution function}
\newacronym{PCM}{PCM}{pulse code modulation}
\newacronym{PPM}{PPM}{pulse position modulation}
\newacronym{DPPM}{DPPM}{differential \nestedgls{PPM}}
\newacronym{U-PAM}{U-PAM}{unipolar \nestedgls{PAM}}
\newacronym{ISI}{ISI}{inter-symbol interference}
\newacronym{DCO-OFDM}{DCO-OFDM}{\nestedgls{DC} biased optical \nestedgls{OFDM}}
\newacronym{ACO-OFDM}{ACO-OFDM}{asymetrically clipped optical \nestedgls{OFDM}}
\newacronym{NDC-OFDM}{NDC-OFDM}{non-DC biased \nestedgls{OFDM}}
\newacronym{PAM-DMT}{PAM-DMT}{pulse amplitude modulated discrete multitone}
\newacronym{U-OFDM}{U-OFDM}{unipolar \nestedgls{OFDM}}
\newacronym{eU-OFDM}{eU-OFDM}{enhanced unipolar \nestedgls{OFDM}}
\newacronym{FFT}{FFT}{fast Fourier transform}
\newacronym{IFFT}{IFFT}{inverse \nestedgls{FFT}}
\newacronym{DAC}{DAC}{digital-to-analog converter}
\newacronym{ADC}{ADC}{analog-to-digital converter}
\newacronym{EOC}{EOC}{electrical-to-optical converter}
\newacronym{OEC}{OEC}{optical-to-electrical converter}
\newacronym{PDF}{PDF}{probability density function}
\newacronym{CP}{CP}{cyclic prefix}
\newacronym{IM}{IM}{index modulation}
\newacronym{SIM}{SIM}{subcarrier \nestedgls{IM}}
\newacronym{DCO-OFDM-SIM}{DCO-OFDM-SIM}{\nestedgls{DCO-OFDM} with \nestedgls{SIM}}
\newacronym{ACO-OFDM-SIM}{ACO-OFDM-SIM}{\nestedgls{ACO-OFDM} with \nestedgls{SIM}}
\newacronym{THP}{THP}{Tomlinson-Harashima Precoder}
\newacronym{LRU}{LRU}{line-replaceable unit}
\newacronym{ACI}{ACI}{adjacent channel interference}
\newacronym{TDD}{TDD}{time division duplex}
\newacronym{MMSE}{MMSE}{minimum mean square error}
\newacronym{CDMA}{CDMA}{code division multiple access}
\newacronym{FWHM}{FWHM}{full width at half maximum}
\newacronym{PM}{PM}{permutation modulation}
\newacronym{PC}{PC}{permutation combinatory}
\newacronym{SS}{SS}{spread spectrum}
\newacronym{PSK}{PSK}{phase shift keying}
\newacronym{BPSK}{BPSK}{binary \nestedgls{PSK}}
\newacronym{QPSK}{QPSK}{quadrature \nestedgls{PSK}}
\newacronym{ICaI}{ICaI}{inter carrier interference}
\newacronym{SE-SIM}{SE-SIM}{spectrally efficient \nestedgls{SIM}}
\newacronym{OFDM-SIM}{OFDM-SIM}{\nestedgls{OFDM} with \nestedgls{SIM}}
\newacronym{APM}{APM}{amplitude and phase modulation}
\newacronym{QSM}{QSM}{quadrature \nestedgls{SM}}
\newacronym{LLR}{LLR}{log-likelihood ratio}
\newacronym{DM}{DM}{dual mode}
\newacronym{DSM}{DSM}{differential \nestedgls{SM}}
\newacronym{DSIM}{DSIM}{differential \nestedgls{SIM}}
\newacronym{CSI}{CSI}{channel state information}
\newacronym{CPU}{CPU}{central process unit}
\newacronym{VoD}{VoD}{video on demand}
\newacronym{EMI}{EMI}{\nestedgls{EM} interference}
\newacronym{FRP}{FRP}{fiber reinforced plastic}
\newacronym{SMX}{SMX}{spatial multiplexing}
\newacronym{JT}{JT}{joint transmission}
\newacronym{SP}{SP}{single point}
\newacronym{FFR}{FFR}{fractional frequency reuse}
\newacronym{CoMP}{CoMP}{coordinated multi-point}
\newacronym{BYOD}{BYOD}{bring your own device}
\newacronym{3D}{3D}{three dimensional}
\newacronym{ICeI}{ICeI}{inter-cell-interference}
\newacronym{FSK}{FSK}{frequency shift keying}
\newacronym{CBE-SM}{CBE-SM}{complex-bipolar encoded \nestedgls{SM}}
\newacronym{AH}{AH}{antenna hopping}
\newacronym{PMF}{PMF}{probability mass function}
\newacronym{AWG}{AWG}{arbitrary waveform generator}
\newacronym{FBE-SM}{FBE-SM}{fractional bit encoded \nestedgls{SM}}
\newacronym{STSK}{STSK}{space-time shift keying}
\newacronym{ESM}{ESM}{enhanced \nestedgls{SM}}
\newacronym{TRX-SM}{TRX-SM}{transceiver \nestedgls{SM}}
\newacronym{ADR}{ADR}{angle diversity receiver}
\newacronym{FLIM}{FLIM}{flexible \nestedgls{LED} index modulation}
\newacronym{GSSK}{GSSK}{generalized \nestedgls{SSK}}
\newacronym{GPSSK}{GPSSK}{generalized pulse position modulated \nestedgls{SSK}}
\newacronym{MA-GSM}{MA-GSM}{multiple active \nestedgls{GSM}}
\newacronym{PEP}{PEP}{pairwise error probability}
\newacronym{APEP}{APEP}{average pairwise error probability}
\newacronym{ABEP}{ABEP}{average \nestedgls{BEP}}
\newacronym{ABER}{ABER}{average \nestedgls{BER}}
\newacronym{BEP}{BEP}{bit error probability}
\newacronym{FEC}{FEC}{forward error correction}
\newacronym{TSM}{TSM}{transmit \nestedgls{SM}}
\newacronym{R-SM}{R-SM}{receive \nestedgls{SM}}
\newacronym{iid}{i.i.d.}{independent identically distributed}
\newacronym{CSIT}{CSIT}{\nestedgls{CSI} at the transmitter}
\newacronym{SC}{SC}{single carrier}
\newacronym{MC}{MC}{multi carrier}
\newacronym{TCM-SM}{TCM-SM}{trellis coded modulation \nestedgls{SM}}
\newacronym{D-SM}{D-SM}{differential \nestedgls{SM}}
\newacronym{DSTSK}{DSTSK}{differential space-time shift keying}
\newacronym{PSM}{PSM}{precoding aided \nestedgls{SM}}
\newacronym{PLS}{PLS}{physical layer security}
\newacronym{F-GSSK}{F-GSSK}{flexible \nestedgls{GSSK}}
\newacronym{Q-SM}{Q-SM}{quadrature \nestedgls{SM}}
\newacronym{E-SM}{E-SM}{enhanced \nestedgls{SM}}
\newacronym{V-SM}{V-SM}{virtual \nestedgls{SM}}
\newacronym{TR-SM}{TR-SM}{transceive \nestedgls{SM}}
\newacronym{SVD}{SVD}{singular value decomposition}
\newacronym{CSIR}{CSIR}{channel state information at the \nestedgls{RX}}
\newacronym{JSM}{JSM}{joint transmitter-receiver \nestedgls{SM}}
\newacronym{GLIM}{GLIM}{generalized \nestedgls{LED} index modulation}
\newacronym{e-GLIM}{e-GLIM}{extended \nestedgls{GLIM}}
\newacronym{DFT}{DFT}{discrete Fourier transform}
\newacronym{IDFT}{IDFT}{inverse \nestedgls{DFT}}
\newacronym{ZF}{ZF}{zero-forcing}
\newacronym{MAP}{MAP}{maximum-a-posteriori-probability}
\newacronym{TD-SM}{TD-SM}{time domain \nestedgls{SM}}
\newacronym{FD-SM}{FD-SM}{frequency domain \nestedgls{SM}}
\newacronym{RC}{RC}{repetition coding}
\newacronym{STBC}{STBC}{space-time block coding}
\newacronym{P/S}{P/S}{space-time block coding}
\newacronym{CN}{CN}{condition number}
\newacronym{eTD-SM}{eTD-SM}{enhanced \nestedgls{TD-SM}}
\newacronym{FDE}{FDE}{frequency domain equalization}
\newacronym{IChI}{IChI}{inter channel interference}
\newacronym{MAC}{MAC}{media access control}
\newacronym{MCS}{MCS}{modulation and coding scheme}
\newacronym{AMC}{AMC}{adaptive modulation and coding}
\newacronym{CQI}{CQI}{channel quality indicator}
\newacronym{RB}{RB}{resource block}
\newacronym{PDCCH}{PDCCH}{physical downlink control channel}
\newacronym{TDMA}{TDMA}{time division multiple access}
\newacronym{SMA}{SMA}{SubMiniature version A}
\newacronym{SPCD}{SPCD}{spectrum}
\newacronym{modem}{modem}{modulator-demodulator}
\newacronym{PLC}{PLC}{powerline communications}
\newacronym{PoE}{PoE}{power over Ethernet}
\newacronym{IATA}{IATA}{The International Air Transport Association}
\newacronym{LSE}{LSE}{The London School of Economics and Political Science}
\newacronym{bpcu}{bpcu}{bits per channel use}
\newacronym{MSE}{MSE}{mean squared error}
\newacronym{IRS}{IRS}{intelligent reconfigurable surfaces}
\newacronym{RIS}{RIS}{reconfigurable intelligent surfaces}
\newacronym{FPGA}{FPGA}{field-programmable gate array}
\newacronym{SDS}{SDS}{software-defined surface}
\newacronym{MBM}{MBM}{media-based modulation}
\newacronym{NOMA}{NOMA}{non-orthogonal multiple access}
\begin{document}

\frontmatter

\mainmatter
\include{p01}

\chapter[Ch1]{Optical Wireless Communications Using Intelligent Walls}

\author[]{Anil Yesilkaya}
\author[]{Hanaa Abumarshoud}
\author[]{Harald Haas$^*$}
%\author[3]{Author IV}

\address[1]{\orgdiv{LiFi Research and Development Centre (LRDC), Department of Electronic \& Electrical Engineering},
\orgname{The University of Strathclyde}, \street{204 George Street}, 
\postcode{G1 1XW}, 
\city{Glasgow }, \country{United Kingdom}
\thanks{This research has been supported in part by EPSRC under Established Career Fellowship Grant EP/R007101/1, Wolfson Foundation and European Commission’s Horizon 2020 research and innovation program under grant agreement 871428, 5G-CLARITY project. This is the pre-peer reviewed version of the following book chapter: "Optical Wireless Communications Using Intelligent Walls" in \emph{Intelligent Reconfigurable Surfaces (IRS) for Prospective 6G Wireless Networks}, Wiley-IEEE Press, Nov 2022, ch. 12, pp. 233-274, ISBN: 978-1-119-87525-3, which has been published in final form at \href{https://www.wiley.com/en-gb/Intelligent+Reconfigurable+Surfaces+(IRS)+for+Prospective+6G+Wireless+Networks-p-9781119875253}{https://www.wiley.com/en-gb/Intelligent+Reconfigurable+Surfaces+(IRS)+for+Prospective+6G+Wireless+Networks-p-9781119875253}. This chapter may be used for non-commercial purposes in accordance with Wiley Terms and Conditions for Use of Self-Archived Versions.}}%

%\address[2]{\orgdiv{II Author Organization Division Name}, \orgname{Organization Name}, \postcode{Postal Code}, \countrypart{Part of the Country},  \city{City Name}, \street{Street Name}, \country{Country}}%

\address*{Corresponding Author: Harald Haas; \email{harald.haas@strath.ac.uk}}

\maketitle% This tag is required to print author and address in the output

\begin{abstract}{Abstract}
This chapter is devoted to discussing the integration of intelligent reflecting surfaces (IRSs), or intelligent walls, in optical wireless communication (OWC) systems. IRS technology is a revolutionary concept that enables communication systems to harness the surrounding environment to control the propagation of light signals. Based on this, specific key performance indicators could be achieved by altering the electromagnetic response of the IRSs. In the following, we discuss the background theory and applications of IRSs and present a case study for an IRS-assisted indoor light-fidelity (LiFi) system. We then highlight some of the  challenges related to this emerging concept and elaborate on future research directions. 
\end{abstract}


%##########################################################################################################
\section{Introduction} 
The current surge in the number of mobile devices and the emerging \gls{IoT} and \gls{IoE} applications are posing an unprecedented demand on wireless connectivity \cite{6770233}. Mobile device density in service areas is approaching the region of more than one wireless device per square meter, which indicates that enhancing the capacity of wireless networks will become even more crucial in the foreseeable future. The use of conventional \gls{RF} small cells in such dense deployments proves challenging due to severe inter-cell \gls{CCI}. This is mainly attributed to the fact that \gls{RF} signals  travel beyond the cell boundaries. Moreover, there is a general consensus that the limited \gls{RF} spectrum resources are not sufficient to future-proof wireless communications. Therefore, other parts of the \gls{EM} spectrum have  been explored for wireless connectivity. Particularly, \gls{OWC} presents a promising solution to satisfy the demands of future wireless networks. The main advantages offered by \gls{OWC} are the freely available resources in the optical band of the \gls{EM} spectrum which is in tera Hertz,  with the visible light  band alone being  almost 2,600 times larger than the full \gls{RF} spectrum \cite{hycvppai2001}. Utilising the entire optical spectrum is not practically feasible due to the limitations of the transceivers front-ends and the necessary electro-optical (EO) and opto-electrical (OE) conversions. However, advancements in solid-state lighting and semiconductor devices are continuously leading to  better  capabilities, which means higher electrical bandwidth and narrower spectral emissions.

\gls{OWC} technologies include \gls{FSO} communication, \gls{UV} communication, \gls{OCC}, \gls{VLC} and \gls{LiFi}. \Gls{FSO} communication systems provide point-to-point links over relatively large transmission distances and are mainly used for enabling low-cost wireless backhaul links  indoors and outdoors.  \gls{UV} communication systems provide  long-distance \gls{NLoS} wireless connectivity through atmospheric scattering. \gls{UV} links are robust to link blockage and, thus, do not require perfect alignment between the transmitter and the receiver.  \gls{OCC} utilizes  cameras   as receivers and operates in the visible light spectrum. \gls{OCC} provides connectivity for  low-rate, short-range \gls{LoS} applications. \gls{VLC} also  utilizes visible light for wireless connectivity by employing  a \gls{PD} as a receiver and offers  high-speed, short-range \gls{LoS} connectivity. The term \gls{LiFi} refers to a network solution that utilizes wireless optical links, mainly \gls{VLC} and \gls{IR}, to provide  bidirectional  connectivity with  mobility support and seamless coverage.

\gls{LiFi} arises as a promising solution to add a new dimension to spectrum heterogeneity in future wireless networks \cite{7360112,HAAS2020443}. A \gls{LiFi} network constitutes multiple interconnected short-range optical \glspl{AP}, referred to as attocells. Each attocell serves a small number of users within a coverage area of a few  square meters. The  high directionality  of  light signals  in \gls{LiFi} networks allow for extreme cell densification and, hence, provide an effective solution to wireless coverage, and thus ultra-high capacity, in extremely dense deployments.

\gls{LiFi} is envisioned to play a significant role in future  cellular networks and \gls{IoT} applications, including smart healthcare provisioning, smart infrastructure management, high-precision manufacturing,  self-driven vehicles and remote robots operations, to count a few. These applications rely on  smart autonomous  operations and reliable high-speed connectivity that facilitate real-time interactions between different entities. Until recently, enhancing the communication capabilities of \gls{LiFi} systems was primarily focused on the development of transmission and reception capabilities in the face of undesirable uncontrollable channel conditions. However, breakthrough advancements in programmable meta-surfaces resulted in a paradigm shift in the way wireless signal propagation is dealt with: from fully uncontrollable to tunable and customized wireless channel engineering. The recent rise of revolutionary \gls{IRS} technology means that the environment itself can be programmed in order to enhance the performance of wireless communication systems. This enhancement comes in terms of spectral efficiency, energy efficiency, link reliability and security. Specifically, an intelligent surface constitutes a number of meta-surfaces that are artificially engineered in order to allow full manipulation of the incident \gls{EM} waves. The \gls{EM} response of these meta-surfaces can be altered on a macroscopic level to control the amplitude and phase of the incident beams. Based on this, it is possible to effectively control wireless signals to achieve the desired performance gains. 

This chapter discusses the use of \gls{IRS} or ``intelligent walls" in the context of \gls{LiFi} systems. Specifically, we discuss the background  of this emerging technology and present some of its applications in Section \ref{sec:background}. A case study for high-performance \gls{IRS}-enabled \gls{LiFi} systems is presented in Section \ref{sec:casestudy}, while Section \ref{sec:challenges} discusses some of the related challenges and future research directions.  

\section{Optical IRS:  Background \& Applications} \label{sec:background}
In this section, we provide a comprehensive literature review of the integration of \glspl{IRS} in \gls{OWC} and shed light on the  different applications of this emerging concept.  

\subsection{IRS from the Physics Perspective}
The \gls{EM} response of a surface is typically determined by the material it is made of as well as its geometry. Surface reflection behaviour can be classified into one of three responses: specular, diffuse, or glossy. Perfectly smooth surfaces act as mirrors and reflect light in a specular manner  according to Snell’s law of reflection.  Rough surfaces, on the other hand, scatter the incident light in all directions. Surfaces with a glossy nature reflect light in a way that contains both specular and  diffuse components. 

\begin{figure}[h]
	\centering
	\includegraphics[width=1\columnwidth]{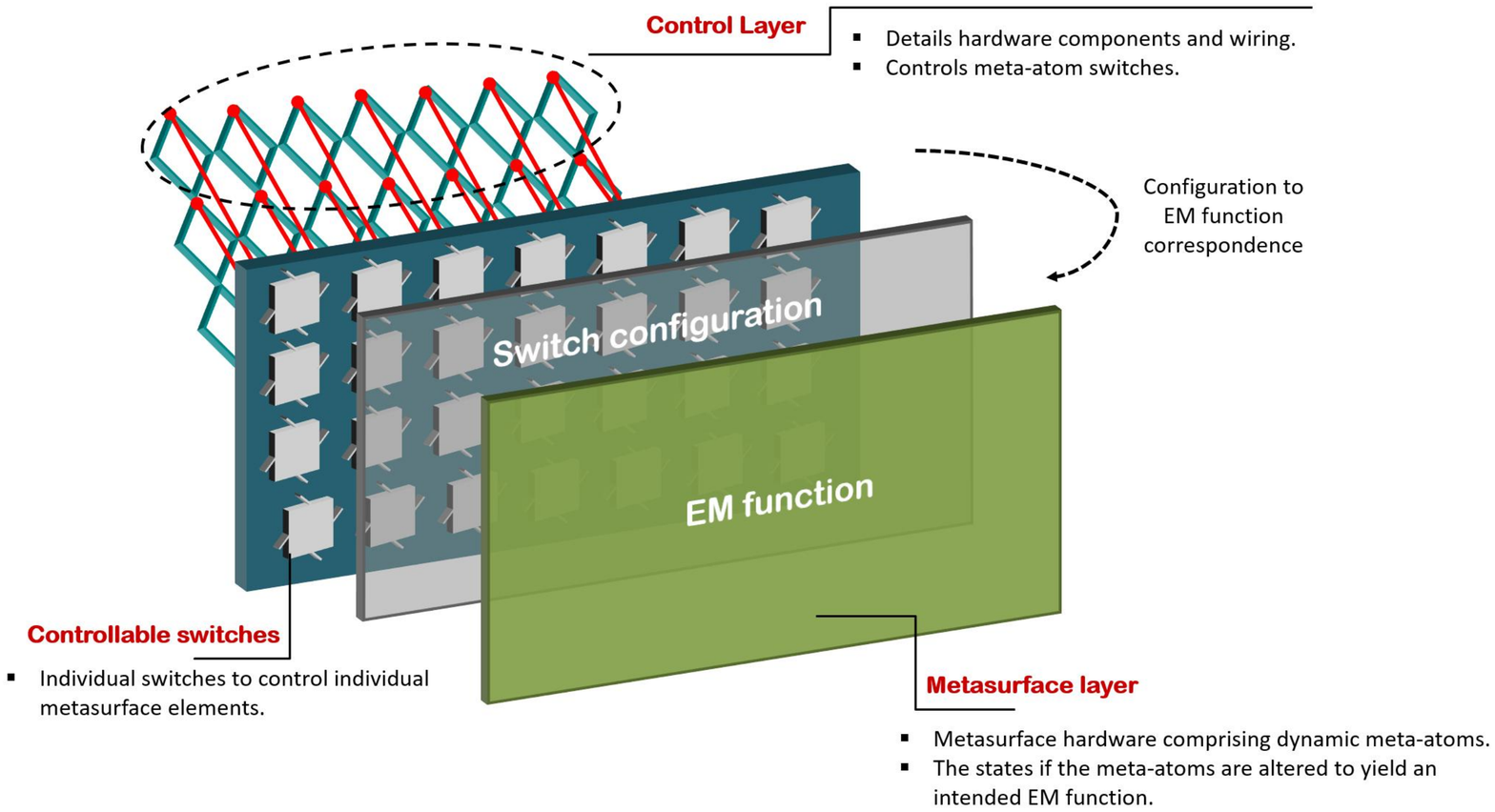}
	\caption{General architecture of an IRS \cite{8466374}.}
	\label{fig:RIS_arch}
\end{figure}

The term \gls{IRS} refers to a surface containing an array of periodically arranged metasurface elements that are engineered to produce a  controlled response to impinging light signals. The \gls{EM} response of each element  of the \gls{IRS} array can be adjusted by tuning the surface impedance through electrical voltage stimulation, which can be controlled via \gls{FPGA} chips. Various \gls{IRS} designs have been presented in the  literature using different numbers of layers as well as different materials, including liquid crystal,  meta-lenses with artificial muscles, doped semiconductors and electromechanical switches. The fundamental architecture of an \gls{IRS} is depicted in Fig. \ref{fig:RIS_arch}. Since the physical layer characteristics  of \glspl{IRS} can be controlled by software, they are also termed \glspl{SDS}. Although the use of metasurfaces in \gls{OWC} have only recently became an active area of research, their capabilities in  light manipulation have already been tested and developed in the field of flat optics \cite{Rubin:21, Zaidi:21}.

Reconfigurable metasurfaces are two-dimensional artificial periodic structures that consist of programmable  space-variant sub-wavelength metallic or dielectric elements,  known as ``meta-atoms". The most distinct feature  of metasurfaces is that their  \gls{EM} characteristics  can be reconfigured  in order to  introduce an  engineered response to the  incident wave-front by  manipulating   the outgoing photons. 
Based on this, the electric and magnetic properties of these structures can be engineered in order to   effectively  control  the  key properties of \gls{EM} waves, namely: amplitude, phase and  polarisation. Based on this,  two categories of phenomena can be observed, namely:   direction-related phenomena and intensity-related phenomena. Direction-related phenomena steer the propagation of the rays through scattering,  reflection,  and  refraction, while intensity-related phenomena affect  the signal power in the form of amplification,  attenuation,  or absorption. 
In the following, we list four of the most important light manipulation functionalities that can be realized by metasurfaces: 
\begin{itemize}
\item \textbf{Refractive Index Tuning:} the refractive index of metasurfaces can be tuned to change the behaviour of light rays as they pass through the material. Controlling  the imaginary  part of the refractive index affects how the material amplifies or absorbs the light. The real part can be positive or negative. With a negative  refractive index,  it is possible to reverse the  phase velocity of the light-wave and bend it in a direction that is impossible with
a positive index \cite{Suzuki:18}. It is also possible to realize near-zero index materials which allow perfect wave transmission in one direction \cite{refractive}. 
 
\item \textbf{Anomalous Reflection:} metasurfaces make it possible to break the law of reflection and reflect the light with an angle that is different to the angle of incidence. Based on this, the light rays impinging on a metasurface element can be 
steered   into desired directions \cite{reflection}. This is done by employing different reflecting phases at different meta-elements in the surface to tune the phase distribution over the metasurface so 
that the reflected waves interfere constructively in the desired
direction.

\item \textbf{Signal Amplification and Attenuation: } metasurfaces can be tuned so as to provide light amplification, attenuation or even complete absorbance \cite{9354893}. The amplitude control can be achieved by varying the conversion efficiency of each of the meta-elements via structurally birefringent meta-atoms.

\item \textbf{Wavelength Decoupling: } metasurfaces offer the possibility of  engineering  a wavelength-specific \gls{EM} response to decouple and independently  control different wavelengths \cite{OAM}. Based on this, it is possible to allow signals within a specific wavelength range to be reflected in a certain direction while absorbing signals with other wavelength values as in the  multiplexer/de-multiplexer metasurface in \cite{OAM} which decomposes the light signal into multiple channels. 
\end{itemize}
\begin{figure}[h]
	\centering
	\resizebox{1\linewidth}{!}{\includegraphics{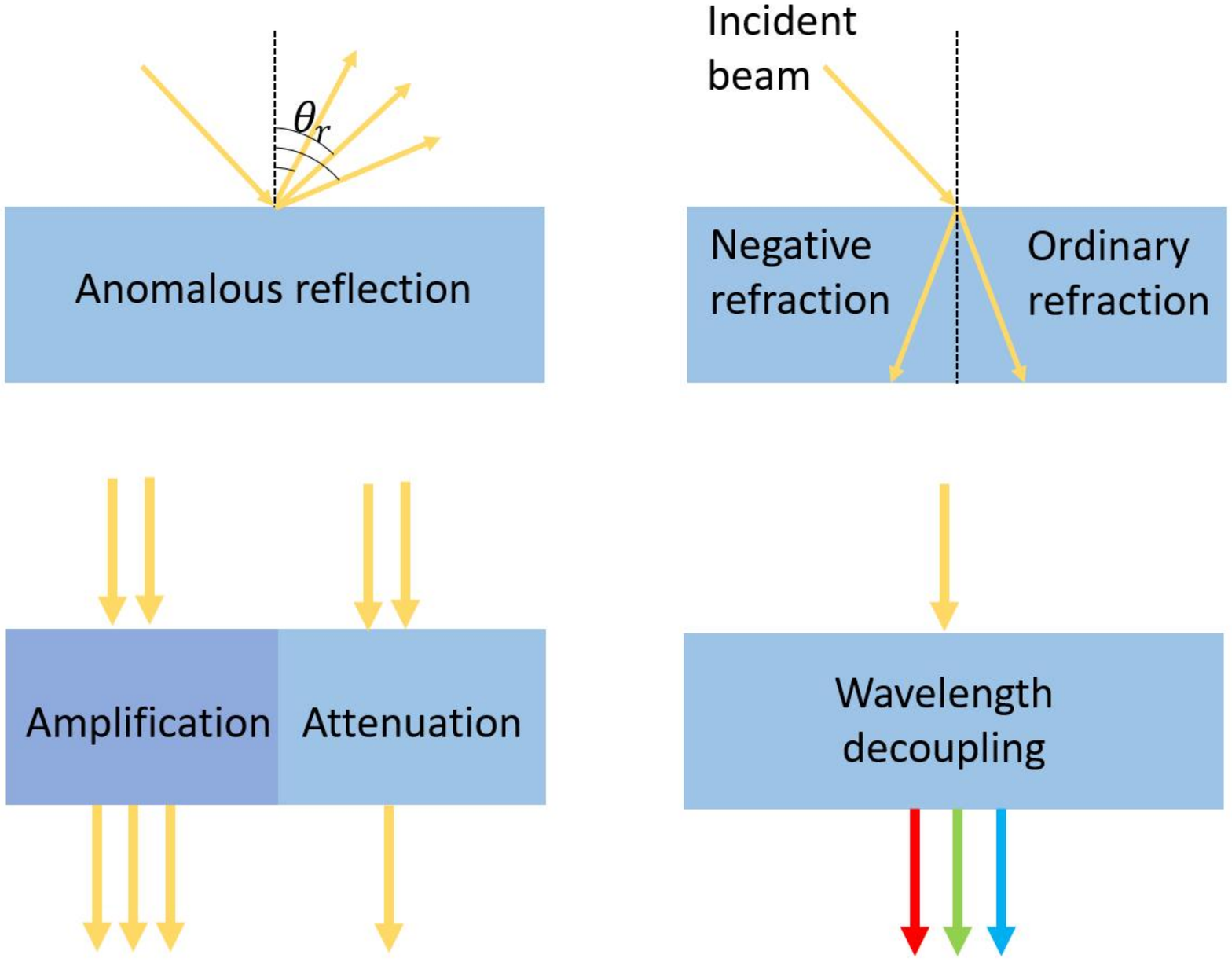}}
	\caption{Metasurfaces light manipulation functionalities. }
	\label{fig:functionalities}
\end{figure}
 \begin{comment}
 \begin{figure}
     \centering
     \begin{subfigure}[b]{\linewidth}
         \centering
         \includegraphics[width=0.5\textwidth]{figures/Refl.pdf}
         \caption{Efficiency of anomalous reflection (the dots represent various experimental  results and the line represents numerical estimation) as reported in \cite{reflection}.}
     \end{subfigure}
     %\hfill
     \begin{subfigure}[b]{\linewidth}
         \centering
         \includegraphics[width=0.5\textwidth]{figures/Ampl.pdf}
         \caption{Transmittance versus externally applied electric field for different liquid crystal  molar concentration in weighted percentage (wt\%) \cite{9354893}.}
     \end{subfigure}
     \hfill
     \begin{subfigure}[b]{\linewidth}
         \centering
         \includegraphics[width=0.5\textwidth]{figures/Atten.pdf}
         \caption{Transmittance versus molar concentration for different types of liquid crystal based metasurfaces \cite{9354893}.}
     \end{subfigure}
        \caption{Some of the reported capabilities of light manipulation functionalities.}
        \label{fig:capabilities}
\end{figure}
\end{comment}
Light manipulation functionalities are illustrated in Fig. \ref{fig:functionalities}. Although the use of metasurfaces in \gls{OWC} have only recently become an active area of research, their capabilities to realize light manipulation have already been tested and developed in the field of flat optics.  
%Typical metasurfaces consist of space-variant sub-wavelength resonators that introduce a gradual or abrupt engineered response to the  incident wave-front and thus manipulate the outgoing photons. 
Most of the research efforts in metasurface fabrication focus on engineering the phase profile of a wave-front since it is a  critically important design feature for many applications. An example of metasurface-enabled  optical phase control is the beam deflector presented  in \cite{ARX2016}. The beam deflector consists of a  super-cell with fifteen nano-disks, and the diameter of each of the nano-disks can be changed to  produce a phase shift from $0$ to $2\pi$ with $\pi/7$ phase increments. The proposed design offers full $2\pi$-phase control and a near-unity transmission efficiency at an operating wavelength of $715$ nm.  
Widening the scope of metasurfaces applications  requires the control of more than just the phase of the waveform. For this reason,  research efforts have pushed towards achieving simultaneous control of multiple signal parameters. For example, a metasurface that can control both the amplitude and phase of light signals was proposed in \cite{OSM2019}. The amplitude is controlled by varying the conversion efficiency of the  meta-atoms, while the phase is
controlled by altering the in-plane orientation of the meta-atoms. These two degrees of freedom enable complete and independent control of the  amplitude and phase of the optical signal. 
Similarly, the work in \cite{AHB2015} showed  the possibility of achieving independent phase and polarisation control with  sub-wavelength spatial resolution and an experimentally measured efficiency of  up to $97\%$. 
Another case of combining multiple functions in one metasurface is the multiplexer/de-multiplexer demonstrated in \cite{OAM} in which wavelength, polarisation,  and space multiplexing capabilities were  integrated into one single ultra-thin metasurface in order to decompose the light signal into multiple channels.

The aforementioned examples demonstrate that it is possible to effectively  manoeuvre the physico-chemical characteristics of metasurfaces in the visible light spectrum, which opens the door for many \gls{IRS} applications in  \gls{OWC} systems, as will be discussed in the following sections.  

\subsection{IRS Applications in OWC}
In the following, we discuss some potential applications for the integration of \gls{IRS} in \gls{OWC} systems. 

\subsubsection{Reflection for Blockage Mitigation}
Optical signals do not  penetrate through solid objects and they typically exhibit increased  attenuation and less diffraction compared to \gls{RF} signals. As a result, 
the optical link quality primarily relies on the existence of a \gls{LoS} path between the transmitter and the receiver. If the \gls{LoS} path is obstructed, i.e., due to the existence of a blocking entity,  it is highly   likely that the communication link quality will significantly deteriorate.  Due to the dependency on \gls{LoS}, \gls{OWC} systems typically require perfect alignment between the transceiver front-ends so as to ensure a reliable transmission. 
Ensuring such alignment is easier in  cases with no or low mobility, i.e., in \gls{VLC} systems with static users.  However, the case is more complicated in \gls{LiFi} networks which are required to fully support high user mobility and ubiquitous connectivity.  One of the key challenges in  such networks  is that the user terminal randomly changes its position and orientation which could lead to obstruction of the \gls{LoS} link between the user and the \gls{LiFi} \gls{AP}.

Along with the \gls{LoS} component,   receivers also perceive multi-path components that result from the  diffusion and reflection of light signals throughout the surrounding environment. Since both the direct \gls{LoS} signal and the reflected signals carry the same data, they can be  added at the receiver to maximize the total received power. In a typical indoor environment, however, \gls{NLoS} signals are totally uncontrolled and nearly isotropic in their spatial distribution and are,  therefore,  weak at their average intensity.   As a result, the effect of  multi-path reflections on the received signal intensity is typically negligible \cite{kwzz1801}. The use of \gls{IRS} mounted on the walls of indoor spaces with \gls{LiFi} coverage means that the reflection of the light signals could possibly be controlled so as to add significantly to the received signal at the mobile device and compensate for the loss of the \gls{LoS} component in the event of a blockage. Moreover, using \gls{IRS} in indoor \gls{LiFi} systems can enhance the coverage of the \gls{LiFi} \glspl{AP} by mitigating the dead-zone problem for cell-edge users. Since the coverage of an indoor \gls{LiFi} \gls{AP} is typically confined within the beam width of the transmitted light, non-uniform coverage distribution is typically exhibited. The reduced link reliability at the cell edge means that users moving in the proximity of the \gls{LiFi} \glspl{AP} will need to perform frequent horizontal handovers, i.e. between neighbouring \gls{LiFi} \glspl{AP}, or vertical handovers, between the \gls{LiFi} network and the \gls{RF} network. With the use of wall-mounted \glspl{IRS}, users close to the wall can receive strong reflected signals that allow for meaningful \gls{SNR} \cite{9614037}. This implies that a cell-edge user in the vicinity of the \gls{IRS} can still connect to the \gls{LiFi} \gls{AP} despite having a weak \gls{LoS} path.

The concept of using wall-mounted \gls{IRS} in indoor \gls{OWC} systems was investigated in \cite{9276478}. Two types of \glspl{IRS}, namely: intelligent metasurface reflectors  and intelligent mirror arrays reflectors, were studied.  
For the metasurface reflectors, a macroscopic model for the metasurface elements is adopted  to abstract them as anomalous reflective rectangular blocks. It is assumed that the phase gradient is kept constant over each metasurface element and that the phase discontinuity of each metasurface element  can be tuned independently of the other elements to reflect the incident light in a specific direction, i.e., the detector's center. For mirror array reflectors, identical rectangular mirrors with two rotational degrees of freedom are used. In the aforementioned work, it was shown that the focusing and steering capabilities of \gls{IRS} are proportional to the number of reflecting elements up to a specific number of elements. Furthermore, it was shown that using a $25$ cm × $15$ cm reflector on the wall can enhance the received power by up to five times compared to the direct \gls{LoS} link. The presented results imply that  \gls{IRS} technology provide a promising solution to compensate for \gls{LoS} blockage in \gls{LiFi} systems, particularly when the communication link is susceptible to blockage because of the  movement of the user device or other objects.  In  other words, by leveraging \gls{IRS} we  can  enable a high-performance operation  for  mobile devices under  mobility conditions and link blockages as well as random device orientations, making \gls{LiFi} resilient to these probabilistic factors.

\subsubsection{Enhanced Optical MIMO}
\Gls{MIMO} configurations  are used to provide  wireless communication systems with multiple transmit/receive paths so as to  enable the  simultaneous transmission of multiple streams of data signals. With the application of appropriate signal processing, \gls{MIMO} systems can enhance spectral efficiency and received signal quality.  The application of \gls{MIMO} systems in \gls{OWC} is an attractive solution  due to the existence of large numbers of \glspl{LED} in a single luminaire and  many luminaires in various indoor settings. In the context of \gls{OWC}, \gls{MIMO} systems can be achieved by  deploying multiple \gls{LED}s on the \gls{AP} side and/or multiple \gls{PD}s on the user side \cite{7314864}.

While \gls{MIMO} configurations have the potential to improve the performance of  \gls{OWC} systems, the achieved performance enhancement is highly dependant on the condition of the \gls{MIMO} channel matrix. If the \gls{MIMO} channel has a strong spatial correlation, the rank of the channel matrix will be low. In a multiplexing \gls{MIMO} system, this means that  only a few \gls{MIMO} sub-channels are actually usable while  the remaining sub-channels are deemed unreliable for transmitting and receiving data. In a diversity \gls{MIMO} system, the high channel correlation implies that the   characteristics of different diversity sub-channels are very similar. Consequently, if one of them is unreliable, the remaining sub-channels are likely to be unreliable as well, which diminishes the diversity gain. In a \gls{SM} system, part of the information signal is carried on the index of the activated \gls{LED} \cite{ybth1901,yppph2001}. As a result, high channel correlation  makes it hard to distinguish the difference between various spatial symbols, which results in high \glspl{BER} \cite{ybmph1701,ycphp1801}. As a result, it is clear that establishing a well-conditioned \gls{MIMO} channel matrix in \gls{OWC} systems is essential to ensure that various \gls{MIMO} configurations can be utilized to achieve the desired advantages. 

The \gls{MIMO} channel conditions in \gls{OWC} are different to \gls{RF}. The \gls{RF} wireless channels  comprise multiple \gls{NLoS}  paths with small-scale fading, which results in \gls{MIMO} channels with weak spatial correlation. On the other hand, \gls{OWC} channels do not exhibit fading  characteristics and rely mainly on front-end characteristics and positions. In the case of similar or symmetrical  positions of multiple \gls{PD}s with respect to the \gls{AP}, the corresponding \gls{MIMO} sub-channels will have a strong spatial correlation, which forms a major challenge in \gls{OWC} \gls{MIMO} systems \cite{fh1301,pytsh1901}. One of the possible solutions to tackle the high channel correlation in optical \gls{MIMO} is to carefully align the locations of the \gls{LED}s and the \gls{PD}s such that they create uncorrelated channel paths as was proposed in \cite{meh1101}. While this solution could be feasible in fixed point-to-point communications, it is not always possible in \gls{LiFi} systems that support high user mobility. 

\Gls{IRS} technology provides an effective solution to improve poorly conditioned optical \gls{MIMO} systems by creating controllable multi-path channels. The reflection coefficients of the \gls{IRS} elements can be optimized such that the rank of the \gls{MIMO} channel matrix is enhanced and the observed sub-channels are sufficiently distinguishable. The higher spatial decorrelation will enable  higher spectral efficiency and boost the achievable data rate of optical \gls{MIMO} systems \cite{9614037}. The rank improvement capabilities of \glspl{IRS} were studied in \cite{oz}, where a multiplexing gain was achieved even when the \gls{LoS} \gls{MIMO} channel is rank-deficient.  It is noted that enhancing the \gls{MIMO} channel rank requires joint optimisation of the reflection coefficients of the \gls{IRS} elements and the transmit precoding matrix, which is a non-convex optimisation problem \cite{9417270}.

\subsubsection{Media Based Modulation}
The concept of \gls{MBM} can be considered as a special case of  \gls{IM} \cite{4382913}. While conventional \gls{IM} systems employ multiple transmitters to carry distinct data symbols on the indices of the activated transmitters, \gls{MBM} can be realized  with a single transmitter aided with  \gls{IRS}. In \gls{MBM}, an \gls{IRS} array  can be used to provide  an additional dimension for data modulation by controlling the \gls{EM} response of each of its reflecting elements. \gls{MBM} offers some advantages  compared to  conventional \gls{IM} including lower cost and higher flexibility \cite{8758978,ris-mod}. 

The basic idea of \gls{MBM} is explained in the following section. The transmission system comprises of the transmitting \gls{AP} and an \gls{IRS}  equipped with multiple reflecting elements. Each reflecting element receives the data signal from the \gls{AP}  and reflects it to the intended receiver with a specific reflection coefficient. Different reflection coefficients  correspond to different observed channel realisations at the receiving \gls{PD}, so we can carry data bits on the state of the reflecting element. %As a result, controlling the reflection coefficients  of the reflecting elements can result in  a spectral efficiency enhancement of $N_r$ bits per symbol.  For an  $M$-ary modulated transmitted signal, the total achievable spectral efficiency  is $\log_2(M)+2^{\lfloor \log_2(R^{N_r}) \rfloor}$,  where $N_r$ is the number of \gls{IRS} reflecting elements and $R$ in the number of distinct reflection states. 
At the receiver, recovering the intended data symbols involves decoding the $M$-ary modulated symbol as well as  estimating the channel state of each \gls{IRS} reflecting element. This necessitates the knowledge of the \gls{CSI} of all the reflecting elements which could be obtained using passive pilot based channel estimation in which each \gls{IRS} passively reflects the pilot sequences sent by
the user to the \gls{AP} to estimate its channel coefficients \cite{9120452}.

In theory, the spectral efficiency of \gls{MBM} can be enhanced by increasing the number of \gls{IRS} reflecting elements, as well as the number of possible reflection states. However, the enhancement is practically limited by  the  degree of decorrelation between the sub-channels created by different \gls{IRS} elements. This is because the  Euclidean distance between the  \gls{MBM} constellation points depends on the values of the possible channel realisations. As a result, the desired performance enhancement can be only realized if these combinations are clearly distinguishable at the receiving terminal.

\subsubsection{Enhanced Optical NOMA}
\Gls{NOMA} is a spectral efficient multi-user access technique that has been widely investigated for use in \gls{OWC} systems \cite{8713381,7972998,9400380}. In \gls{NOMA}, signals of different users are multiplexed in the power domain by assigning distinct power levels to different users' signals. This process is referred to as superposition coding  and the power allocation coefficients are usually  determined based on the users' channel conditions. The basic principle  is that users with more favourable channel conditions are allocated lower power levels than those with weaker channel conditions.  \Gls{SIC} is then performed at the receiver side  to decode and subtract the signals with higher power levels first until the desired signal is extracted. \gls{NOMA} has been shown to provide promising performance gain in \gls{OWC} due to the high \gls{SNR} and the somewhat deterministic nature of the optical  channel, which facilitates  the acquisition of \gls{CSI} of the users with less overhead compared to \gls{RF} fading channels \cite{8352627}.

The operating principle of \gls{NOMA} is based on having  distinct channels for different users. However, this is not always guaranteed in \gls{OWC} due to the high correlation between the optical channels. In fact, the characteristics of the optical channel imply that it is likely that multiple users experience  similar channel conditions, hindering the feasibility of  \gls{NOMA}. \gls{IRS} technology offers the possibility to overcome this limitation and create distinct channels for  different users by dynamically  altering  the perceived channel gains. Moreover, by carefully designing the reflection coefficients of the \gls{IRS} elements, it is possible to create better conditions for effective power allocation which, in turn, leads to successful \gls{SIC} and high reliability. Moreover, \gls{IRS}-enabled \gls{NOMA} offers  enhanced user fairness compared to conventional \gls{NOMA} systems. This is because, in conventional \gls{NOMA}, the decoding order of a user depends on its channel gain value compared to the rest of the users. To this effect, users with a lower decoding order, i.e. a lower channel gain, must always decode their signals with the existence of interference from users with a higher decoding order. Dynamic \gls{IRS} configuration means that it would be possible to change the users decoding order despite their locations, leading to enhanced fairness \cite{9120476}. The work in \cite{abumarshoud2021intelligent} proposed a framework for the joint design of the \gls{NOMA} decoding order, power allocation, and the  reflection coefficients in an \gls{IRS}-assisted \gls{NOMA} \gls{VLC} system. The NP-hard multi-dimensional optimisation problem was solved by a heuristic technique with the objective of enhancing the \gls{BER} performance. The presented results showed that optimized \gls{IRS} leads to  higher link reliability, particularly when the links are subject to the adverse effects of link blockage and random device orientation.

\subsubsection{Enhanced PLS}
The fact that light signals cannot penetrate through opaque objects makes \gls{OWC} systems particularly secure in confined spaces, provided that eavesdroppers are located outside the room \cite{PLS,9524909,pychp2001,ycesphp2001,spkyph2101}. This means that \gls{OWC} systems provide inherent \gls{PLS} compared to \gls{RF} systems where the signals can easily be intercepted by eavesdroppers from behind a wall. Nevertheless, \gls{OWC} links are susceptible to eavesdropping by malicious users that exist within the coverage area of the \gls{AP} of interest.

\gls{IRS} technology offers a solution to enhance the security of such systems by implementing \gls{IRS}-assisted \gls{PLS} techniques. Several adaptive methods can be utilized to tune the optical properties of the environment so as to enhance the link reliability for legitimate users while degrading the reception quality for  potential eavesdroppers. Moreover, the \gls{IRS} elements can be tuned to produce friendly jamming signals by creating  randomized multi-path reflections  to produce artificial noise at the eavesdropper without affecting the transmission of the legitimate user.  Another possibility to use \gls{IRS} to boost the \gls{PLS} is to employ secure beamforming in \gls{MIMO} systems. Multiple reflecting elements can be configured to produce a precoding matrix so that the signal can be decoded only at the intended receiver location to reduce the probability of signal interception. 

In order to realize \gls{IRS} based \gls{PLS},  the secrecy performance of such systems, i.e., secrecy capacity and secrecy outage probability, must be analyzed in relation to   the locations and capabilities of \gls{IRS}. The work in \cite{qian2021secure} proposed a framework for maximising the data security of an \gls{IRS}-assisted \gls{VLC} system. An array of wall-mounted mirrors was utilized and the orientation of each mirror was adjusted so as to maximize the achievable secrecy capacity. It was found that the \gls{IRS} channel gain is highly affected by any change in the mirrors' orientation, which makes it hard to find the optimal combination of orientation angles. In order to overcome  this issue,  the orientation optimisation problem was converted into a reflected spot position finding problem. As a result, the secrecy capacity can be maximized by finding the optimal position of the reflected spots for each mirror, which greatly reduces the complexity. 
 
%##########################################################################################################
\section{Case Study: High Performance IRS-Aided Indoor LiFi} \label{sec:casestudy}
\subsection{Channel Modelling} 
In this section, the adopted methodology for obtaining the channel properties of an indoor \gls{LiFi} network, where the system performance is enhanced via the deployment of \glspl{IRS}, will be presented. Then, the effect of \glspl{IRS} on the optical link budget, \gls{CIR}, \gls{CFR} and other important channel parameters will also be detailed.

Unlike conventional \gls{RF} based wireless communication systems, the operation wavelengths are in the nm region in \gls{LiFi}, which makes the wavelengths comparable with atoms and molecules that comprise the surrounding materials. The penetration characteristics of the \gls{EM} propagation in the \gls{VL} and \gls{IR} spectra become weak due to this relative comparability. As a result, the optical channel becomes highly dependent on the surface geometry as well as the coating characteristics. In order to capture the optical channel in a realistic manner, we propose a non-sequential \gls{MCRT} based channel modelling technique, which is able to capture the realistic \gls{TX}, \gls{RX}, environment geometry, and coating material characteristics. Note that the simplest implementation of an indoor \gls{LiFi} system is to utilize off-the-shelf \glspl{LED} and \glspl{PD} at the \gls{TX} and \gls{RX}, respectively. Thus, the information will be transmitted from the \gls{TX} to \gls{RX} incoherently by using the subtle changes encoded onto instantaneous light intensity, which could be detected at the \gls{RX}. Consequently, the measurement of incoherent irradiance for given scenarios will be the main target in our simulations.
\begin{figure}[!t]
	\centering
	\includegraphics[width=0.8\columnwidth]{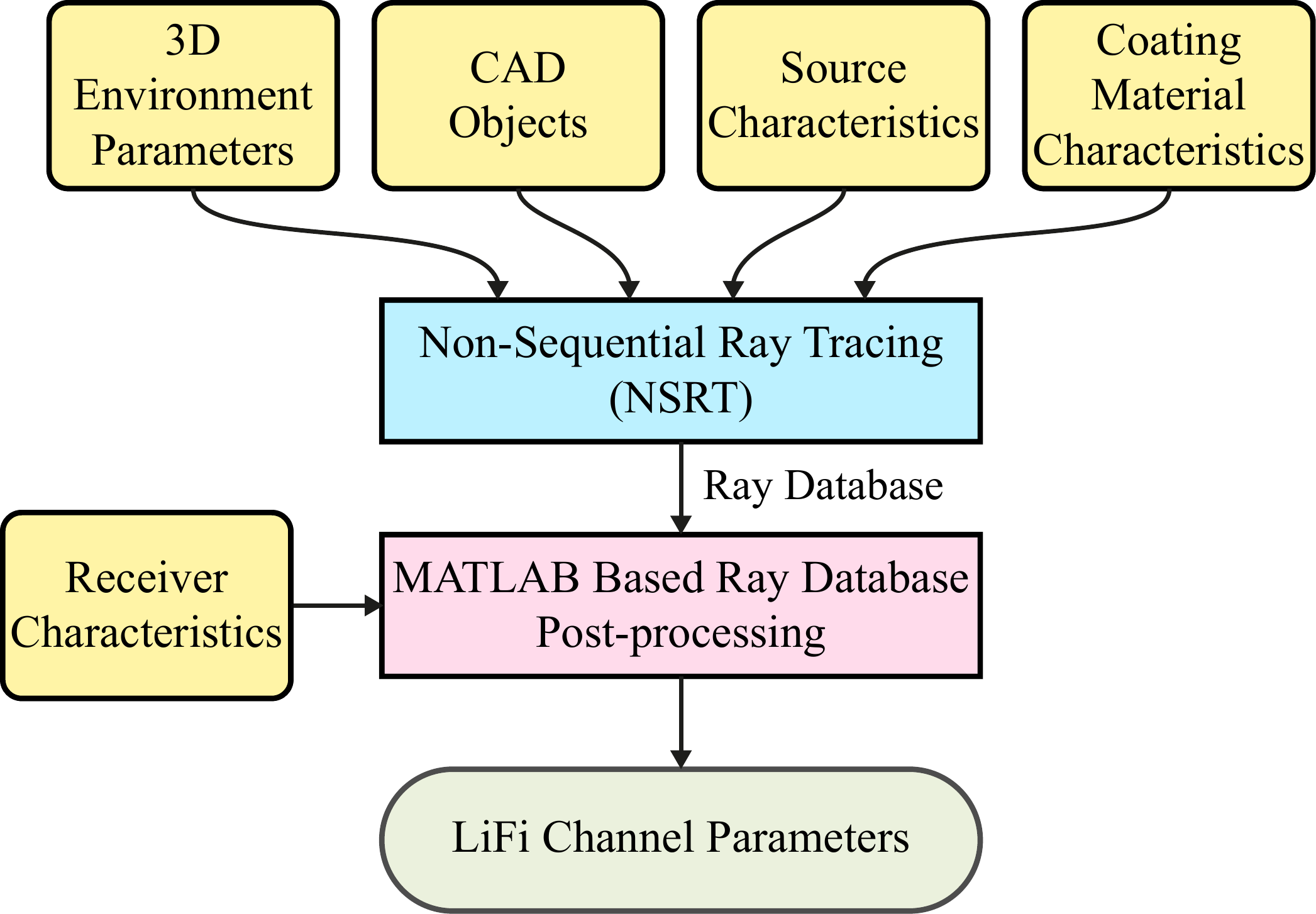}
	\caption{Block diagram for MCRT based LiFi channel modelling environment.}
	\label{fig:block_diagram}
\end{figure}

The proposed \gls{MCRT} simulation toolbox, which is depicted in Fig. \ref{fig:block_diagram}, amalgamates the geometrical \gls{NSRT} capabilities of Zemax OpticStudio version 21.3.2 \cite{zemaxopticstudio} with our custom MATLAB based computation and post-processing libraries. The accuracy of the ray-tracing in Zemax Opticstudio for only \gls{LoS} and \gls{LoS}-plus-\gls{NLoS} cases have been reported to match the real-world measurements closely by \cite{80211bb_2,eumhhj2001,eumhhj2002,emjhhu2101}. Accordingly, the \gls{MSE} between the Zemax OpticStudio simulations and real-world measurements of $0.7\%$, $1\%$ and $1.5\%$ are achieved for \gls{LoS}, only first-order \gls{NLoS} and higher-order \gls{NLoS} reflection scenarios, respectively \cite{80211bb_2}. Therefore, a flexible and highly accurate optical channel modelling compared to the recursive method \cite{bkklm9301,barry94} based calculations is obtained. It is important to note that the recursive method lacks the ability to model complex geometrical shapes, sources, receivers and coating materials, which is crucial in our meta-material based \gls{IRS} application. As can be seen from Fig. \ref{fig:block_diagram}, the environment parameters, \gls{CAD} object models, as well as source and coating material characteristics are inputted into our simulation environment. Since the main purpose of  \gls{LiFi} is to achieve broadband information transmission and illumination by using non-imaging off-the-shelf components simultaneously, the \gls{NSRT} method is adopted in our simulations.

\subsubsection{Generation of the Indoor Environment}
\begin{figure}[!t]
	\centering
	\includegraphics[width=0.85\columnwidth]{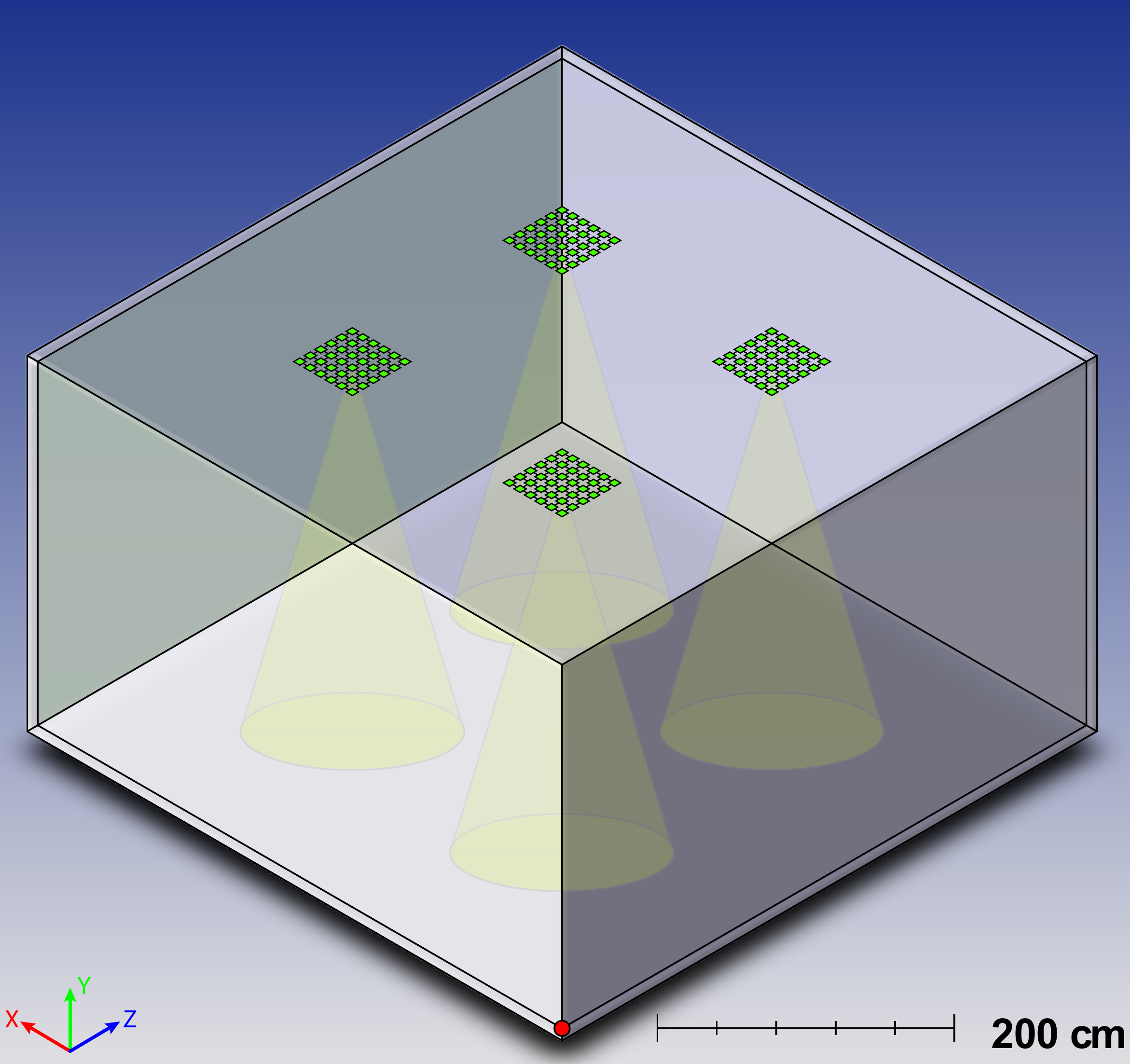}
	\caption{Isometric view of the considered scenario. The global origin point of the simulation environment is indicated by the red point.}
	\label{fig:system_description_1}
\end{figure}
A typical indoor \gls{LiFi} application scenario within a cuboid-shaped room, which represents a typical indoor room structure for homes, offices, hospitals etc., with dimensions of $5\times 5 \times 3~\text{m}$ is adopted in this work. In Fig. \ref{fig:system_description_1}, the scenario under investigation is depicted, where four luminaires serve as optical \glspl{AP}. Each optical \gls{AP} is placed at the corners of a square, with sides measuring $2$ m, that is located at the center of the ceiling. To harvest the benefits of the \gls{IRS} functionality, the side walls are assumed to be covered with meta-material based wallpaper, which is depicted only for a single wall in Fig. \ref{fig:system_description_1} for the sake of simplicity. Thus, the reflection characteristics of the side walls will be altered dynamically by using \glspl{IRS} to enhance the achievable \gls{SNR} as well as the average error performance of a mobile \gls{UE}.

The global positions and orientations of the transmit \gls{LED} array based luminaires, receive \glspl{PD} and \glspl{IRS} are defined by $3\times 1$ vectors, which are defined \gls{w.r.t.} the global origin point $O(0,0,0)$. Note that the global origin of the system is depicted by the red dot in Fig. \ref{fig:system_description_1}. The location vector of each element could be given in the format of $\mathbf{v}=(v_x,v_y.v_z)$, where $x$, $y$ and $z$ axis components of the vector are denoted by $v_x$, $v_y$ and $v_z$, respectively. Similarly, the orientation vectors of each object are given by $\mathbf{o}=(o_x, o_y, o_z)$, where the rotation \gls{w.r.t.} $x$, $y$ and $z$ axis, in other words, within the $yz$, $xy$ planes, are given by $o_x$, $o_y$ and $o_z$, respectively.

\subsubsection{Source Characterization}
The application of metasurfaces in \gls{LiFi} requires the development of optical models of the transmit \glspl{LED}, receive \glspl{PD}, and reflective surfaces. These models need to be  as realistic as possible to capture the system performance. The accurate selection of the operation wavelengths is of the highest importance since the achievable system performance could change based on the spectral response profiles of the transmitters, receivers, and coating materials of the walls, ceiling and floor.

The generic indoor luminaires in the illumination market typically  consist of a concave mirror, an \gls{LED} array as the base,  and a diffuser. Since our aim is to use the luminaires both for communication and illumination purposes, a $6\times 6$ \gls{LED} chip array is used as the base without the mirror and diffuser structures. To take both \gls{VL} and \gls{IR} band characteristics into consideration, an off-the-shelf OSRAM GW QSSPA1.EM High Power White LED \cite{gwqsspa1.em} and OSRAM SFH 4253 High Power IR LED \cite{sfh4253} chips with a continuous radiated spectrum between the limits of $0.383$ $\micro$m - $0.780$ $\micro$m and $0.770$ $\micro$m - $0.920$ $\micro$m, respectively, are adopted in our simulations. The origin of the luminaires are  assumed to be the center point of the \gls{LED} array. Each \gls{LED} chip within the luminaires is set to be radiating $1$ W optical power, which yields $36$ W per luminaire. The separation between each \gls{LED} chip in the array is $10$ cm, and the dimensions of the luminaires are chosen to be $60\times 60$ cm. It is important to note that both the nominal wattage and dimensions of the luminaires are chosen to replicate generic commercial \gls{LED} flat panel products \cite{LEDVANCE}.
 
\begin{figure}[!t]
	\centering
	\begin{subfigure}[t]{.5\columnwidth}
		\includegraphics[width=\columnwidth]{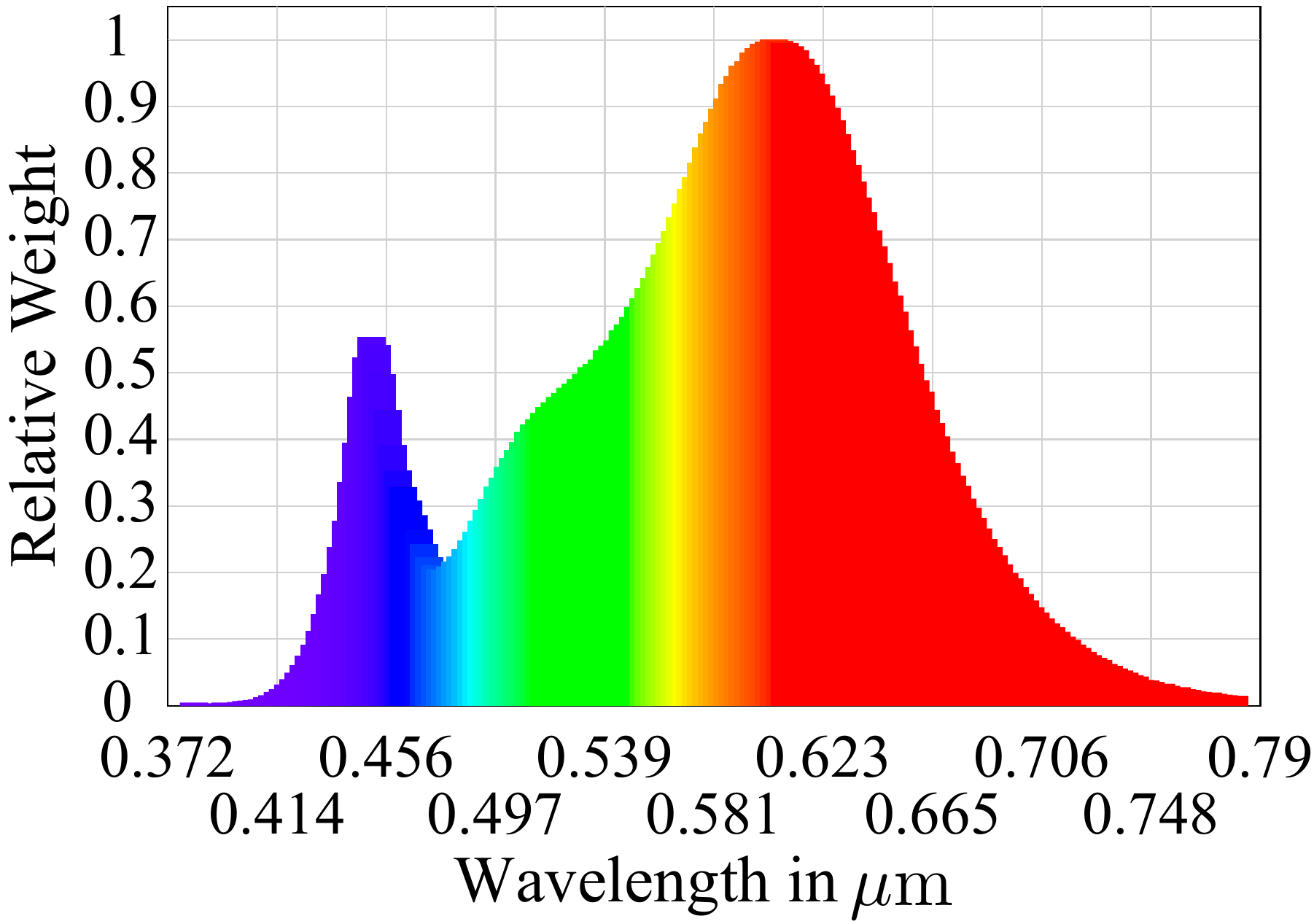}
		\caption{OSRAM GW QSSPA1.EM}
		\label{fig:spec_VL}
	\end{subfigure}~
	\begin{subfigure}[t]{.5\columnwidth}
		\includegraphics[width=\columnwidth]{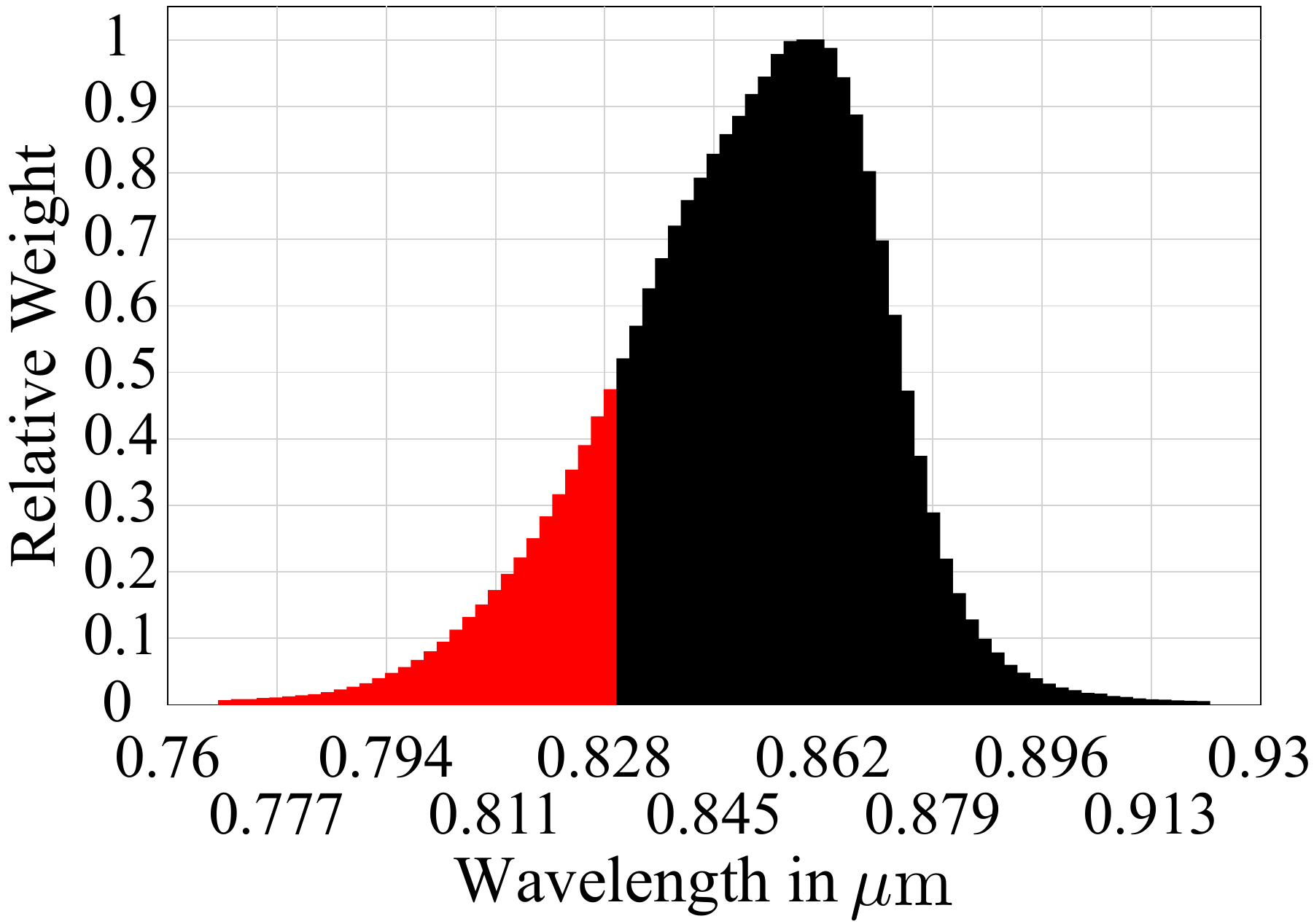}
		\caption{OSRAM SFH 4253}
		\label{fig:spec_IR}
	\end{subfigure}
	\caption{Relative radiometric color spectrum of; (a) \glsentrytext{VL} $[0.382~0.780]~\micro\text{m}$ and (b) \glsentrytext{IR} $[0.770~0.920]~\micro\text{m}$ band \glsentrytext{LED} chips that used in the MCRT simulations.}
	\label{fig:source_spectral}
\end{figure}

The realistic radiometric spectral characteristics of the sources are defined by using the spectrum files (.spcd) in Zemax OpticStudio within our simulation environment. Thus, the related spectral distribution values for the adopted sources are obtained after processing the data-sheet values, which report the real-world measurements results. Then, the spcd files are obtained accordingly and fed into the simulation environment, which contain the relative spectral distribution coefficients \gls{w.r.t.} the corresponding wavelength. The coefficients in the spcd file determine the density of a ray with the given wavelength among all the other rays with various wavelengths in \gls{MCRT}. Thus, the optical power dedicated to the given wavelength will become directly proportional to the measured relative coefficients in ray-tracing simulations, which will replicate the real source spectral characteristics in the simulation environment. The relative radiometric color spectrum plots for the adopted \gls{VL} and \gls{IR} sources are given in Figs. \ref{fig:spec_VL} and \ref{fig:spec_IR}, respectively. As can be seen from the figures, the spectrum distribution function of the white \gls{LED}, $f_\text{w}(\cdot)$, which is plotted against the wavelength, $\lambda$, has two peaks at $\lambda_1 = 0.450~\micro\text{m}$ and $\lambda_2=0.604~\micro\text{m}$. This is primarily due to the manufacturing process of the \glspl{LED}, where a yellow color-converting phosphor coating ($\lambda_2$) is superimposed onto the blue source ($\lambda_1$) to achieve white light emission. The \gls{CCT} of our adopted white \gls{LED} model becomes $3000$ K, which sits on the warmer side of the scale. 
\begin{figure}[!t]
	\centering
	\begin{subfigure}[t]{.5\columnwidth}
		\includegraphics[width=1\columnwidth]{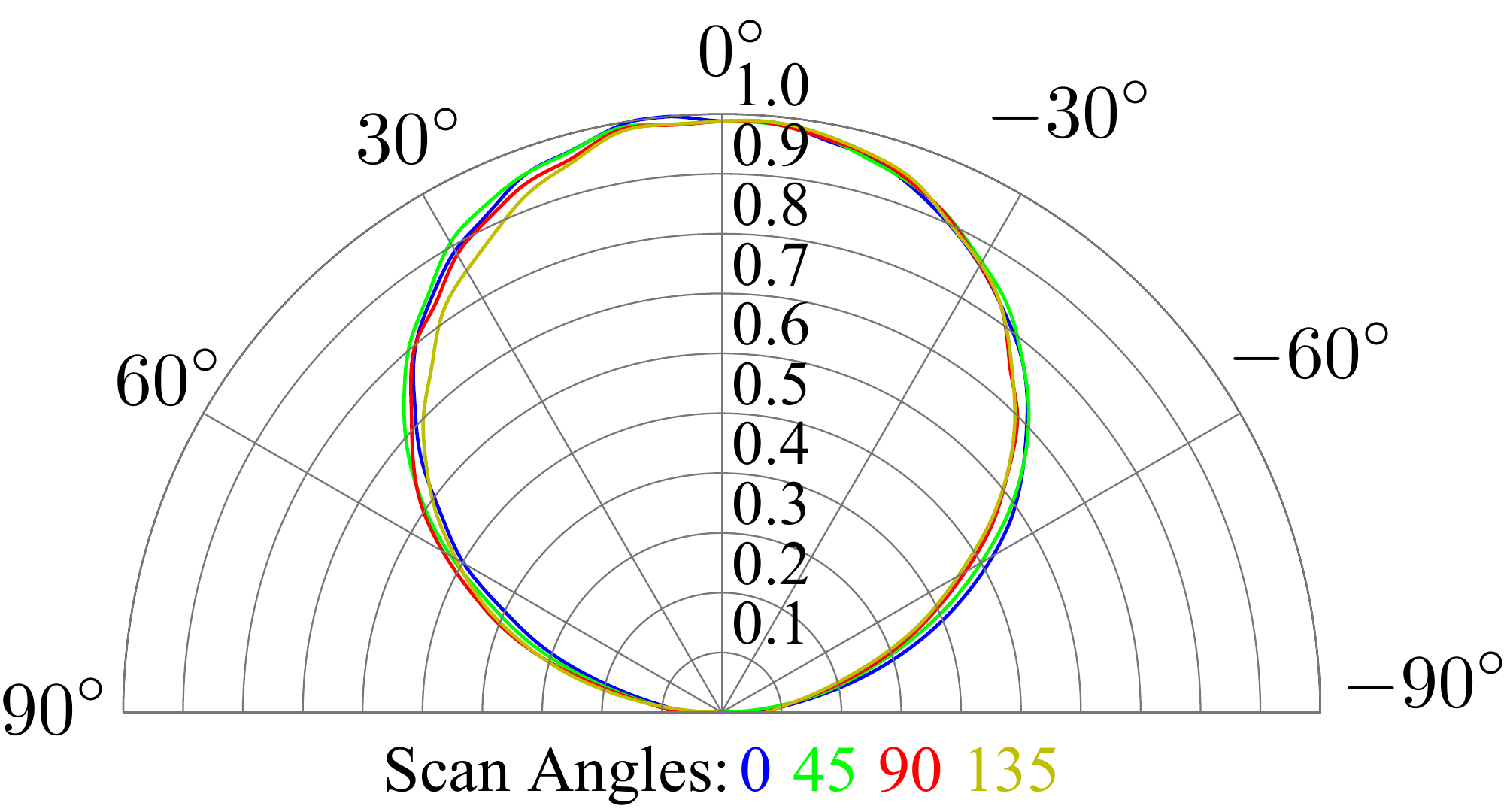}
		\caption{OSRAM GW QSSPA1.EM}
		\label{fig:direct_VL}
	\end{subfigure}~
	\begin{subfigure}[t]{.5\columnwidth}
		\includegraphics[width=1\columnwidth]{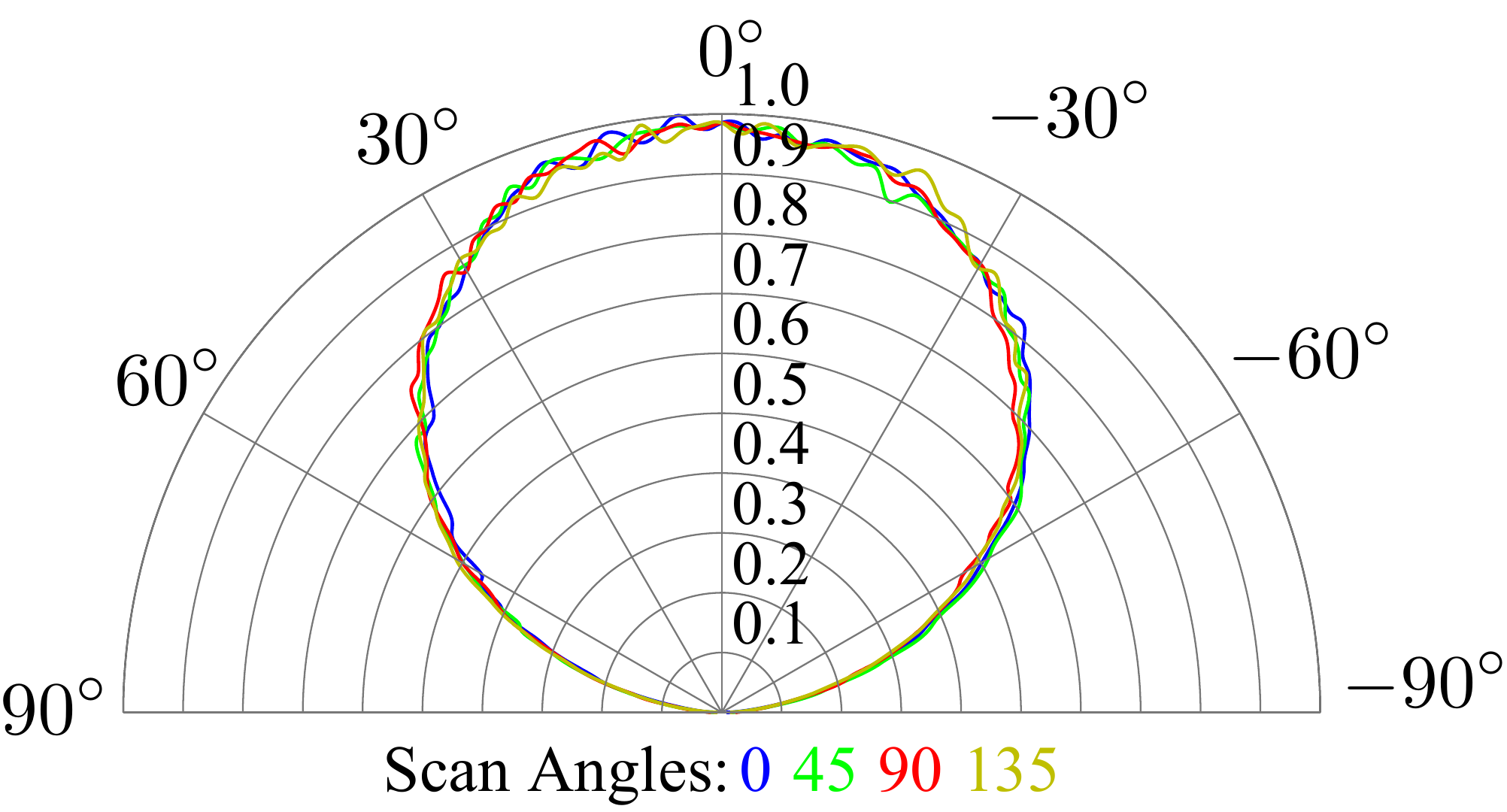}
		\caption{OSRAM SFH 4253}
		\label{fig:direct_IR}
	\end{subfigure}
	\caption{Source directivity plots of the \glsentrytext{VL} (left) and \glsentrytext{IR} (right) band \glsentrytext{LED} chips for the azimuthal angles $[0~ 45~ 90~ 135]$.}
	\label{fig:source_angular}
\end{figure}
Similarly, the relative radiometric distribution function for the \gls{IR} \gls{LED}, $f_\text{i}(\cdot)$, is also depicted against the wavelength in Fig. \ref{fig:spec_IR}, where there is a single peak value at the $\lambda_3=0.860~\micro\text{m}$. Another important parameter is the spectral spread of the adopted sources, where the bandwidth value for the \gls{IR} source of $0.150~\micro \text{m}$ shows intrinsic dominant monochromatic characteristics, where most of the power is concentrated within a relatively small range of wavelengths. On the contrary, the white source with the bandwidth value of $0.398~\micro\text{m}$ presents a widespread poly-chromatic profile. The importance of the source chromacity for the optical link budget will be clearer in the following subsections when the coating material reflectances are presented. In our simulation environment, each source profile is represented by $200$ data points, which is the maximum value that is permitted by the simulation environment.

The realistic spatio-angular profile of the sources are also obtained from the measurement results provided by the manufacturer. The source model files with a large number of ray recordings, which are obtained by real-world goniometer measurements, are utilized to represent the adopted source spatio-angular characteristics. The main advantage of this method is the ability to model complex sources without requiring knowledge of inner opto-electrical and quantum effects dominated working principles. The radiation patterns of the \gls{VL} and \gls{IR} band \glspl{LED} are given by the directivity plots in Fig. \ref{fig:direct_VL} and \ref{fig:direct_IR}, respectively. In these polar plots, the normalized radiant intensity distribution of the sources are plotted against the polar angle \gls{w.r.t.} a source located in the $-y$ direction, refer to Fig.\ref{fig:system_description_1}. Furthermore, the diferent colors in the plots represent the various azimuth angle scans, $0^\circ$, $45^\circ$, $90^\circ$ and $135^\circ$, for the spherical coordinate system. It can be inferred from the figure that both sources have a Lambertian-like emission pattern with a strong $y$--axis symmetry, which corresponds to the semi-angle of half power of $\Phi_\text{h}=60^\circ$. Compared to the ideal diffuse (Lambertian distributed) emitter model in \cite{kb9701} with point source and receiver assumptions, the spatial domain characteristics for sources, receivers and coating materials are considered in our simulations. It is also important to determine the number of rays that will be traced in the \gls{MCRT} based channel modelling approach, since it will determine the resolution of the simulations. Moreover, the statistical significance of the transmitter, receiver and reflection models needs to be maintained in \gls{MCRT} by the utilization of the law of large numbers. Therefore, as a rule of thumb, the number of rays that are traced in \gls{MCRT} based channel modelling applications are generally chosen to be in the order of millions \cite{mu1501,mu2001}. To trace sufficient number of rays and realistically model the adopted sources, the spatio--angular profile of the \gls{IR} band source is represented by $5\times 10^6$ measured rays in our simulations. For the \gls{VL} band, blue and yellow spectra have been represented by $5\times 10^6$ measured rays each, which yields $10^7$ measured rays in total. In the following subsection, the optical modelling for the meta-surfaces as well as the coating materials will be detailed.
%#####################################################################################
\subsubsection{IRS \& Coating Material Characterization}
The coating surface characteristics for the \gls{IRS} aided \gls{LiFi} applications is of crucial importance since the achievable system performance will directly be affected by the total optical power that is transferred from the \gls{TX} to the \gls{RX}. Furthermore, the system reliability is also dictated by higher-order reflections if the direct \gls{LoS} link is not available. As depicted in Fig. \ref{fig:system_description_1}, the meta-material based \glspl{IRS}, which are envisaged to be implanted on the wallpaper, have \quot{ON} and \quot{OFF} states. Please note that the \quot{ON} and \quot{OFF} states of the meta-materials represent two indoor application scenarios; (i) with \gls{IRS} and (ii) without \gls{IRS} deployment, respectively. When the \glspl{IRS} are in the \quot{ON} state, the electrical/mechanical/electro-mechanical meta-surfaces are activated and they act as micro-mirrors, which means that the reflectance of the wall increases. Since the manufacturing process of the meta-materials is complex and costly, the small grid of meta-surfaces are envisaged to be utilized on the wallpaper instead of being produced as a large single sheet of meta-surface. Therefore, depending on the density of the meta-materials, the average reflectance value of the wallpaper when the \glspl{IRS} are activated could vary. 

To model the \gls{IRS} implanted wallpaper as realistically as possible, the relative reflectance of the side walls is assumed to be $90\%$ in our simulations when the \glspl{IRS} are \quot{ON}. The remaining $10\%$ of the incident optical power is assumed to be absorbed by the coating material of the meta-surfaces to take real-life imperfections into consideration. Furthermore, the reflection characteristics of the \glspl{IRS} are chosen to follow the Phong reflection model with $75\%$ specular and $25\%$ diffuse reflection components, since the wallpaper and meta-material mixture contains both rough and shiny surface structures. Note that the diffuse reflections are scattered into $\nu$ rays with Lambertian distribution. In other words, the \gls{BSDF} and the resultant intensity could be given by $1/\pi$ and $\cos(\theta_\text{s-s})$, for each scattered ray, respectively. The parameter $\theta_\text{s-s}$ denotes the angle between the specular component and the scattered rays. In every reflection (bounce) of a light ray, $25\%$ of the $90\%$ of the incident optical power is equally divided among all the scattered rays, due to the adopted fractions of reflection and scattering coefficients. Similarly, $75\%$ of the $90\%$ of the incident optical power is allocated to the specular component. Thus, the number of the total rays in the system that must be traced after the $\kappa^\text{th}$ reflection becomes $n_\text{R}=(\nu+1)^\kappa$, where $0\leq \kappa \leq \kappa_\text{max}$ represents the arbitrary number of reflections in the system. The parameter $\kappa_\text{max}$ is the maximum number of reflections that are considered in the \gls{MCRT} simulations. On the contrary, diffuse reflection characteristics with no specular components, which stem from the rough nature of the surfaces \cite{gb7901,hjs9801,hja9802,hi0701,mu1501,mu2001}, are adopted as the reflection characteristics of the wallpapers when the \glspl{IRS} are \quot{OFF}. Accordingly, the incident light rays become scattered into $\nu$ rays, which follow a Lambertian distribution. Lastly, the reflection profile of the wallpaper is directly dependent on the adopted surface coating material properties of the side walls.
\begin{figure}[!t]
	\centering
	\begin{subfigure}[t]{.45\columnwidth}
		\includegraphics[width=\columnwidth]{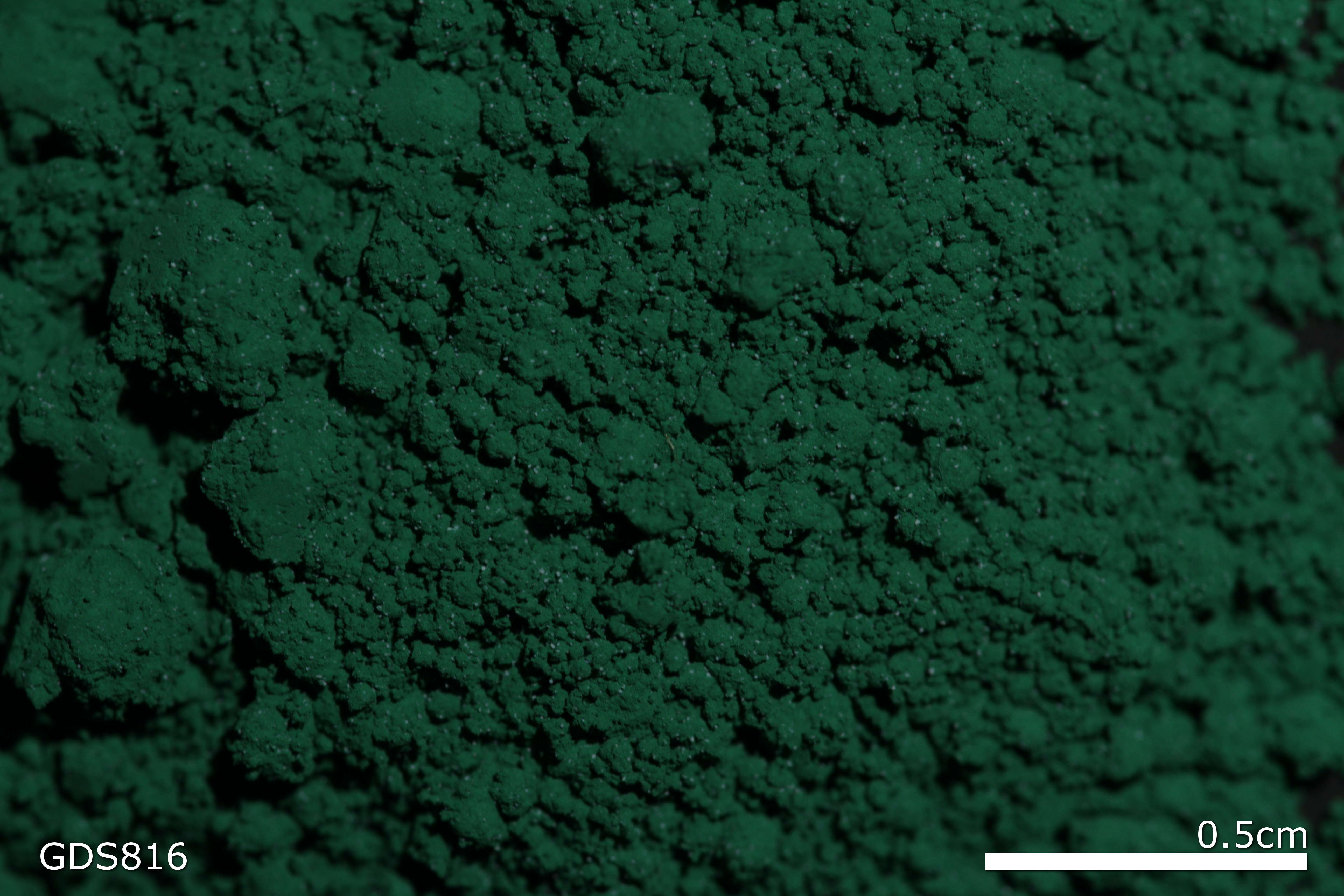}
		\caption{Cobalt Green Paint Pigment}
		\label{fig:mat_walls}
	\end{subfigure}~
	\begin{subfigure}[t]{.45\columnwidth}
		\includegraphics[width=\columnwidth]{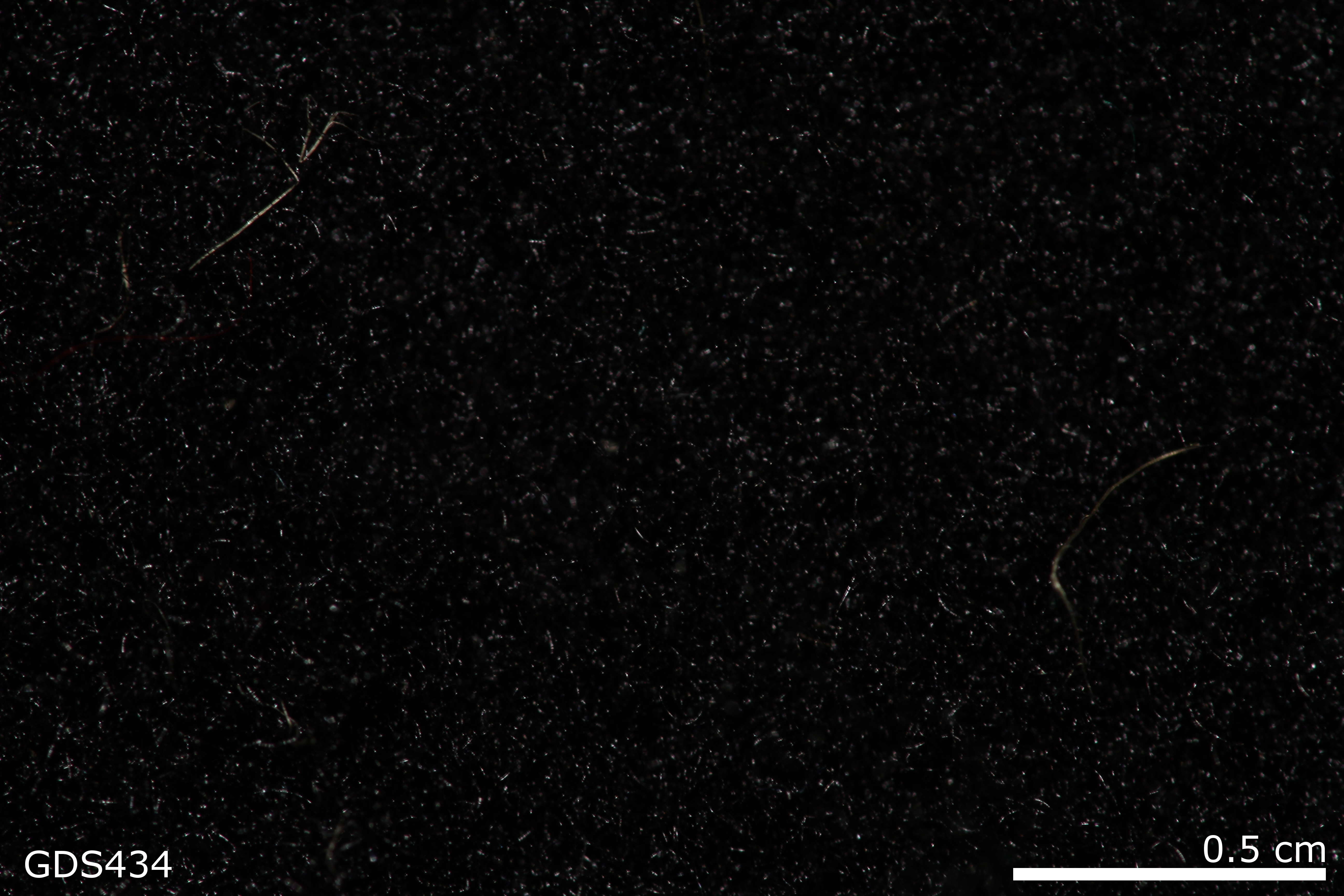}
		\caption{Black Polyester Pile Carpet}
		\label{fig:mat_carpet}
	\end{subfigure}
	\caption{The materials that are used in our simulations \cite{USGS}.}
	\label{fig:coating}
\end{figure}

In order to capture the realistic surface coating characteristics of the walls when the \glspl{IRS} are OFF, the measurement based spectral reflectance data had to be included in our simulations. Accordingly, the wall coating material which introduces relatively low reflectance in both the \gls{VL} and \gls{IR} spectra, \quot{Cobalt Green Paint Pigment}, was chosen as the main coating for the side walls as well as the ceiling. The main reason behind the selection of this material is to have a relatively small reflection contributions when \glspl{IRS} are not deployed. Hence, we can simulate and report the achievable rate performance difference between \gls{IRS} \quot{OFF} and \quot{ON} states, which correspond to the worst and best case scenarios, respectively. Furthermore, the floor of the room is assumed to be covered with a \quot{Black Polyester Pile Carpet}, in order to mimic a realistic indoor scenario. The measured relative reflectivity weight values for the chosen materials \gls{w.r.t.} the measurement spectra are obtained from the \gls{USGS} High Resolution Spectral Library Version 7 \cite{USGS}.
\begin{figure}[!t]
	\centering
	\includegraphics[width=0.58\columnwidth]{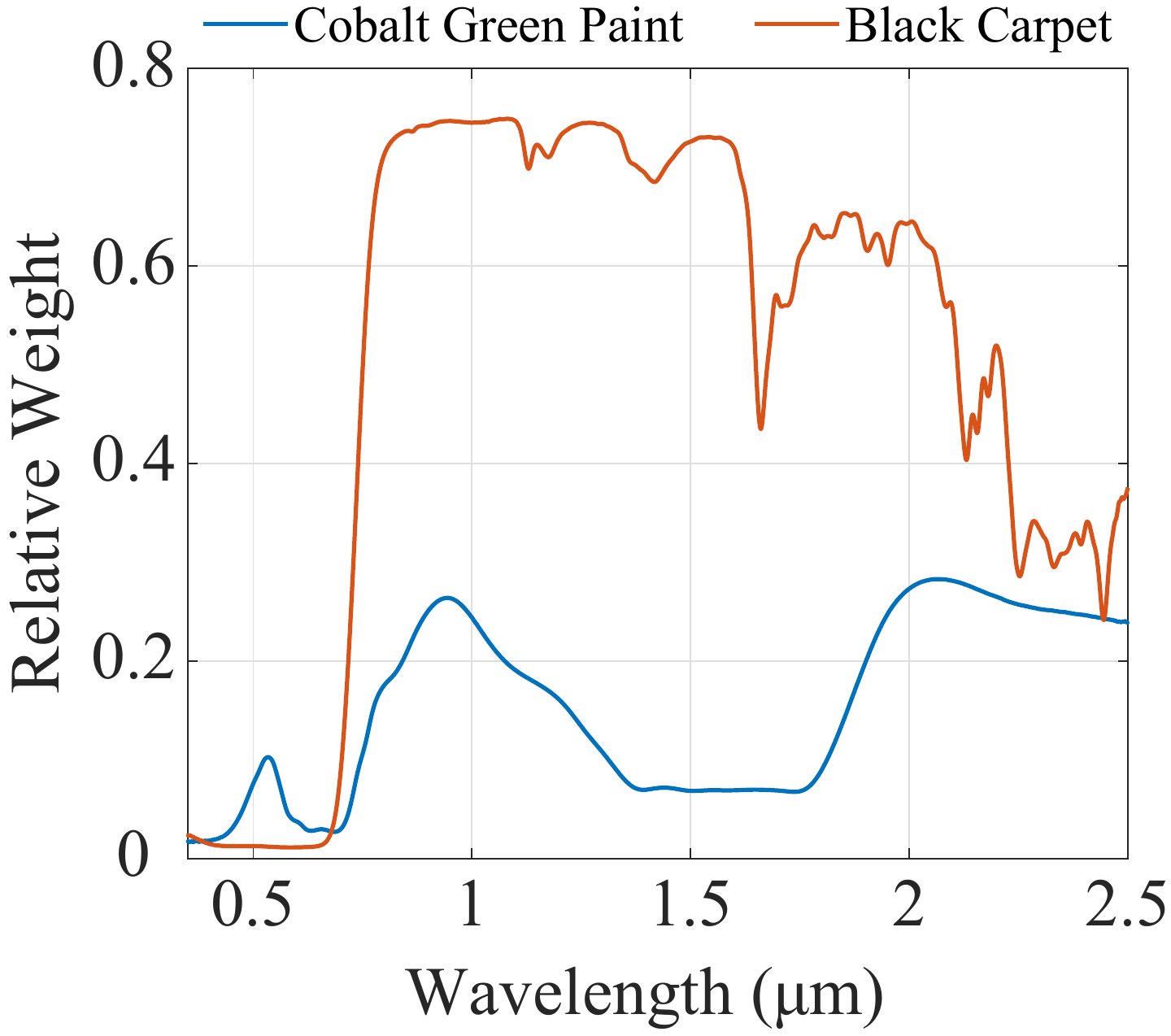}
	\caption{Relative spectral reflectivity values of the coating materials, which are adopted for IRS aided LiFi channel modelling simulations.}
	\label{fig:material_spectrum}
\end{figure}
\begin{figure}[!t]
	\centering
	\begin{subfigure}[t]{.5\columnwidth}
		\includegraphics[width=\columnwidth]{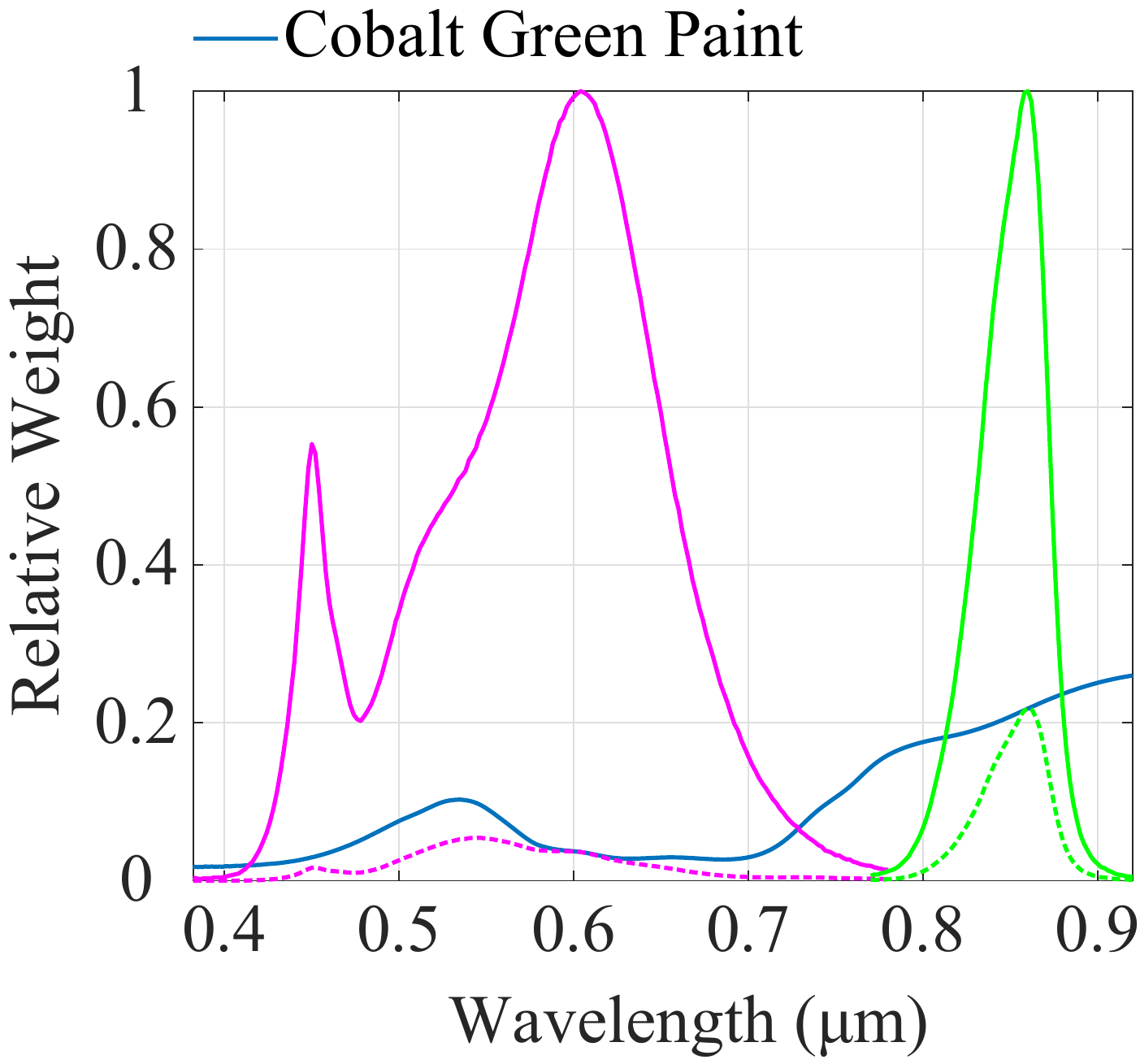}
		\caption{}
		\label{fig:spec_paint}
	\end{subfigure}~
	\begin{subfigure}[t]{.5\columnwidth}
		\includegraphics[width=\columnwidth]{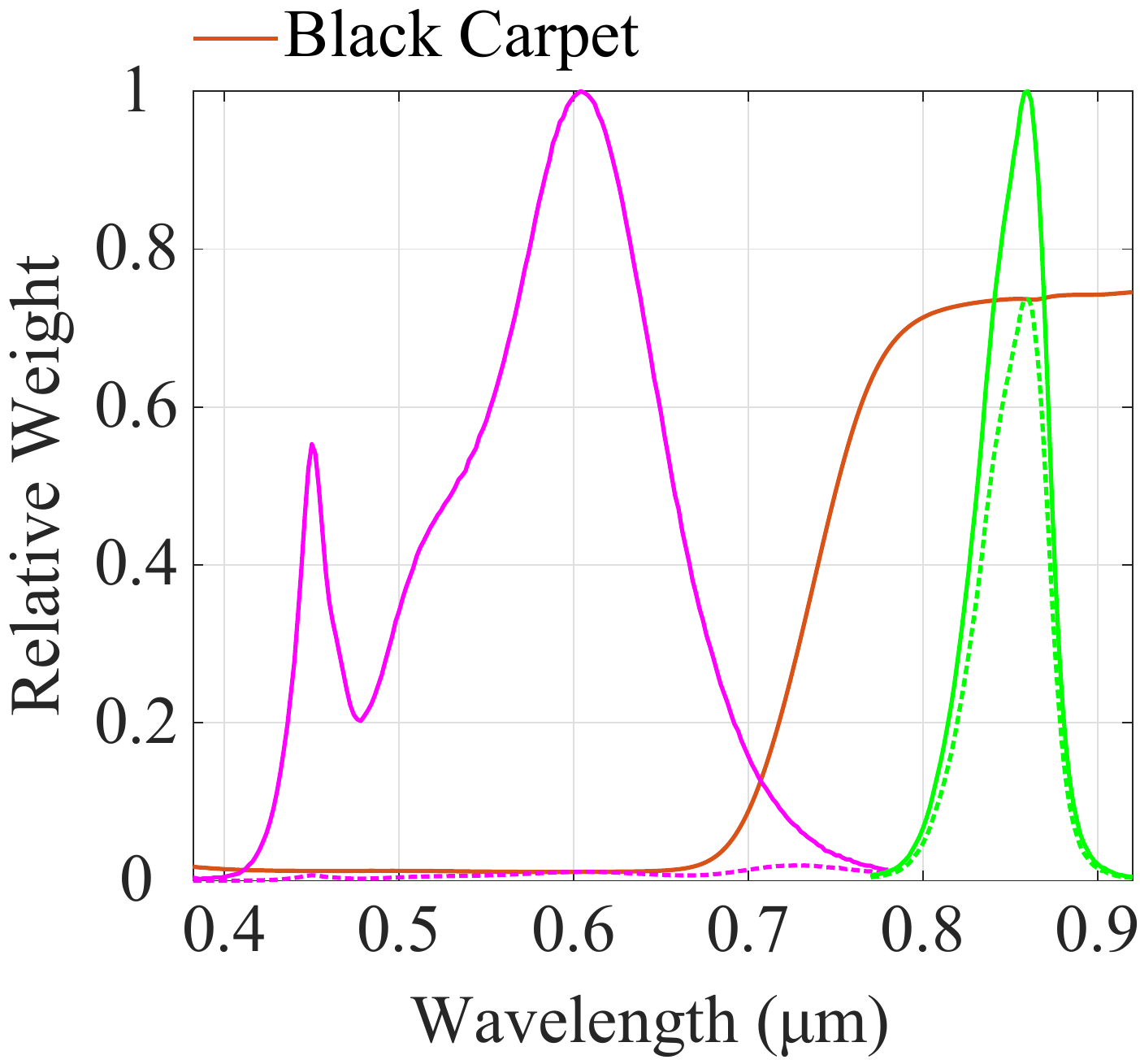}
		\caption{}
		\label{fig:spec_carpet}
	\end{subfigure}
	\caption{Relative spectral reflectivity values of the chosen materials with respect to the adopted VL and IR band source characteristics. The $f_\text{v}(\lambda)$ and $f_\text{i}(\lambda)$ are depicted as magenta and green solid lines, respectively. The resultant distribution of the sources after multiplication with the material characteristics are given by dotted lines under the respective curves.}
	\label{fig:spec_coating}
\end{figure}
The visual representations of the adopted coating materials are given in Fig. \ref{fig:coating}. Furthermore, the measured relative reflectivities of the adopted coating materials against the measurement spectrum, $0.35-2.5~\micro$m, are depicted for $2151$ data points in Fig. \ref{fig:material_spectrum}. In Figs. \ref{fig:spec_paint} and \ref{fig:spec_carpet}, the portion of the reflectivity spectrum that corresponds to the adopted source characteristics, $f_\text{v}(\lambda)$ and $f_\text{i}(\lambda)$, is given for the chosen materials, respectively. Accordingly, the effective source spectral distribution after the multiplication of the source characteristics and the coating material reflectivity values are plotted by dotted lines under the respective source plot. It can be seen from Figs. \ref{fig:spec_paint} and \ref{fig:spec_carpet} that the reflectivity characteristics of the black carpet are approximately $0.01$ and $0.75$ for the \gls{VL} and \gls{IR} bands, respectively. Similarly, the cobalt green paint has a relative reflectance value of $0.05$ and $0.2$ in the \gls{VL} and \gls{IR} bands, respectively. It is important to note that the reflectivity values of the coating material could significantly change within the spectra of each source as shown by Figs. \ref{fig:spec_paint} and \ref{fig:spec_carpet}. Thus, representing such fluctuations with an average value, as implemented by the recursive method based techniques \cite{bkklm9301,barry94}, would introduce a significant error in optical channel characterization. In our simulation environment, each coating material is represented by $200$ spectral data points, as this is the maximum number that is allowed by our simulation environment. In the next subsection, the receiver characterization will be detailed.
%##################################################################################################
\subsubsection{Receiver Characterization}
Although other \gls{MCRT} based toolkits are also proposed in the literature, the techniques in \cite{mu1501,mu2001,mup1501} are not capable of capturing the optical channel completely since they are unable to reflect the receiver spatio-angular and spatio-spectral characteristics within their calculations. Similar to the source modelling procedure, spatial, angular, and spectral parameterization is needed for realistic detector modelling.

\begin{figure}[!t]
	\centering
	\begin{subfigure}[t]{.5\columnwidth}
		\includegraphics[width=\columnwidth]{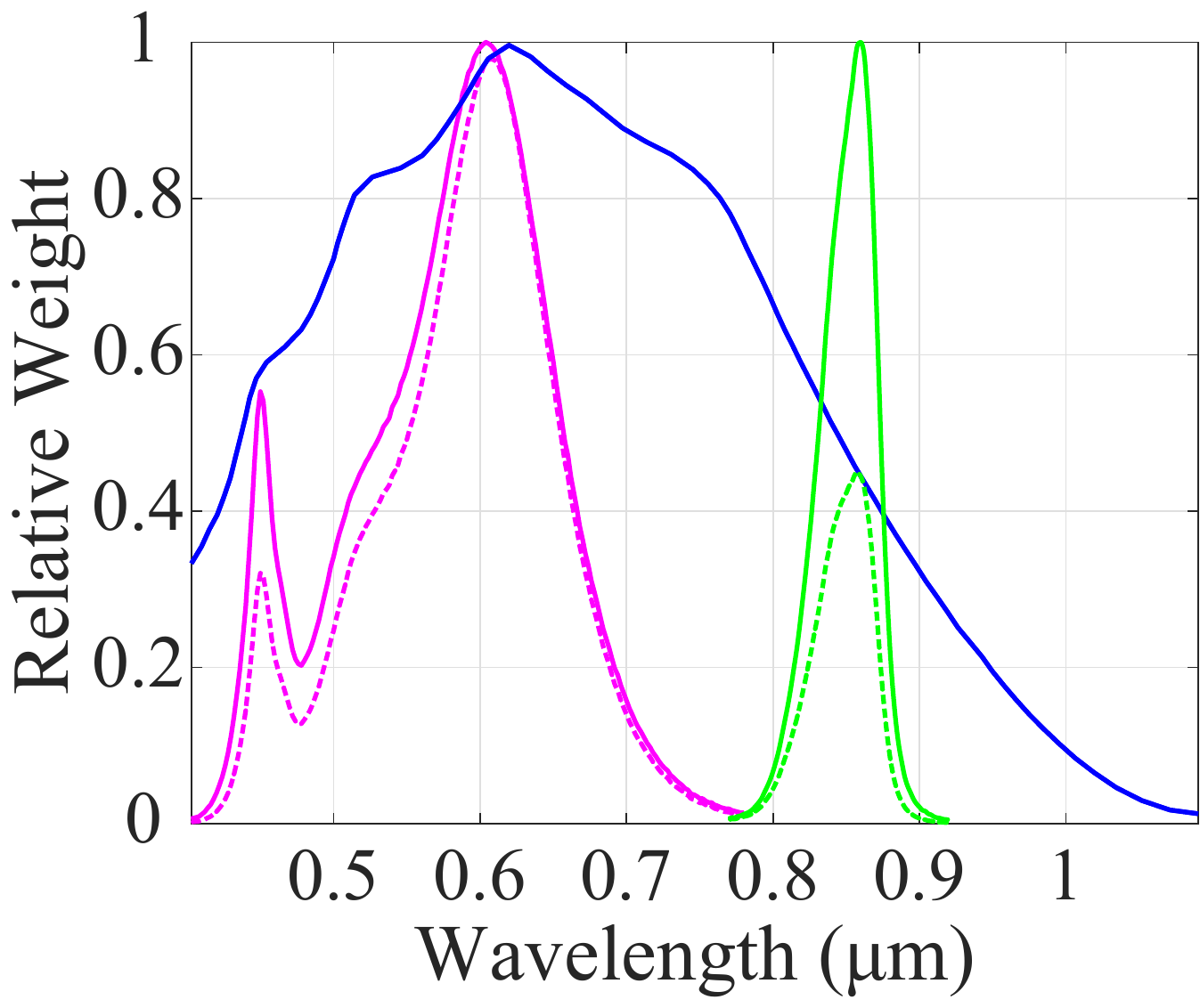}
		\caption{OSRAM SFH 2716 with $\lambda_\text{v}=0.62$ $\micro$m}
		\label{fig:det_spec_VL}
	\end{subfigure}~
	\begin{subfigure}[t]{.5\columnwidth}
		\includegraphics[width=\columnwidth]{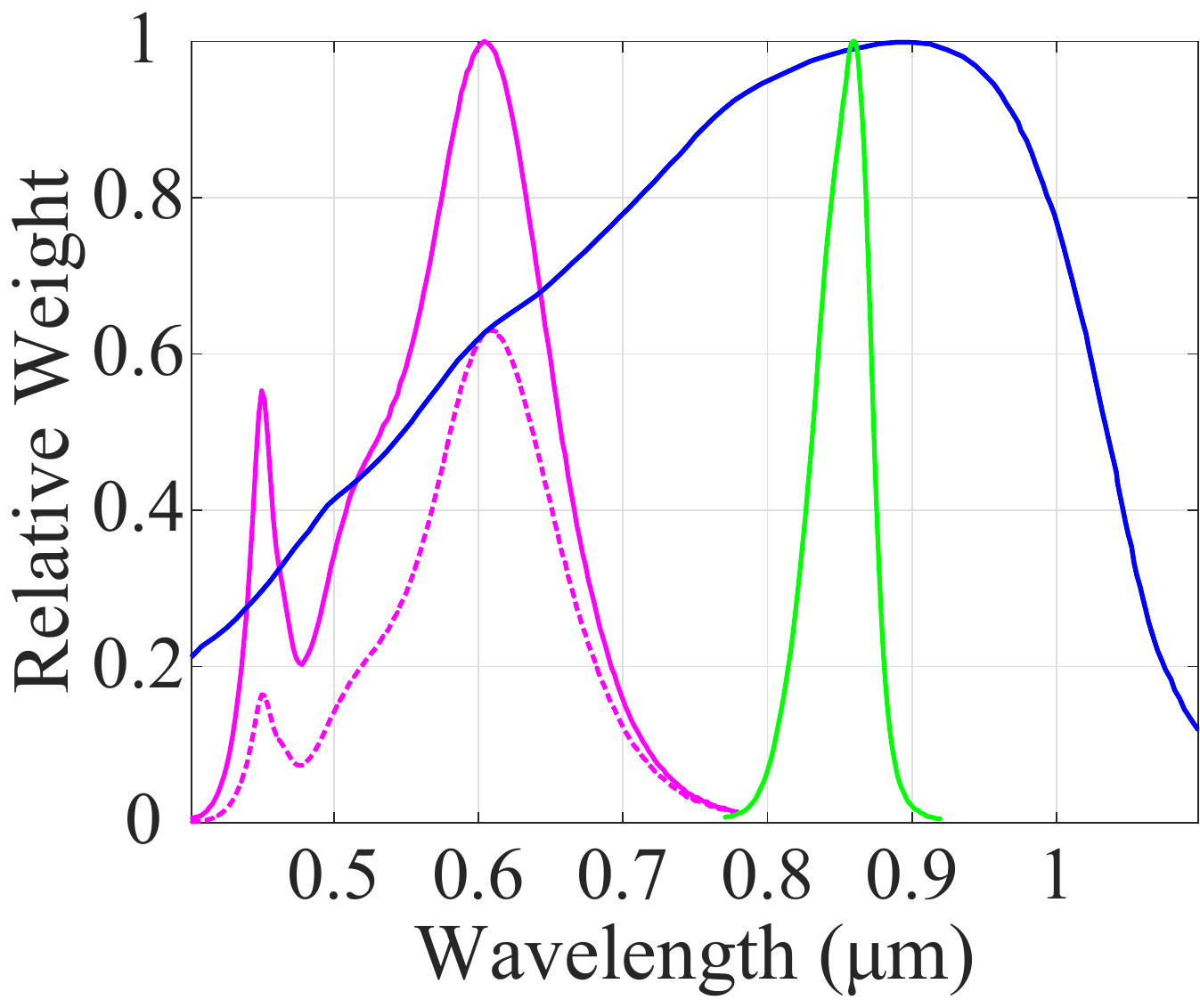}
		\caption{OSRAM SFH 2704 with $\lambda_{\text{i}}=0.9$ $\micro$m}
		\label{fig:det_spec_IR}
	\end{subfigure}~
	\caption{The relative spectral response curves for the adopted detectors, $g_\text{v}(\lambda)$ and $g_\text{i}(\lambda)$ (solid blue) with the relative spectral distributions for the OSRAM GW QSSPA1.EM, $f_\text{v}(\lambda)$ (solid magenta) and OSRAM SFH 4253, $f_\text{i}(\lambda)$ (solid green). The overall spectral response for \gls{VL} and \gls{IR} bands are plotted as dotted lines under the respective source spectral distribution plots.}
	\label{fig:detector_spectral}
\end{figure}

It is important to emphasize that two different source spectra; \gls{VL} and \gls{IR} band emission characteristics are adopted in this work. Therefore, the spectral responsivity curves at the \gls{RX} must match the intended sources. Accordingly, two silicone PIN \glspl{PD}; (i) OSRAM SFH 2716 with peak sensitivity at $\lambda_\text{v}=0.62~\micro$m and (ii) OSRAM SFH 2704 with the peak sensitivity at $\lambda_\text{i}=0.9~\micro$m are adopted as the \gls{VL} and \gls{IR} band receivers, respectively. Hence, non-imaging bare \gls{PD} models, without any front-end optics, that are rectangular shaped with $1~\text{cm}^2$ active area are generated in our simulation environment. The relative responsivity functions of the SFH 2716, $g_\text{v}(\cdot)$, and SFH 2704, $g_\text{i}(\cdot)$, plotted against the wavelength, $\lambda$, are given in Figs. \ref{fig:det_spec_VL} and \ref{fig:det_spec_IR}, respectively. Accordingly, the solid blue lines depict the relative spectral response weights of each detector, where the magenta and green colors are $f_\text{v}(\lambda)$ and $f_\text{i}(\lambda)$, respectively. Moreover, the overall spectral response for \gls{VL} and \gls{IR} bands, which are calculated by $f_\text{v}(\lambda)g_\text{v}(\lambda)$ and $f_\text{i}(\lambda)g_\text{i}(\lambda)$, are plotted as dotted lines under the \gls{VL} and \gls{IR} source spectral distribution plots, respectively. As can be seen from Figs. \ref{fig:det_spec_VL} and \ref{fig:det_spec_IR} each detector closely matches their intended sources and severely attenuates the out-of-band signals. For instance, the OSRAM SFH 2716 detector responsivity curve matches the OSRAM GW QSSPA1.EM emission spactra with an average responsivity of approximately $0.7$. However, the same detector filters more than $50\%$ of the optical power emitted from OSRAM SFH 4253. In a similar manner, the OSRAM SFH 2704 detector introduces a responsivity value of approximately $0.95$ for the OSRAM SFH 4253 source, where it filters more than $40\%$ of the optical power from the OSRAM GW QSSPA1.EM source.

\begin{figure}[!t]
	\centering
	\begin{subfigure}[t]{.43\columnwidth}
		\includegraphics[width=\columnwidth]{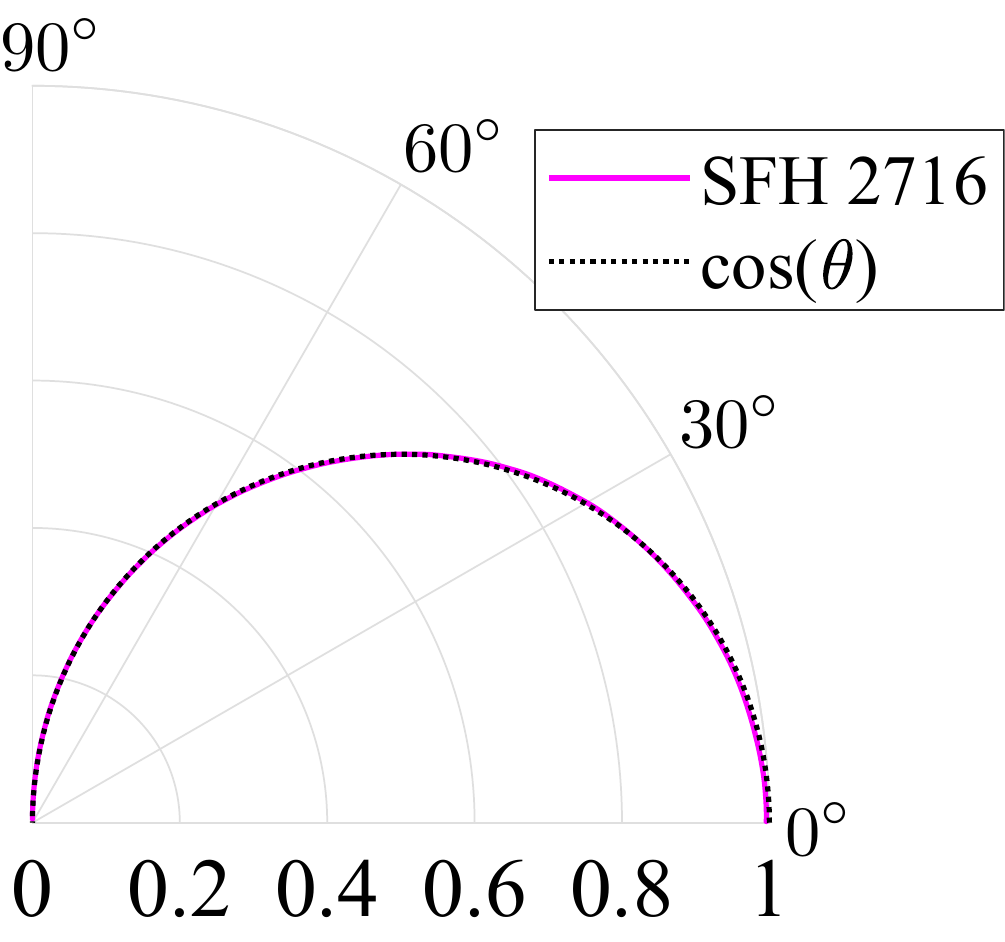}
		\caption{OSRAM SFH 2716}
		\label{fig:det_direct_VL}
	\end{subfigure}~
	\begin{subfigure}[t]{.43\columnwidth}
		\includegraphics[width=\columnwidth]{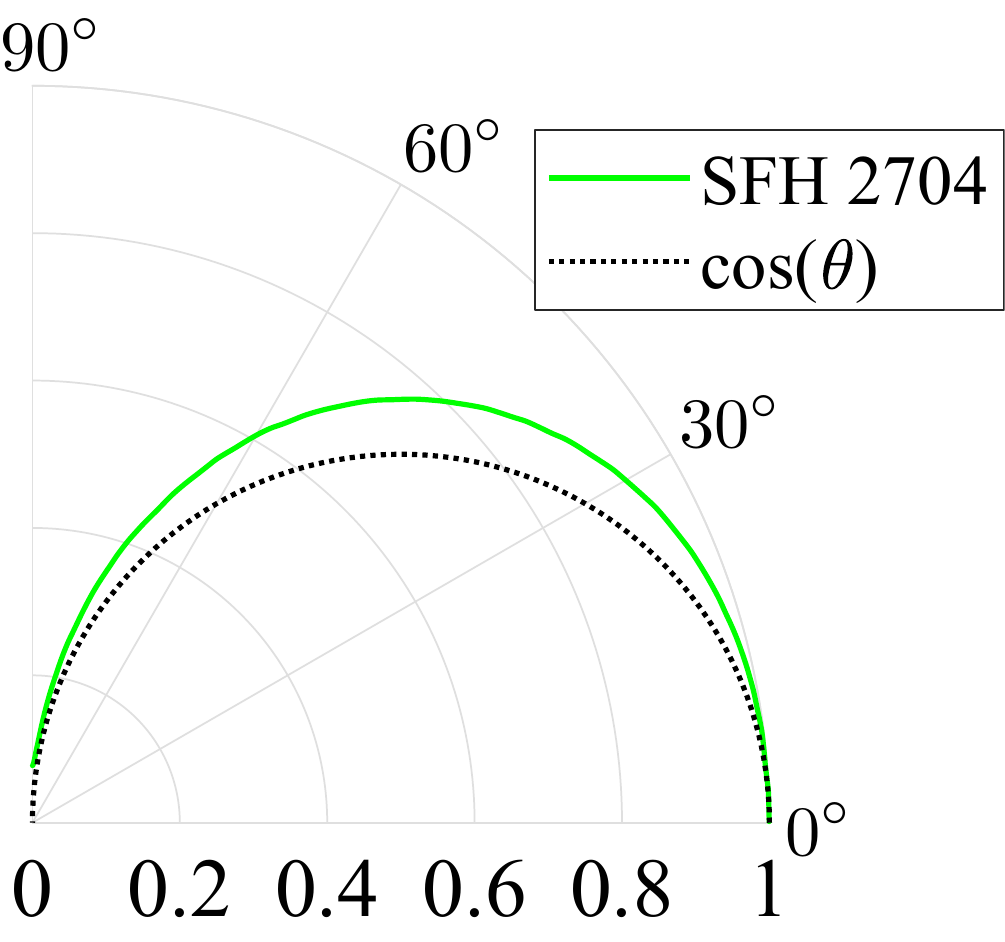}
		\caption{OSRAM SFH 2704}
		\label{fig:det_direct_IR}
	\end{subfigure}
	\caption{Relative angular responsivity characteristic plots for the adopted \glsentrytext{VL} (left) and \glsentrytext{IR} (right) band detectors. The ideal cosine responsivity curve is given by black dotted line as a benchmark.}
	\label{fig:detector_directivity}
\end{figure}

The relative angular responsivity profile of the chosen detectors, which are obtained via goniometer measurements conducted and reported by the manufacturer, are fed into our simulation environment after multiple processing stages. The relative angular responsivity curves of the OSRAM SFH 2716 and OSRAM SFH 2704 detectors are given in Figs. \ref{fig:det_direct_VL} and \ref{fig:det_direct_IR}, respectively. Furthermore, the ideal cosine angular profile for the receivers, $\cos(\theta)$, which is depicted by a dotted black line is also given as a benchmark. The parameter $\theta$ denotes the angle of incidence of a ray that strikes the detector. It can be seen from the angular profile of the OSRAM SFH 2716 \gls{VL} band photo-detector that it closely follows the ideal cosine detection responsivity. However, this is not the case for the \gls{IR} band SFH 2704 photo-detector, due to the real world geometrical imperfections introduced in the manufacturing process of the receiver micro-chips. Please note that the curves in Figs. \ref{fig:det_direct_VL} and \ref{fig:det_direct_IR} are obtained by the real world measurements. Hence, any real world imperfections and/or manufacturing errors are captured by the measurement results, unlike the ideal cosine curve. It is also important to emphasize that the non-ideal angular characteristic will play a significant role in the \gls{IR} band \glspl{CIR} as the received optical power values will be multiplied with a modified cosine profile of the receiver. In the next subsection, the channel models, which are obtained via \gls{MCRT} simulations by considering a realistic source, \gls{IRS}, coating and receiver characteristics, will be presented.

%###################################################################################################
\subsection{Obtaining the Channel Models}
The multipath \gls{CIR} between source $S$ and receiver $R$ could be expressed by our \gls{MCRT} simulation results as follows:
\begin{align}
	h(t;S,R)=\sum_{i=1}^{i_\text{hit}}P_i \delta(t-t_i),
	\label{eq:CIR}
\end{align}\noindent where the parameters $P_i$, $i_\text{hit}$ and $t_i$ denote the received incoherent irradiance, total number of rays that hit the receiver and the elapsed time for the $i^\text{th}$ ray to reach the receiver, respectively. Note that incoming rays that strike the detector surface introduce various irradiance and time-of-flight values due to the different ray paths, even when there is no \gls{NLoS} path in the system. The main reason behind this is the spatial dispersion of the rays, which emerges due to the realistic \gls{LED} and \gls{PD} geometrical models in our simulations. Unlike the point source and receiver assumptions in the analytical models, the source and receiver models are accurately shaped as their actual micro-chip form factor, which effects the generation and capture of the traced rays. Therefore, to reduce the temporal fluctuations caused by spatial dispersion and ensure the statistical significance, data binning also known as clustering, on $h(t;S,R)$ is applied, which yields the discrete-time optical \gls{CIR} by
\begin{align}
	h[n;S,R]=\sum_{n=0}^{N_\text{b}-1}\widetilde{P}_n \delta(n-t_n), \quad \text{for } n \in \{ 0,1,\cdots,N_\text{b}-1 \}.
	\label{eq:CIR_discrete}
\end{align}
\noindent Accordingly, the number of bins is calculated by 
\begin{align}
	N_\text{b}=\left\lceil \frac{t_\text{L}-t_1}{\Delta w} \right\rceil,
	\label{eq:numbins}
\end{align}
\noindent where the time of arrival for the first and last rays are denoted by $t_1$ and $t_\text{L}$, respectively. Moreover, bin widths are also given by $\Delta w$. The bin edge for the $n^\text{th}$ bin could also be calculated by $t_n=t_1 + n\Delta w$. The cumulative irradiance within the given bin interval is calculated by $\widetilde{P}_n=\sum\limits_{\forall i}P_i$, $\forall i \in \left[ t_n~ t_{n+1} \right]$, if $n = N_\text{b}-1$ and $\forall i \in \left[ t_n~ t_{n+1} \right)$, otherwise. Note from \eqref{eq:CIR_discrete} and \eqref{eq:numbins} that the temporal domain accuracy is directly related to the bin width, $\Delta w$, where the resulting discrete-time \gls{CIR} closely approximates the actual channel when the bin width approaches zero, $\lim\limits_{\Delta w \rightarrow 0} h[n;S,R] \approx h(t;S,R)$.

Other important channel characterization parameters could also be devised by using the \gls{CIR} expression obtained in \eqref{eq:CIR_discrete}. Accordingly, the optical \gls{CFR} is described in terms of the discrete-time \gls{CIR} as follows:
\begin{align}
	H(f;S,R) &= \int\limits_{-\infty}^{\infty}h(t;S,R)e^{-j 2\pi f t}\text{d}t \approx \sum\limits_{n=0}^{N_\text{b}-1}h[n;S,R]e^{-j\frac{2\pi kn}{N}},\nonumber \\
	&\text{for } k \in \frac{\Delta f}{N}\circ\left\{ -\frac{N}{2}, -\frac{N}{2}+1,\cdots,\frac{N}{2}-1 \right\}
	\label{eq:h_freqresp}
\end{align}\noindent where the element-wise multiplication operation is given by $\circ$. It is important to note from the above expression that the continuous time \gls{FFT} and \gls{DFT} of the channel will closely approximate each other as the bin width approaches zero. The sampling frequency is calculated as $\Delta f = 1/\Delta w$. Moreover, the number of subcarriers in the \gls{DFT} operation is also calculated by $N=2^{\lceil \log_2  \left(N_\text{b}\right) \rceil}$. 

Another important parameter is the \gls{DC} channel gain or total optical power of the impulse response, which can also be calculated by using \eqref{eq:h_freqresp},
\begin{align}
	H(0;S,R)&=\int_{-\infty}^{\infty}h(t;S,R)\text{d}t \nonumber \\
&\approx\sum_{n=-\infty}^{\infty}h[n;S,R]=\sum\limits_{n=0}^{N_\text{b}-1}\sum\limits_{\kappa=0}^{\kappa_\text{max}}h^{(\kappa)}[n;S,R].
\end{align}
\noindent The average transmitted and received optical powers could be linked by using the above \gls{DC} channel gain as follows, $P_R=H(0;S,R)P_S$. Similarly, by using the previous expression, the \gls{PL} in decibels could be given by
\begin{align}
	\text{PL} = -10\log_{10}H(0;S,R).
\end{align}

The \gls{RMS} delay spread and mean delay are two important measures to define the multipath richness of the channel, which also indicates the impact of \gls{ISI} on the system performance. Hence, the \gls{RMS} delay spread can be calculated by using the second and zeroth central moments of $h(t;S,R)$ as follows:
\begin{align}
	\tau_\text{RMS}=\sqrt{\frac{\mu_2(\tau_0)}{\mu_0(\tau_0)}}=\sqrt{\frac{\int\limits_{-\infty}^{\infty}(t-\tau_0)^2 h^2(t;S,R)\text{d}t}{\int\limits_{-\infty}^{\infty}h^2(t;S,R)\text{d}t}} = \sqrt{\frac{\sum\limits_{n=0}^{N_\text{b}-1}(n-\tau_0)^2 h^2[n;S,R]}{\sum\limits_{n=0}^{N_\text{b}-1}h^2[n;S,R]}},
	\label{eq:RMS_delayspread}
\end{align}
\noindent where the mean delay is given in terms of the zeroth and first raw moments of $h[n;S,R]$ by
\begin{align}
	\tau_0=\frac{\mu_1(0)}{\mu_0(0)}=\frac{\mu_1(0)}{\mu_0(\tau_0)}=\frac{\int\limits_{-\infty}^{\infty}th^2(t;S,R)\text{d}t}{\int\limits_{-\infty}^{\infty}h^2(t;S,R)\text{d}t} =\frac{\sum\limits_{n=0}^{N_\text{b}-1}nh^2[n;S,R]}{\sum\limits_{n=0}^{N_\text{b}-1}h^2[n;S,R]}.
	\label{eq:mean_delay}
\end{align}

Lastly, the relative power of the \gls{LoS} component compared to the total received optical power is also an important factor to determine the dominance of the \gls{LoS} path. The dominance of the \gls{LoS} component directly indicates higher system reliability and link quality in cases where the direct link is broken. Accordingly, the \quot{flatness factor} of the optical channel is calculated by using the Rician $K$-factor as follows:
\begin{align}
\rho = \frac{K}{K+1}=\frac{P_{\text{LoS}}}{P_{\text{LoS}}+P_{\text{NLoS}}}=\frac{\int\limits_{-\infty}^{\infty}h^{(0)}(t;S,R)\text{d}t}{\sum\limits_{\kappa=0}^{\kappa_\text{max}}\int_{-\infty}^{\infty}h^{(\kappa)}(t;S,R)\text{d}t} =\frac{\sum\limits_{n=0}^{N_\text{b}-1}h^{(0)}[n;S,R]}{H(0;S,R)},
\end{align}
\noindent where the Rician $K$-factor is defined as $K=P_\text{LoS}/P_\text{NLoS}$ \cite{mdcc1701}. The \gls{MCRT} based channel characterization results for \gls{IRS} aided \gls{LiFi} will be provided for both \gls{VL} and \gls{IR} bands and will be presented in the following subsection.
%####################################################################################################
\subsubsection{MCRT Channel Characterization Results}
\begin{figure}[!t]
	\centering
	\includegraphics[width=0.8\columnwidth]{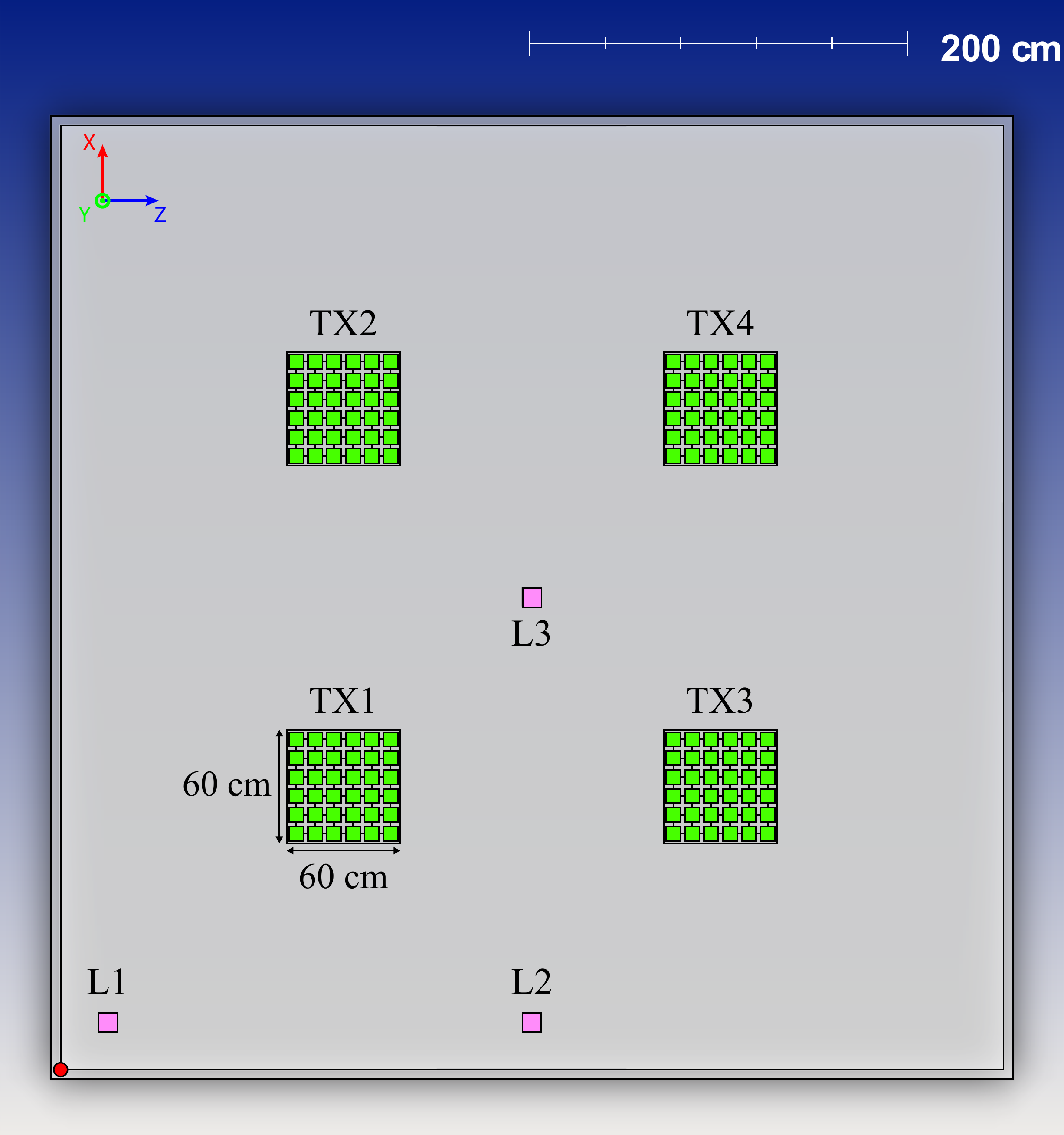}
	\caption{Top view of the considered scenario with transmitter (TX1, TX2, TX3 and TX4) and receiver locations (L1, L2 and L3). The global origin point of the simulation environment is indicated by the red point.}
	\label{fig:system_description_2}
\end{figure}
\begin{table}
\caption{Details of the parameters used in the MCRT simulations.}{%
\begin{tabular}{c|c}
			\hline
			Room Dimensions & $5\times5\times3~\text{m}$ \\ \hline
			LED Luminaire Positions (cm) & \begin{tabular}[c]{@{}c@{}}$\mathbf{p}_{\text{TX1}}=(135,~300,~135)$ \\ $\mathbf{p}_{\text{TX2}}=(335,~300,~135)$ \\ $\mathbf{p}_{\text{TX3}}=(135,~300,~335)$ \\ $\mathbf{p}_{\text{TX4}}=(335,~300,~335)$ \end{tabular} \\ \hline
			IRS Dimensions & $250\times 150~\text{cm}$ \\ \hline
			IRS Positions (cm) & \begin{tabular}[c]{@{}c@{}}$\mathbf{p}_{\text{IRS1}}=(250,~150,~0)$ \\ $\mathbf{p}_{\text{IRS2}}=(500,~150,~250)$ \\ $\mathbf{p}_{\text{IRS3}}=(250,~150,~500)$ \\ $\mathbf{p}_{\text{IRS4}}=(0,~150,~250)$ \end{tabular} \\ \hline
			IRS Orientations ($^\circ$) & \begin{tabular}[c]{@{}c@{}}$\mathbf{o}_{\text{IRS1}}=(0,~0,~0)$ \\ $\mathbf{o}_{\text{IRS2}}=(0,~-90,~0)$ \\ $\mathbf{o}_{\text{IRS3}}=(0,~-180,~0)$ \\ $\mathbf{o}_{\text{IRS4}}=(0,~-270,~0)$ \end{tabular} \\ \hline
			PD Positions (cm) & \begin{tabular}[c]{@{}c@{}}
				$\mathbf{p}_{\text{L1}}=(25,~135,~25)$ \\ $\mathbf{p}_{\text{L2}}=(25,~135,~250)$ \\ $\mathbf{p}_{\text{L3}}=(250,~135,~250)$ \end{tabular} \\ \hline
			%LED Chip $z$-axis Orientation & \begin{tabular}[c]{@{}c@{}}$\alpha_1=-33^\circ$ \\ $\alpha_3=33^\circ$\end{tabular} \\ \hline
			Number of Chips per Luminaire & $36$ ($6\times 6$) \\ \hline
			\begin{tabular}[c]{@{}c@{}}Number of Generated Rays per \\ LED Chip $\lvert$ Luminaire\end{tabular} & \begin{tabular}[c]{@{}c@{}}VL Band: $10\times10^6$ $\lvert$ $360\times 10^6$ \\ IR Band: $5\times 10^6$ $\lvert$ $180\times 10^6$ \end{tabular}  \\ \hline
			Power per Luminaire ($P_S$) & $36$ W \\ \hline
			Model of the LED Chips & \begin{tabular}[c]{@{}c@{}}VL Band: OSRAM GW QSSPA1.EM \\ IR Band: OSRAM SFH 4253 \end{tabular} \\ \hline
			FWHM of the LED Chips & $120^\circ$ \\ \hline
			Model of the PDs & \begin{tabular}[c]{@{}c@{}}VL Band: OSRAM SFH 2716 \\ IR Band: OSRAM SFH 2704\end{tabular} \\ \hline
			Effective Area of the PDs & $1~\text{cm}^2$ \\ \hline
			FWHM of the PDs & \begin{tabular}[c]{@{}c@{}}OSRAM SFH 2716: $120^\circ$ \\ OSRAM SFH 2704: $132^\circ$\end{tabular} \\ \hline
			Coating Materials & \begin{tabular}[c]{@{}c@{}}Cobalt Green Paint, $\nu=5$ \\ Black Carpet, $\nu=5$ \end{tabular} \\ \hline
			Time Resolution (Bin Width, $\Delta w$) & $0.2$ ns \\ \hline
		\end{tabular}}{}
		\label{table:MCRT_param}
\end{table}
In this subsection, the optical channels obtained with the proposed \gls{MCRT} toolkit will be presented for both \gls{VL} and \gls{IR} bands when \glspl{IRS} are ON and OFF. Moreover, the effect of user mobility on the channel parameters will also be investigated. Three mobile \gls{UE} locations; L1, L2 and L3, as depicted in Fig. \ref{fig:system_description_2}, are chosen to generalize the \gls{UE} mobility. Accordingly, the point L1 is located near the corner and \gls{UE} will receive reflections from 2 side walls. Similarly, point L2 is located near the side wall aligned with the center of the room. The \gls{UE} will primarily receive reflections from a single wall. On the contrary, point L3 is located at the center of the room, which will yield a significant \gls{LoS} path but negligible side wall reflections. Note that the average \gls{UE} height is taken as $H_\text{UE}=0.8H_\text{human}$, which is  reported in \cite{ycphp1801}. The parameter $H_\text{human}$ represents the average of the mean female and male height values ($168.75$ cm) for England obtained in 2016 \cite{HSE2016}. Hence, the average height of the mobile user becomes $135$ cm. The complete set of parameters used in the \gls{MCRT} based optical channel simulations could be found in Table \ref{table:MCRT_param}. It is important to note that the interior surfaces of the room are designed to introduce spectrum dependent reflection characteristics. When the \glspl{IRS} are active, the side walls introduce mostly specular reflections, which corresponds to $75\%$ of all reflections Thus, $25\%$ of the all reflections becomes diffuse when \glspl{IRS} are ON. The diffuse reflections are designed to consist of $\nu=5$ scattered rays for every surface in the simulation environment. Note that the trade-off between the time complexity and reflection accuracy in our simulations is directly controlled by the number of scattered rays, $\nu$. Hence, $\nu=5$ is determined to be a sufficient value to model the reflection characteristics accurately without saturating the computational resources. The floor and ceiling always introduce diffuse reflections, similar to the side walls, when \glspl{IRS} are inactive. The parameter \quot{minimum relative ray intensity} in our \gls{MCRT} simulations, which decides when to terminate the trace of a single ray is chosen to be $10^{-5}$ and $10^{-4}$ for \gls{VL} and \gls{IR} band simulations, respectively. Similar to \cite{mu1501,mu2001}, a trace of a single ray is terminated when the optical power of the light ray is decreases to $0.001\%$ and $0.01\%$ of the initial intensity in \gls{VL} and \gls{IR} band simulations, respectively. Note that the overall reflectivity of the coating materials in the \gls{VL} band is significantly lower compared the \gls{IR} band, which is compensated via smaller \quot{minimum relative ray intensity} value. The number of rays generated per \gls{LED} chip is chosen to be $5$ million in our simulation environment. Therefore, a total of $180$ million rays per luminaire are generated in \gls{IR} band simulations. However, since the white \gls{LED} chips consists of yellow and blue components, a total of $360$ million rays per luminaire are generated in our \gls{VL} band simulations. It is also important to note that the receive \glspl{PD} are assumed to capture the light rays that strike the front face of the detector surface. In our simulations, both the transmit luminaires and receive \glspl{PD} are assumed to be orientated towards $-y$ and $+y$ axes directions, respectively. Consequently, the front face of the \glspl{PD} becomes the face that looks towards $+y$ direction, whereas the other face is assumed to be insensitive to the incoming light to model realistic receiver characteristics. Lastly, all four luminaires are assumed to be transmitting the same information in our channel measurement simulations without loss of generality.
\subsubsection{VL Band Results}
\begin{figure*}[p]
	\centering
	\begin{subfigure}[t]{.43\columnwidth}
		\includegraphics[width=\columnwidth]{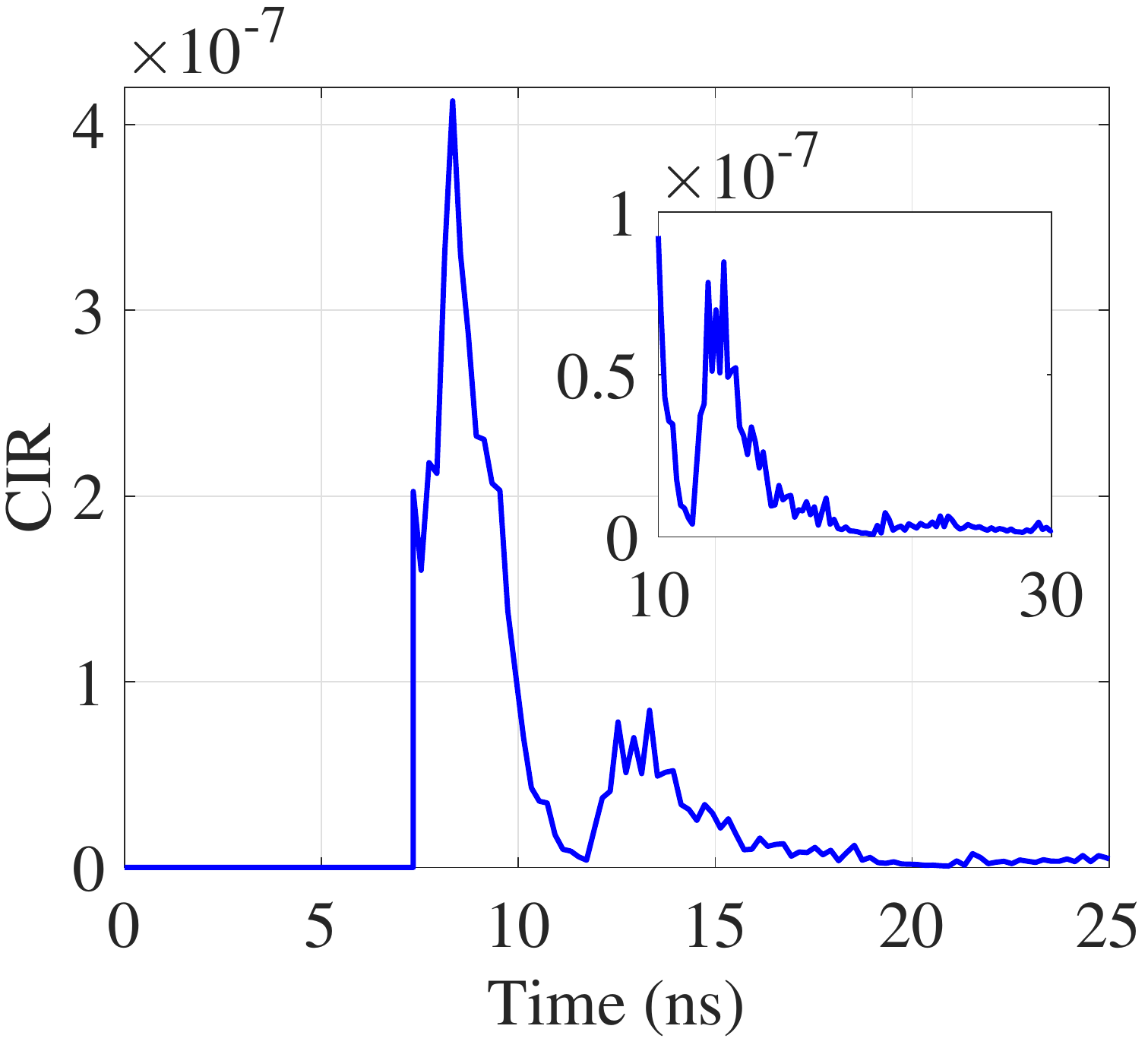}
		\caption{IRS ON $h(t;\forall S,\text{L1})$}
		\label{}
	\end{subfigure} ~
	\begin{subfigure}[t]{.43\columnwidth}
		\includegraphics[width=\columnwidth]{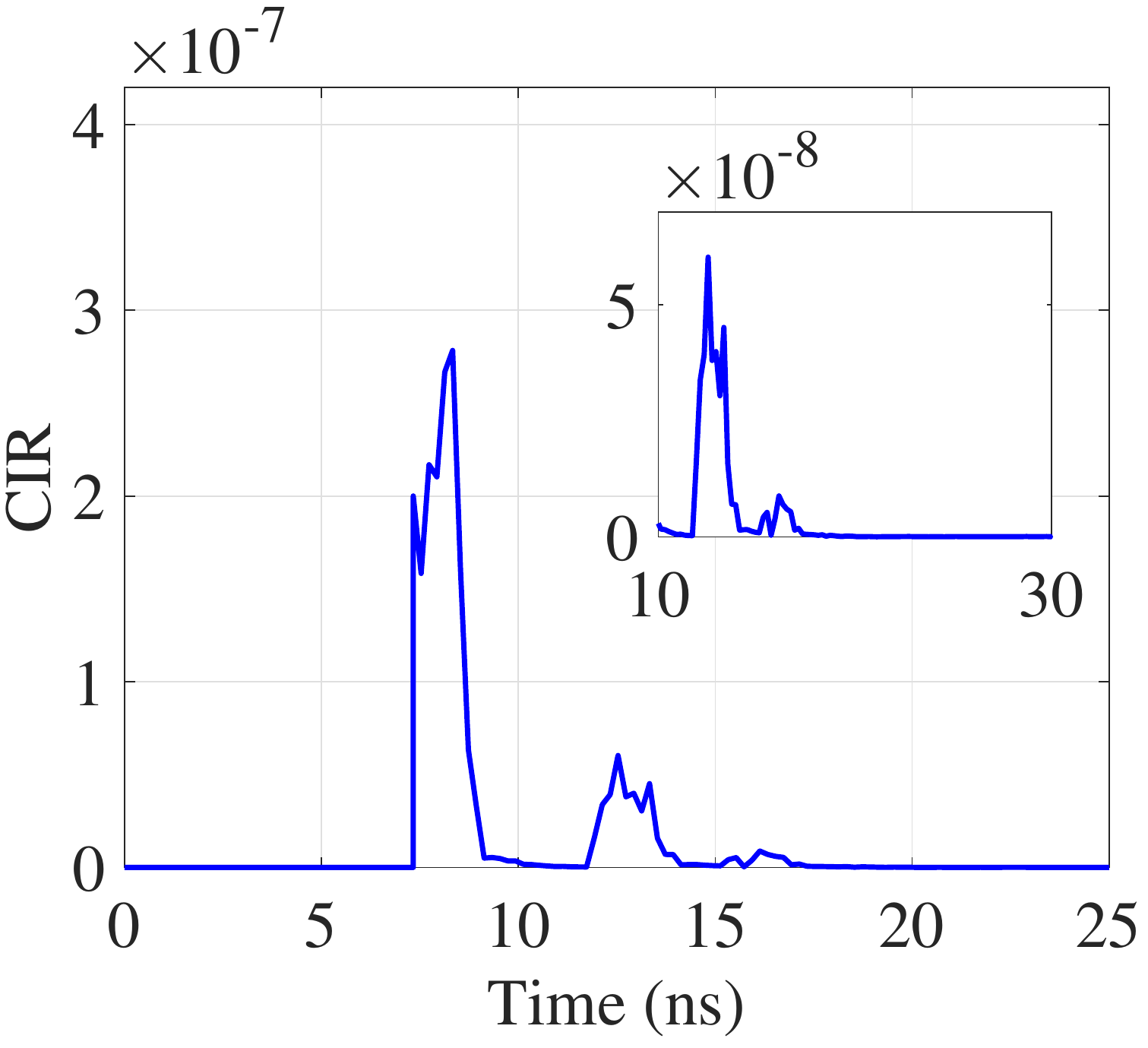}
		\caption{IRS OFF $h(t;\forall S,\text{L1})$}
		\label{}
	\end{subfigure} \\
	\begin{subfigure}[t]{.43\columnwidth}
		\includegraphics[width=\columnwidth]{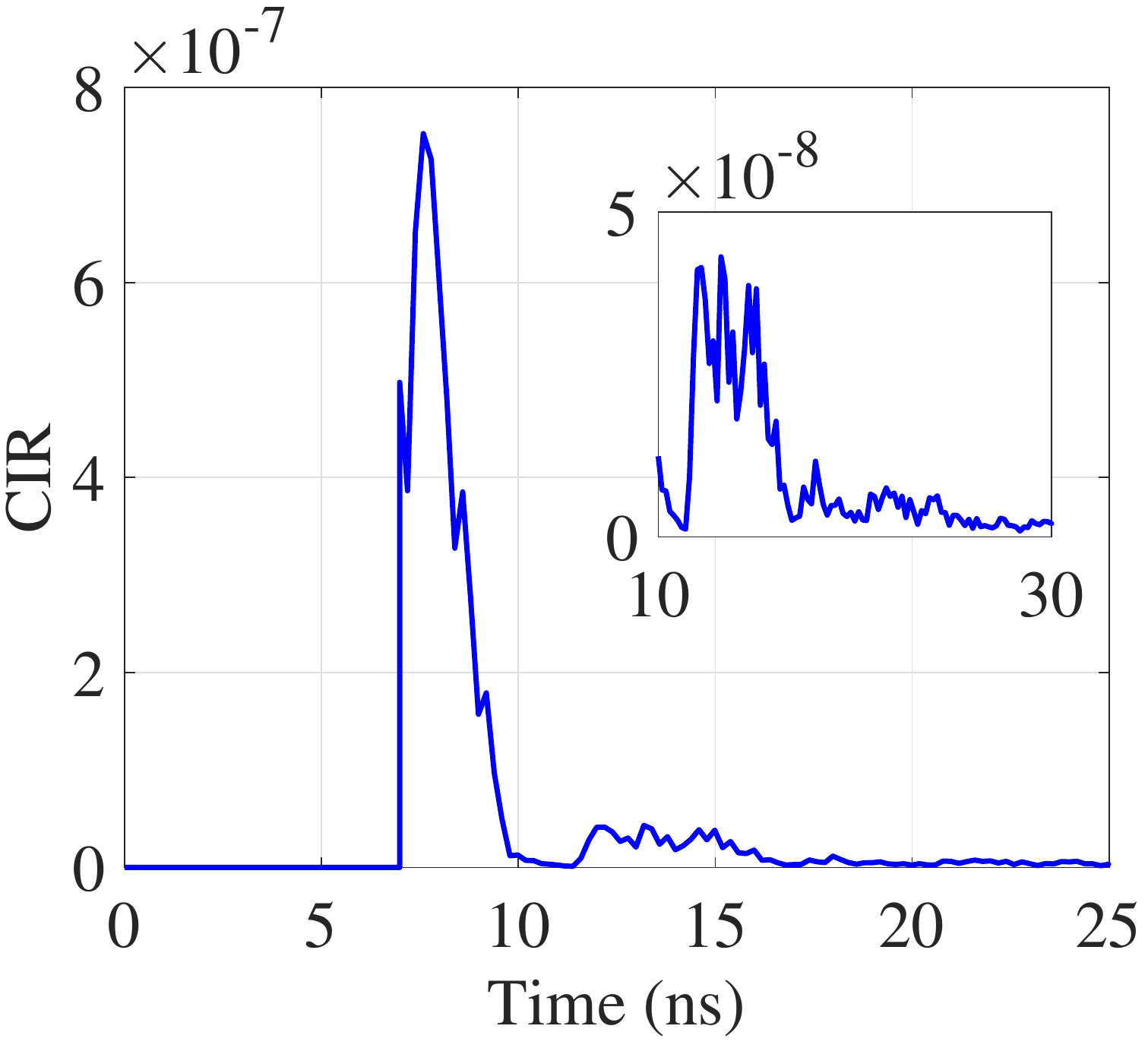}
		\caption{IRS ON $h(t;\forall S,\text{L2})$}
		\label{}
	\end{subfigure} ~
	\begin{subfigure}[t]{.43\columnwidth}
		\includegraphics[width=\columnwidth]{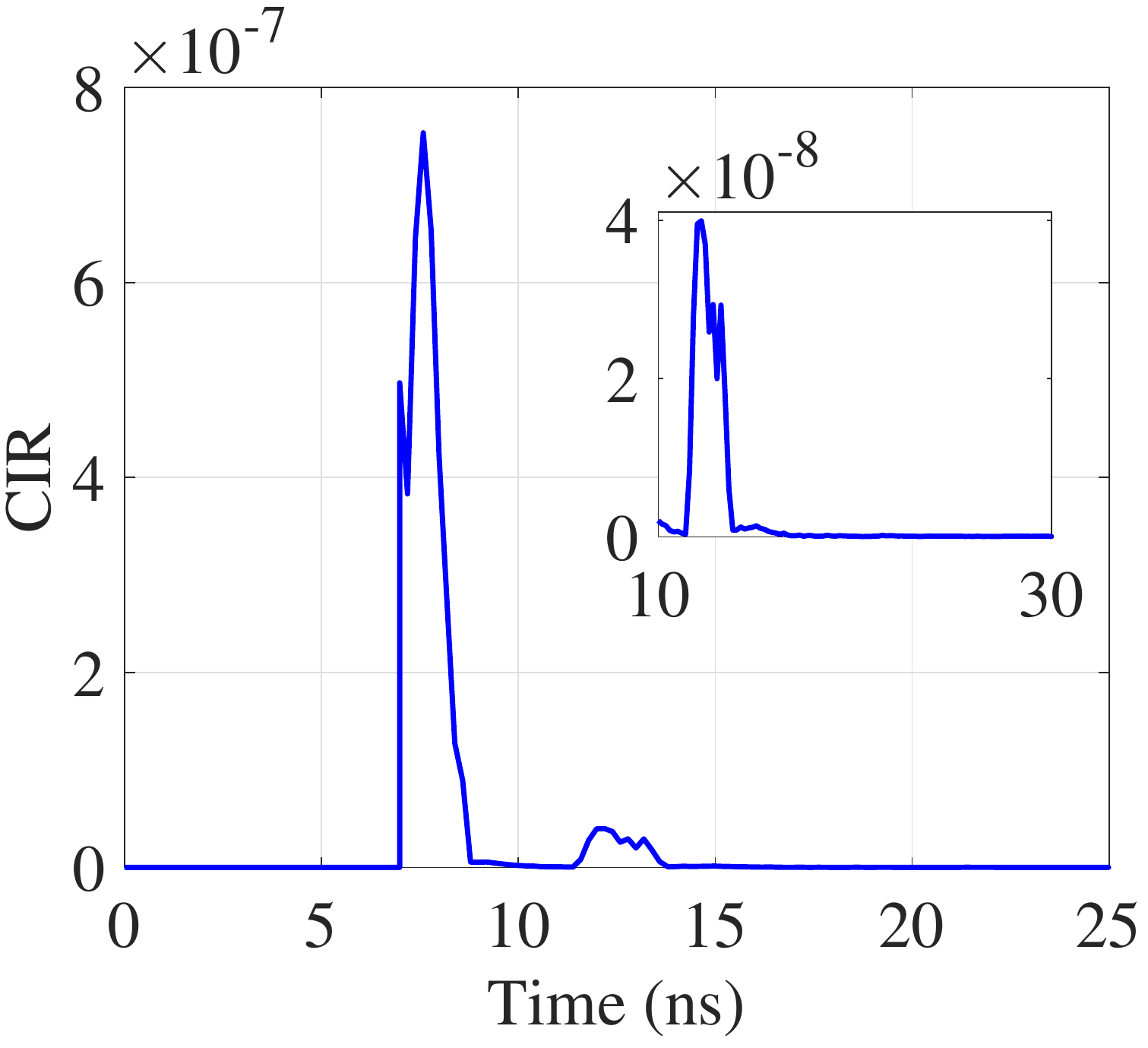}
		\caption{IRS OFF $h(t;\forall S,\text{L2})$}
		\label{}
	\end{subfigure} \\
	\begin{subfigure}[t]{.43\columnwidth}
		\includegraphics[width=\columnwidth]{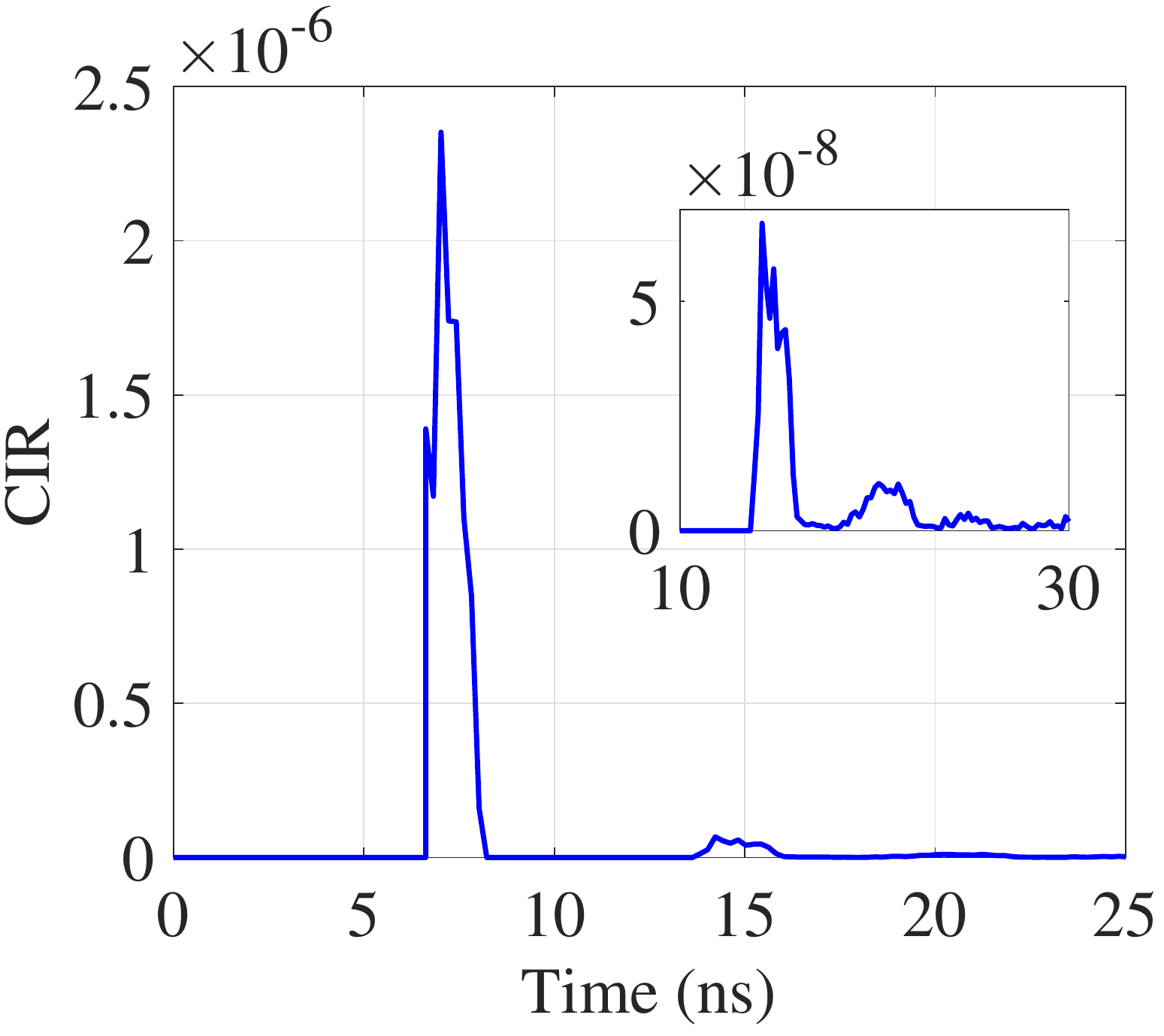}
		\caption{IRS ON $h(t;\forall S,\text{L3})$}
		\label{}
	\end{subfigure} ~
	\begin{subfigure}[t]{.43\columnwidth}
		\includegraphics[width=\columnwidth]{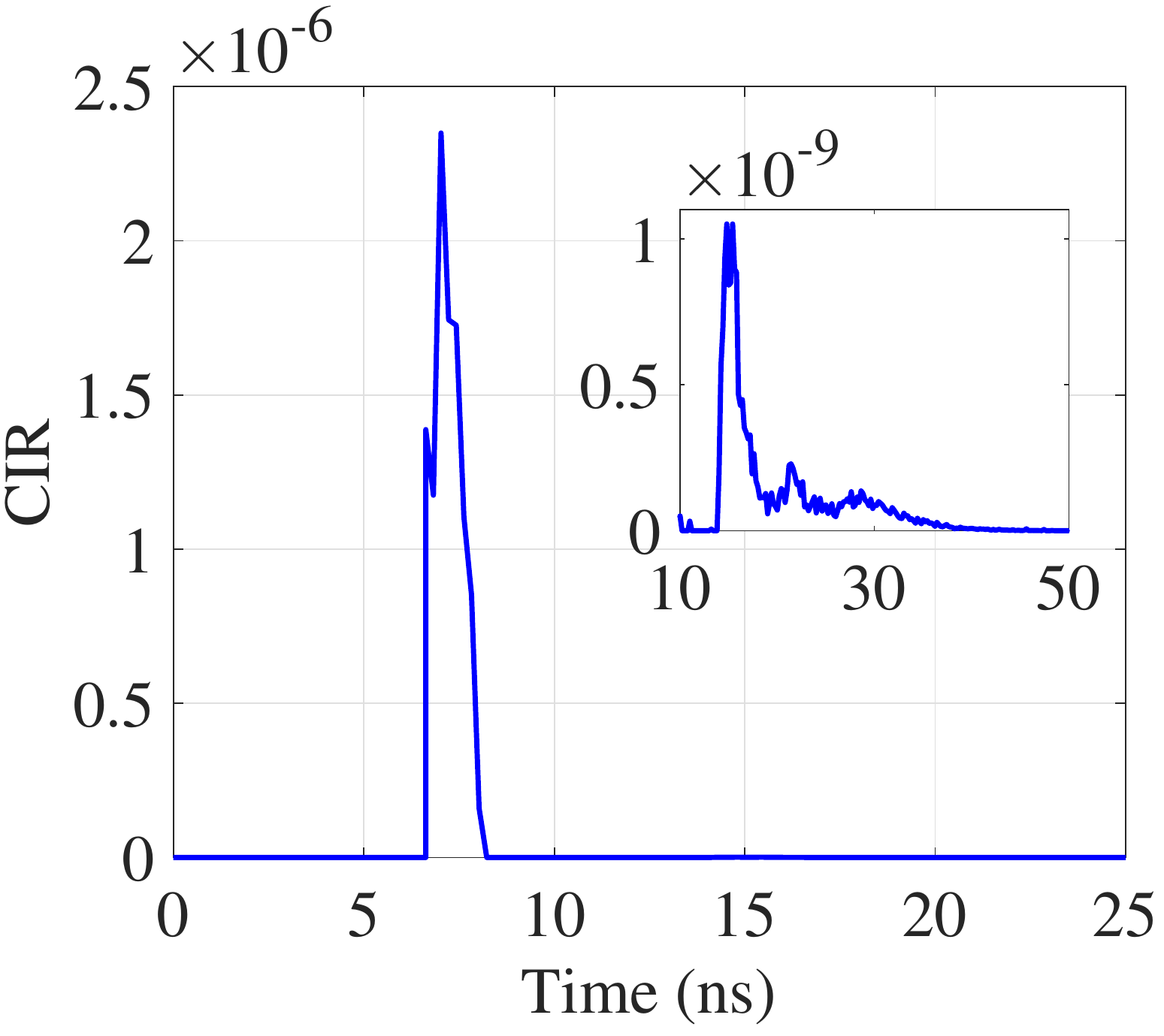}
		\caption{IRS OFF $h(t;\forall S,\text{L3})$}
		\label{}
	\end{subfigure}
	\caption{IRS aided indoor LiFi CIR simulation results for VL band. The results are obtained by the proposed MCRT based simulation technique. Each row represents the UE's location, whereas the columns are the state of the IRSs.}
	\label{fig:CIR_VL}
\end{figure*}

The \gls{CIR} plots of the \gls{VL} channels when the \glspl{IRS} were in both the ON and OFF states are given for \gls{UE} locations L1, L2 and L3 in a $2\times 3$ matrix formation in Fig. \ref{fig:CIR_VL}. The rows of the $2\times 3$ subplot matrix represents the location of the \gls{UE}, and the columns show the states of the \glspl{IRS}. Furthermore, the previously mentioned channel parameters for the same configuration are also given in Table \ref{table:channel_VL}.

For the \gls{UE} location L1, the \gls{CIR} results are depicted by Figs. \ref{fig:CIR_VL}(a) and (b). As can be seen from the figures, the magnitudes of the channel taps are significantly higher in the \gls{IRS} ON state compared to when the  \gls{IRS} is OFF, as expected. More specifically, the peak CIR and \gls{DC} channel gain in the \gls{IRS}-ON state are approximately $48.56\%$ and $141.6\%$ higher compared to the \gls{IRS}-OFF state for point L1, respectively. The main reason behind this significant difference is the enhanced wall reflectivity  coefficients when the \glspl{IRS} are activated. This effect can also be observed by the channel dispersion, where the \gls{RMS} delay spread is approximately $41.97\%$ larger when \glspl{IRS} are active compared to the \gls{IRS}-OFF state for \gls{UE} location L1. It can also be seen from the figures that the contribution of higher order reflections from the side walls, ceiling and floor surfaces yield a very high \gls{LoS} spike (between $6-10$ ns) as well as multiple \gls{NLoS} spikes (between $11-30$ ns) in \gls{IRS}-ON case.
\begin{table}[!t]
\caption{Details of the VL band optical channel parameters when IRSs are ON and OFF.}{%
		\begin{tabular}{|c|c|c|c|c|c|}
			\hline
			\textbf{} & \multicolumn{5}{c|}{\textbf{IRS ON (VL Band)}}  \\ \hline
			$R$ & $\kappa_\text{max}$ & $H[0;\forall S,R]$ & $\tau_0$ (ns) & $\tau_\text{RMS}$ (ns) & $\rho$  \\ \hline
			$\text{L1}$ & $33$ & $4.856\text{E}^{-6}$ & $8.762$ & $1.414$ & $0.402$  \\ \hline
			$\text{L2}$ & $38$ & $6.707\text{E}^{-6}$ & $7.815$ & $0.780$ & $0.615$  \\ \hline
			$\text{L3}$ & $38$ & $1.123\text{E}^{-5}$ & $7.161$ & $0.428$ & $0.934$  \\ \hline
			\textbf{} & \multicolumn{5}{c|}{\textbf{IRS OFF (VL Band)}}  \\ \hline
			$R$ & $\kappa_\text{max}$ & $H[0;S,R]$ & $\tau_0$ (ns) & $\tau_\text{RMS}$ (ns) & $\rho$  \\ \hline
			$\text{L1}$ & $4$ & $2.010\text{E}^{-6}$ & $8.178$ & $0.996$ & $0.967$  \\ \hline
			$\text{L2}$ & $4$ & $4.183\text{E}^{-6}$ & $7.574$ & $0.449$ & $0.983$  \\ \hline
			$\text{L3}$ & $4$ & $1.052\text{E}^{-5}$ & $7.151$ & $0.312$ & $0.998$  \\ \hline
		\end{tabular}}{}
	\label{table:channel_VL}
\end{table}
The maximum number of reflections captured for the given configuration becomes $\kappa_\text{max}=33$ when the \glspl{IRS} are ON, unlike \gls{IRS}-OFF case, where $\kappa_\text{max}=4$. Furthermore, the contribution of the \gls{NLoS} channel power compared to the whole \gls{CIR} is approximately $59.8\%$ and $3.3\%$ when the \glspl{IRS} are ON and OFF, respectively. It is important to emphasize that this significant difference between the \gls{NLoS} channels prove that the deployment of the \glspl{IRS} substantially increases system reliability in cases where the direct \gls{LoS} channel is blocked.

In Figs. \ref{fig:CIR_VL}(c) and (d), the \gls{CIR} results are depicted for \gls{UE} location L2. Accordingly, the peak \gls{CIR} and channel \gls{DC} gain when the \glspl{IRS} are ON becomes $0.11\%$ lower and $60.34\%$ higher, respectively, compared to the case where \glspl{IRS} are OFF. The reason behind the close match among the peak values of the \gls{CIR} consists of the significantly higher \gls{LoS} component that come from all four sources. Compared to location L1, the \gls{DC} channel gain has increased by approximately $32.57\%$ and $106.44\%$ at point L2 when the \glspl{IRS} are ON and OFF, respectively. Similarly, the time dispersion of the channel becomes $73.72\%$ higher in the case where the \glspl{IRS} are ON compared to the case where the \glspl{IRS} are OFF for L2. The \gls{RMS} delay spread has decreased $44.84\%$ and $121.83\%$ from point L1 to L2 when \glspl{IRS} are ON and OFF, respectively. The maximum number of reflections in point L2 also becomes $38$ and $4$ when the \glspl{IRS} are ON and OFF, respectively. The contribution of the \gls{NLoS} path compared to the whole \gls{CIR} is $38.5\%$ and $1.7\%$ for the \gls{IRS}-ON and \gls{IRS}-OFF states, respectively. As can be seen from Fig. \ref{fig:CIR_VL}(c), the activation of \glspl{IRS} creates the first and second tier of reflections, which are attached to the \gls{LoS} component between $7-10$ ns. Moreover, much higher order reflections induced by the \glspl{IRS}, could also be observed in between $13-30$ ns.

In Figs. \ref{fig:CIR_VL}(e) and (f), the \gls{CIR} plots are provided for the mobile \gls{UE} location L3. As can be clearly seen from these figures, the \glspl{CIR} (especially the \gls{LoS} components) are closely matching, as expected. The only difference between two cases is the larger magnitude of the higher order reflections when \glspl{IRS} are active, which could be observed between $12-25$ ns region in both figures. However, the maximum number of reflections captured in our simulations are $38$ and $4$ for the cases \gls{IRS}-ON and \gls{IRS}-OFF, respectively. The reason behind this phenomena is the resolution and accuracy of our \gls{MCRT} simulations, which is able to capture very small irradiance fluctuations even if they have no significance for communication purposes. The reflections captured from the side walls, ceiling and floor becomes significantly lower for location L3 even when the \glspl{IRS} are active, which is due to the large physical distance between \gls{UE} and the side walls. The peak value of the \gls{CIR} and \gls{DC} channel gain are only $0.11\%$ and $14.23\%$ higher when \glspl{IRS} are ON and OFF, respectively. The channel \gls{RMS} delay spread is also increased $125.40\%$ when \glspl{IRS} are activated for point L3. The \gls{NLoS} channel component comprises the $6.6\%$ and $0.2\%$ of the whole \gls{CIR} for \gls{IRS}-ON and \gls{IRS}-OFF cases in location L3, respectively. 
%Consequently, the \gls{VL} link capacity for a \gls{UE}, which experiences \gls{LoS} blockage will enhance $18.12$, $22.65$ and $33$ fold with the aid of \glspl{IRS} for the mobile terminal locations L1, L2 and L3, respectively.
%#########################################################################################################
\subsubsection{IR Band Results}
\begin{figure*}[p]
	\centering
	\begin{subfigure}[t]{.43\columnwidth}
		\includegraphics[width=\columnwidth]{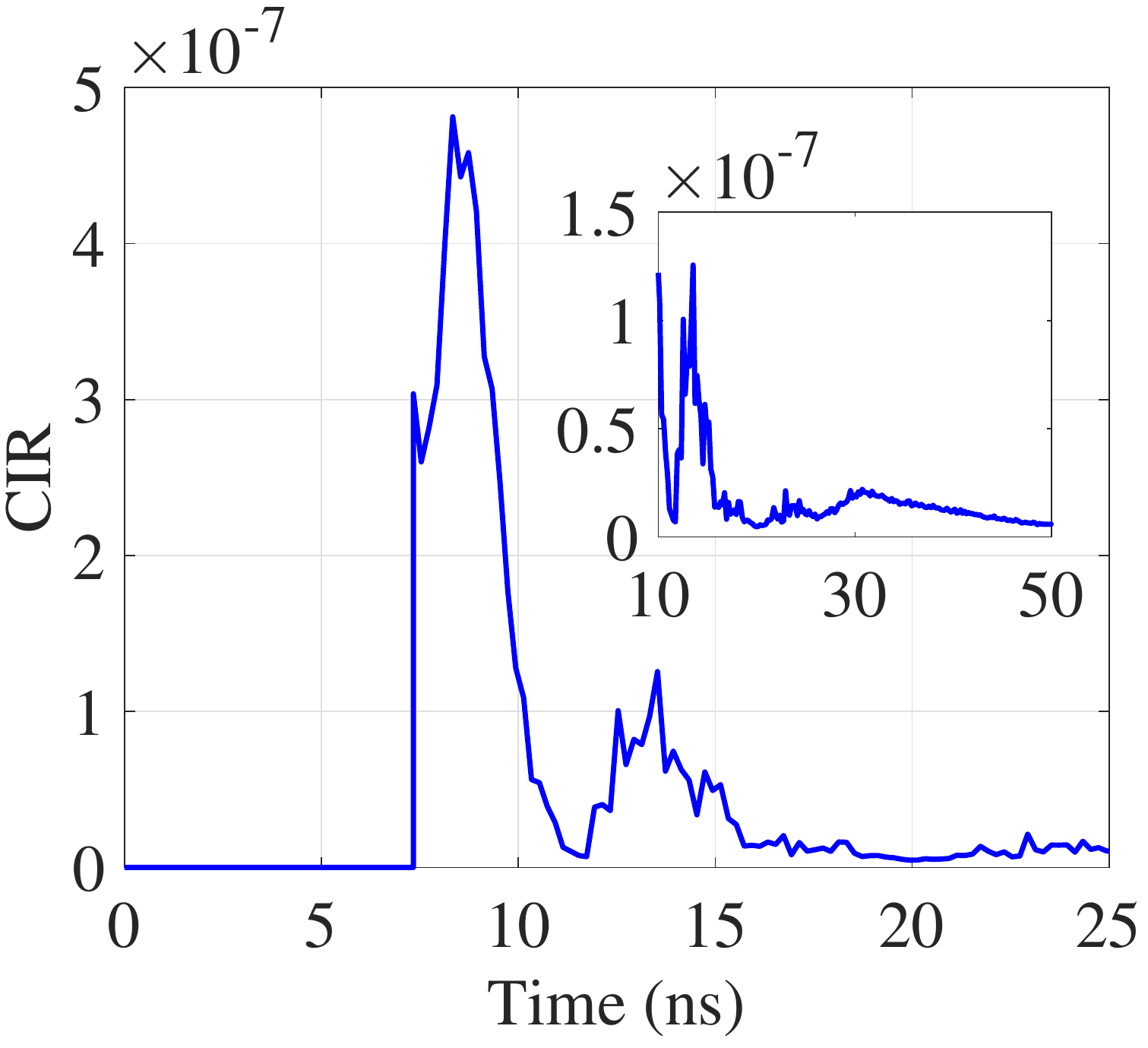}
		\caption{IRS ON $h(t;\forall S,\text{L1})$}
		\label{}
	\end{subfigure} ~
	\begin{subfigure}[t]{.43\columnwidth}
		\includegraphics[width=\columnwidth]{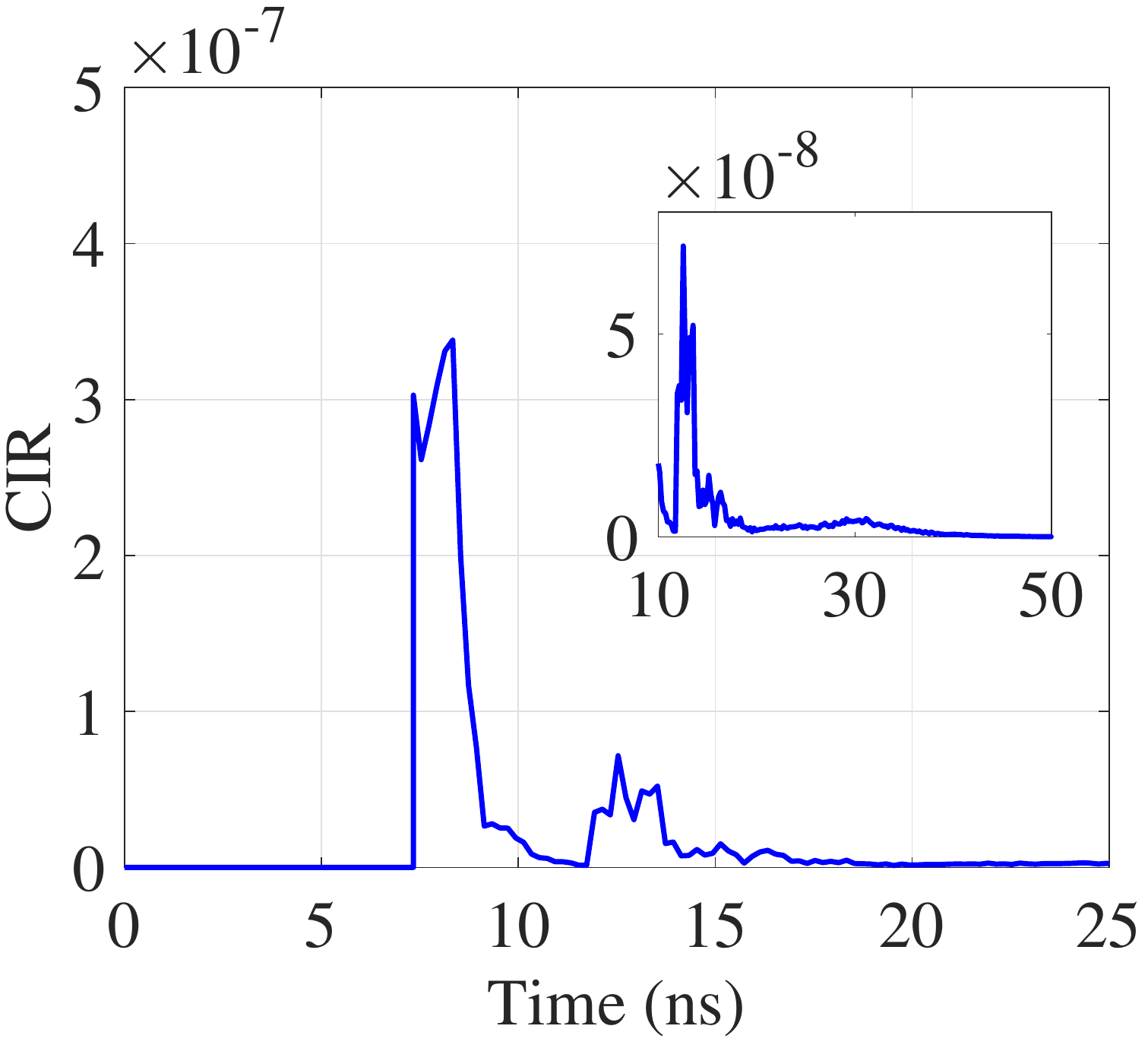}
		\caption{IRS OFF $h(t;\forall S,\text{L1})$}
		\label{}
	\end{subfigure} \\
	\begin{subfigure}[t]{.43\columnwidth}
		\includegraphics[width=\columnwidth]{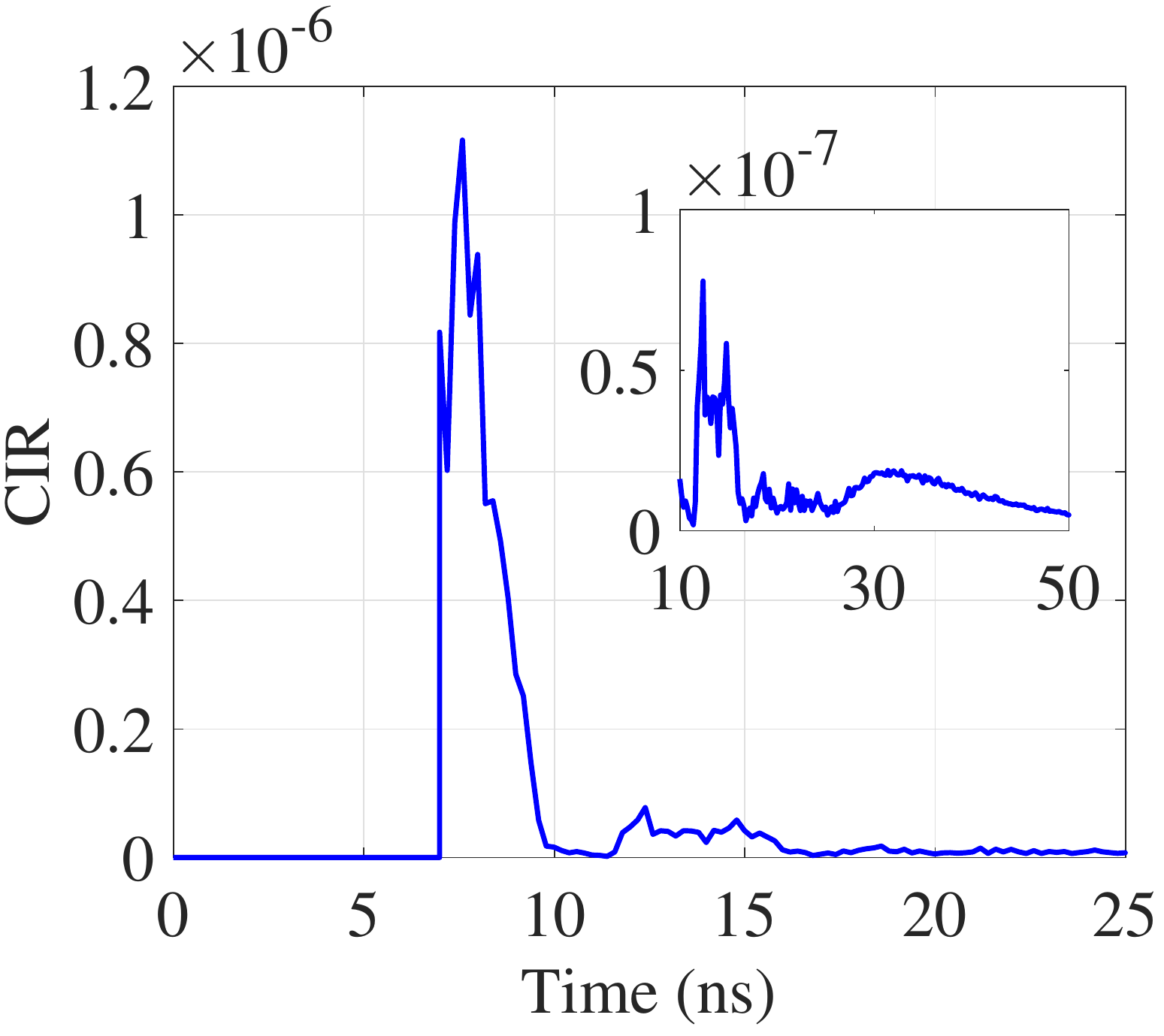}
		\caption{IRS ON $h(t;\forall S,\text{L2})$}
		\label{}
	\end{subfigure} ~
	\begin{subfigure}[t]{.43\columnwidth}
		\includegraphics[width=\columnwidth]{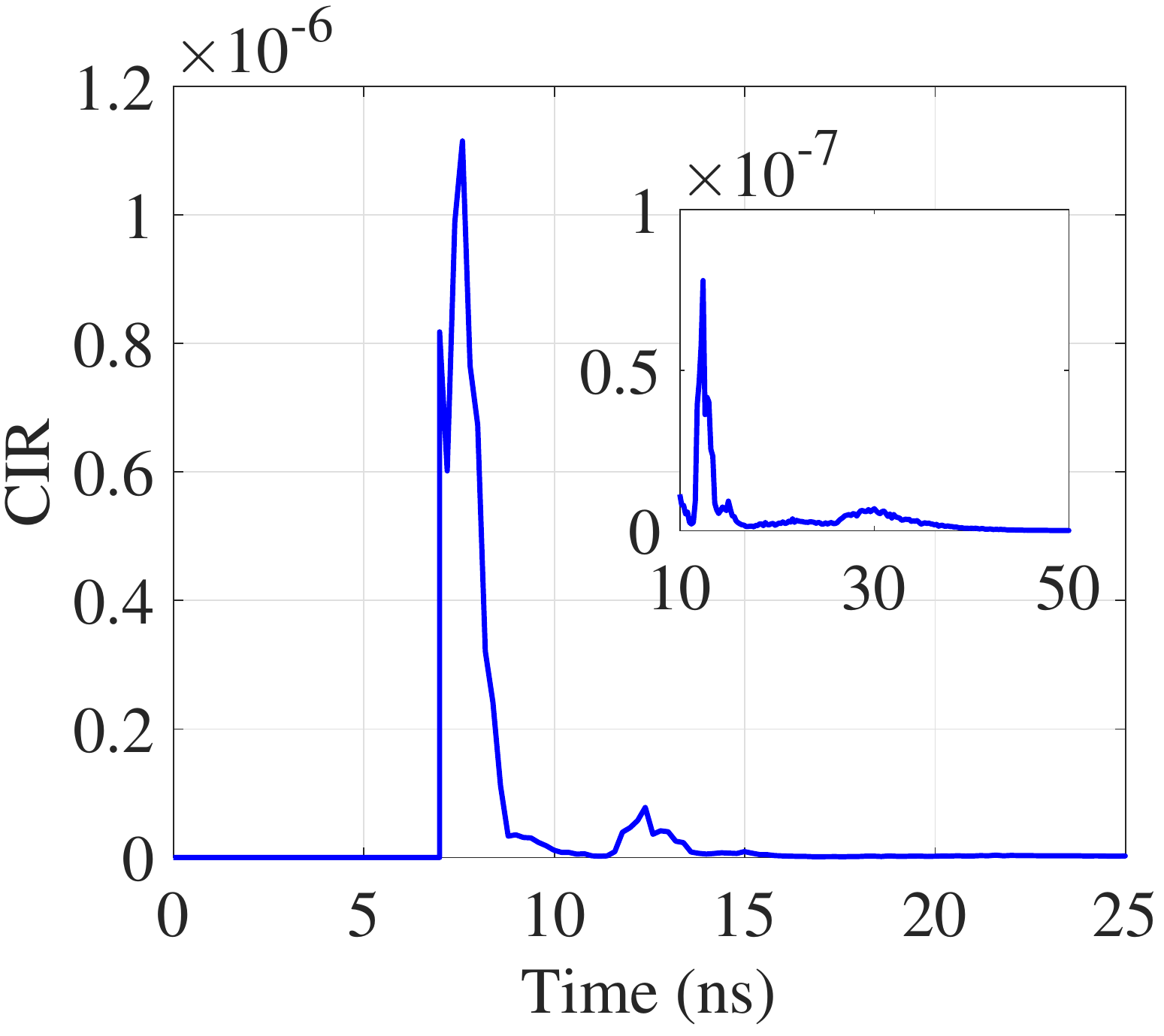}
		\caption{IRS OFF $h(t;\forall S,\text{L2})$}
		\label{}
	\end{subfigure} \\
	\begin{subfigure}[t]{.43\columnwidth}
		\includegraphics[width=\columnwidth]{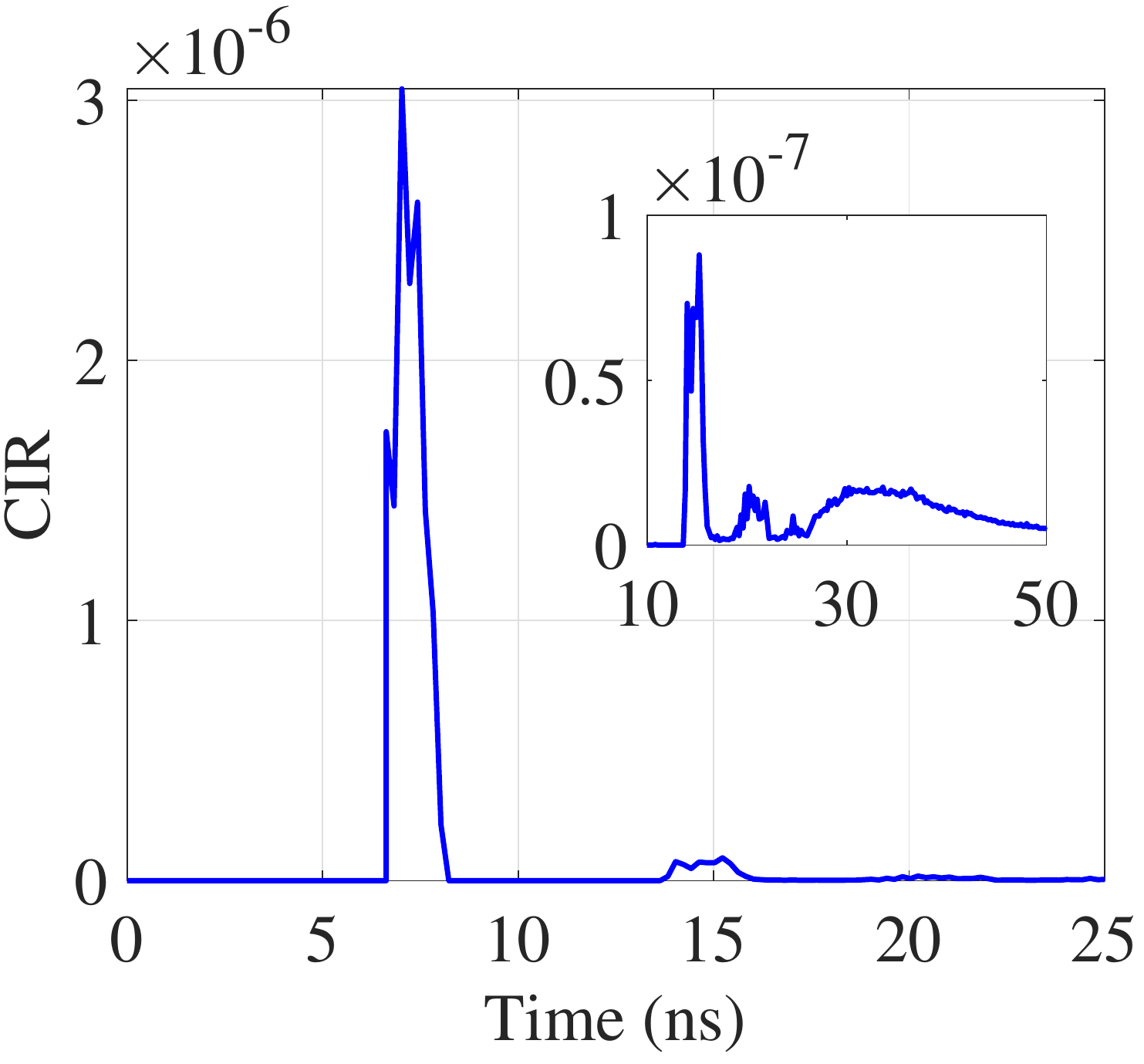}
		\caption{IRS ON $h(t;\forall S,\text{L3})$}
		\label{}
	\end{subfigure} ~
	\begin{subfigure}[t]{.43\columnwidth}
		\includegraphics[width=\columnwidth]{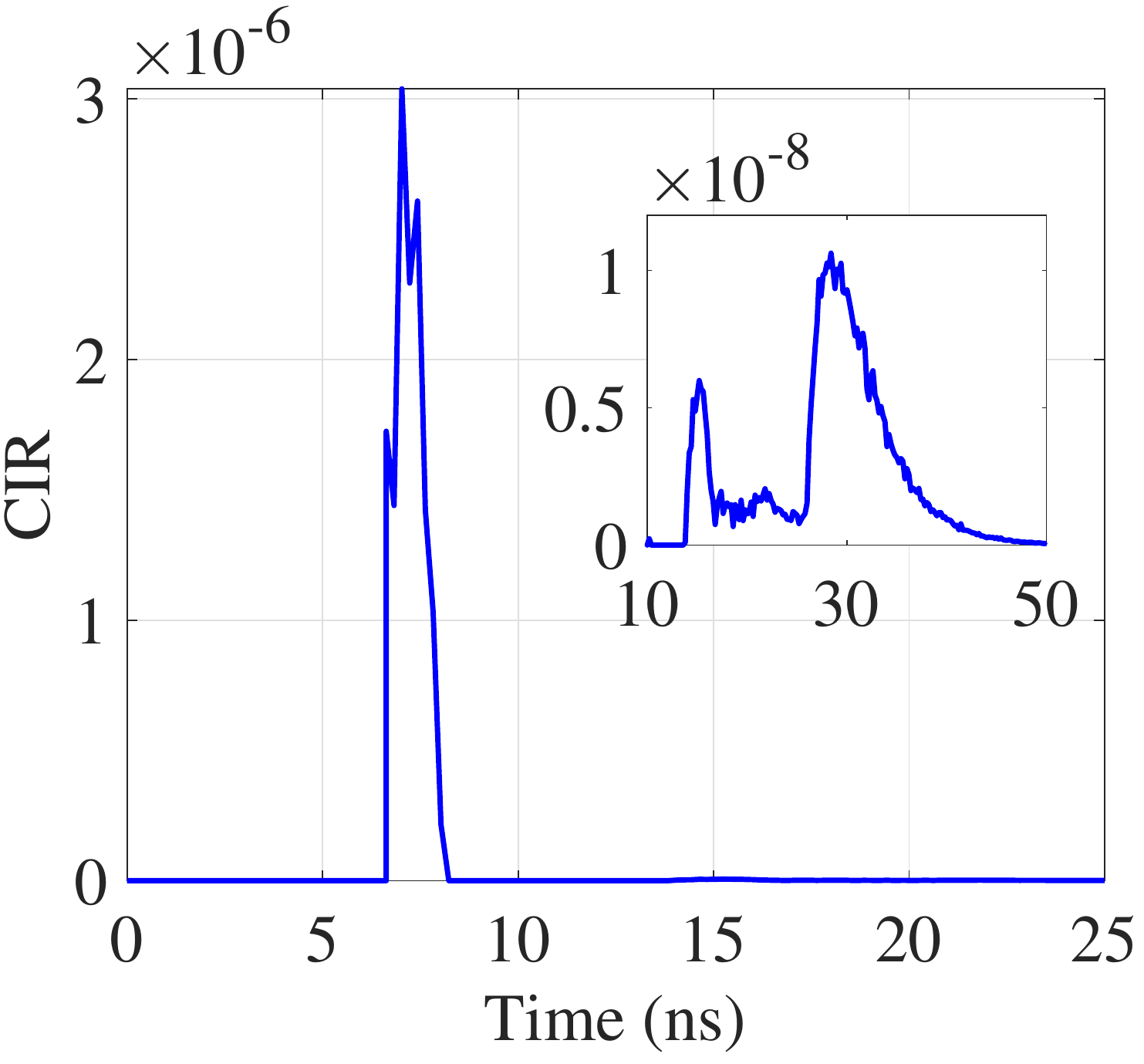}
		\caption{IRS OFF $h(t;\forall S,\text{L3})$}
		\label{}
	\end{subfigure}
	\caption{IRS aided indoor LiFi CIR simulation results for IR band. The results are obtained by the proposed MCRT based simulation technique. Each row represents the UE's location, whereas the columns are the state of the IRSs.}
	\label{fig:CIR_IR}
\end{figure*}

The \gls{IR} band \gls{CIR} plots and related channel parameters are also given in Fig. \ref{fig:CIR_IR} and Table \ref{table:channel_IR}, respectively.

In Figs. \ref{fig:CIR_IR}(a) and (b), the \gls{IR} band \glspl{CIR} plots for the mobile \gls{UE} location L1 when the \glspl{IRS} are ON and OFF, respectively, are depicted. Similar to the \gls{VL} band results, the channel conditions significantly improve with the utilization of the \glspl{IRS} when the mobile user is close to the highly reflective \glspl{IRS} and far from the sources. As can be seen from Fig. \ref{fig:CIR_IR}(a), a very high peak emerged by the first and second order reflections, depicted between $8-12$ ns, which is merged with the \gls{LoS} component. Furthermore, the third and further order reflections, depicted between $13-50$ ns, are also significantly enhanced with the aid of the \glspl{IRS}. The peak value of the \gls{CIR} and \gls{DC} channel gain values increased approximately $42.37\%$ and $161.51\%$ when \glspl{IRS} are activated, respectively. Similarly, an increase of $184.42\%$ in the \gls{RMS} delay spread can also be observed with the employment of the \glspl{IRS} compared to the case when \glspl{IRS} are OFF. The number of captured bounces from the side walls, ceiling and the floor becomes $22$ and $3$ when \gls{IRS}-ON and \gls{IRS}-OFF, respectively. Please note that the difference between the \gls{VL} and \gls{IR} bands stems from the different spectral reflection profiles of the materials as well as the \quot{minimum relative ray intensity} value in our simulations. The fraction of the \gls{NLoS} component compared to the whole \gls{IR} band channel becomes $69.2\%$ and $19.5\%$ for \gls{IRS}-ON and \gls{IRS}-OFF cases, respectively. The reason behind the higher \gls{NLoS} component in the \gls{IR} band compared to its \gls{VL} band counterpart is the higher reflectivity values of the coating materials in the \gls{IR} band (Fig. \ref{fig:material_spectrum}). Lastly, the maximum number of bounces considered in location L1 also becomes $22$ and $3$ for the \gls{IRS}-ON and \gls{IRS}-OFF cases, respectively.
\begin{table}[!t]
\caption{Details of the IR band optical channel parameters when IRSs are ON and OFF.}{%
		\begin{tabular}{|c|c|c|c|c|c|}
			\hline
			\textbf{} & \multicolumn{5}{c|}{\textbf{IRS ON (IR Band)}}  \\ \hline
			$R$ & $\kappa_\text{max}$ & $H[0;\forall S,R]$ & $\tau_0$ (ns) & $\tau_\text{RMS}$ (ns) & $\rho$  \\ \hline
			$\text{L1}$ & $22$ & $8.486\text{E}^{-6}$ & $9.165$ & $3.433$ & $0.308$  \\ \hline
			$\text{L2}$ & $25$ & $1.125\text{E}^{-5}$ & $7.875$ & $1.838$ & $0.529$  \\ \hline
			$\text{L3}$ & $28$ & $1.630\text{E}^{-5}$ & $7.201$ & $0.843$ & $0.845$  \\ \hline
			\textbf{} & \multicolumn{5}{c|}{\textbf{IRS OFF (IR Band)}}  \\ \hline
			$R$ & $\kappa_\text{max}$ & $H[0;S,R]$ & $\tau_0$ (ns) & $\tau_\text{RMS}$ (ns) & $\rho$  \\ \hline
			$\text{L1}$ & $3$ & $3.245\text{E}^{-6}$ & $8.146$ & $1.207$ & $0.805$  \\ \hline
			$\text{L2}$ & $3$ & $6.699\text{E}^{-6}$ & $7.547$ & $0.601$ & $0.888$  \\ \hline
			$\text{L3}$ & $3$ & $1.427\text{E}^{-5}$ & $7.175$ & $0.374$ & $0.965$  \\ \hline
		\end{tabular}}{}
	\label{table:channel_IR}
\end{table}

The \gls{IR} band channel impulse response plots and the related parameters are also provided for \gls{UE} location L2 in Figs. \ref{fig:CIR_IR}(c)-(d) and Table \ref{table:channel_IR}, respectively. Similar to the \gls{VL} band results, the impact of the \glspl{IRS} could be seen from Fig. \ref{fig:CIR_IR}(c) in the first and second order reflections ($7-10$ ns) as well as the higher order reflections ($12-50$ ns). However, the effect of the \glspl{IRS} are not as dominant as in location L1 due to the dominant \gls{LoS} link ($0^\text{th}$ order reflection). The peak value of \gls{CIR} and \gls{DC} channel gain when \gls{IRS}-ON are $0.12\%$ lower and $67.94\%$ higher compared to \gls{IRS}-OFF case, respectively. Again, the reason behind this is the effect of the \glspl{IRS} on the higher order reflections, which also indicate that the system capacity will be higher when \glspl{IRS} are activated if the direct \gls{LoS} link is blocked. The \gls{RMS} delay spread also increases around $205.82\%$ when the \glspl{IRS} are activated, which shows the significance of the higher order reflections. Compared to location L1, the \gls{DC} channel gain increases by $32.57\%$, whereas the \gls{RMS} delay spread decreases by $46.46\%$ compared to L2 when the \gls{IRS}-ON, respectively. For cases where the \glspl{IRS} are OFF, the \gls{DC} channel gain increase is approximately $106.44\%$ and the \gls{RMS} delay spread decrease is around $50.21\%$ from point L1 to L2, respectively. Moreover, only the \gls{NLoS} path contributions compared to the whole channel becomes $47.1\%$ and $11.2\%$ for \gls{IRS}-ON and \gls{IRS}-OFF cases, respectively, when the \gls{UE} is located at L2. The number of bounces encountered in our \gls{MCRT} simulations becomes $25$ and $3$ for \gls{IRS}-ON and \gls{IRS}-OFF cases, respectively, when the \gls{UE} is at L2.

In Figs. \ref{fig:CIR_IR}(e) and (f), the \gls{CIR} plots for mobile \gls{UE} location L3 when \glspl{IRS} are ON and OFF, respectively, are depicted. The peak value of the \gls{CIR} and \gls{DC} channel gain are increased $0.16\%$ and $14.24\%$ when the \glspl{IRS} are activated for \gls{UE} location L3, respectively. As can be seen from the channel parameters as well as Figs. \ref{fig:CIR_IR}(e) and (f), the effect of the \glspl{IRS} on the main lobe, between $6-8$ ns, is minimal due to the dominant \gls{LoS} power. However, the optical power emerges due to the higher order reflections, between $14-50$ ns, is significantly higher in \gls{IRS}-ON case compared to \gls{IRS}-OFF. Furthermore, the \gls{RMS} delay spread of the channel increases by more than $125.4\%$ when the \glspl{IRS} are activated at location L3. The maximum number of bounces captured in our simulations for the case when the \glspl{IRS} are ON and OFF becomes $28$ and $3$, respectively, when the \gls{UE} is at L3. The \gls{NLoS} contribution against the whole channel optical power also increased from $3.5\%$ to $15.5\%$ when the \glspl{IRS} are activated for a mobile user located at L3. 
%Therefore, the \gls{IR} link capacity for a \gls{UE} under the \gls{LoS} blockage will become almost $3.55$, $4.2$ and $4.4$ fold higher when the \glspl{IRS} are utilized for L1, L2 and L3, respectively.
%####################################################################################################
\subsection{The Achievable Rates for IRS-aided LiFi}
In this subsection, the achievable rate curves for the \gls{IRS}-aided indoor \gls{LiFi} application scenario will be provided. The maximum achievable system capacity for \gls{LiFi} systems could simply be calculated as follows \cite{kb9701,hycvppai2001}:
\begin{align}
 C(R) = \frac{1}{2}B\log_{2}\left(1 + \frac{4P_S^2H^2(0;\forall S,R)}{\sigma_w^2}\right),
 \label{eq:capacity}
\end{align}
\noindent where the average transmitted optical power, \gls{DC} channel gain, the effective noise power in the electrical domain and communication network's effective bandwidth are denoted by $P_S$, $H(0;S,R)$, $\sigma_w^2$ and $B$, respectively. Note that the effective noise term at the \gls{UE} consists of the addition of shot and thermal noises, where the shot noise emerges as a result of ambient light sources and the information bearing signal itself. In cases where high ambient light power at the \gls{PD} is significantly larger than the transmit signal power, the shot noise becomes signal independent. Therefore, the high intensity shot noise at the \gls{RX} could be modelled as a summation of independent low power Poisson processes, which could be approximated as a zero mean Gaussian distribution. Consequently, the effective noise could be modelled as \gls{AWGN}, $w\sim\mathcal{N}\left(0,\sigma_w^2 \right)$, where $\sigma_w^2 = \sigma_\text{shot}^2 + \sigma_\text{thermal}^2$. A real Gaussian distribution with mean $\mu$ and variance $\sigma^2$ is denoted by $\mathcal{N}(\mu,\sigma^2)$.

By using Tables \ref{table:channel_VL} and \ref{table:channel_IR}, the \gls{DC} channel gain values for both the \gls{VL} and \gls{IR} band could be obtained for the cases where \glspl{IRS} are ON and OFF \gls{w.r.t.} \gls{UE} locations L1, L2 and L3. Also note from Table \ref{table:MCRT_param}, the transmitted optical power is chosen to be $P_S=36$ W per luminaire. The effective channel bandwidth, $B$, for the considered system could be calculated by using the expression, 
\begin{align}
B = \frac{\Delta f (N-1)}{N} = \frac{\left( 2^{\lceil \log_2(N_\text{b}) \rceil}-1 \right)}{2^{\lceil \log_2(N_\text{b}) \rceil}\Delta w} \approx \Delta f \quad \text{if } N \gg 64.
\label{eq:capacity_2}
\end{align}
\begin{figure}[!t]
	\centering
	\begin{subfigure}[t]{.7\columnwidth}
		\includegraphics[width=\columnwidth]{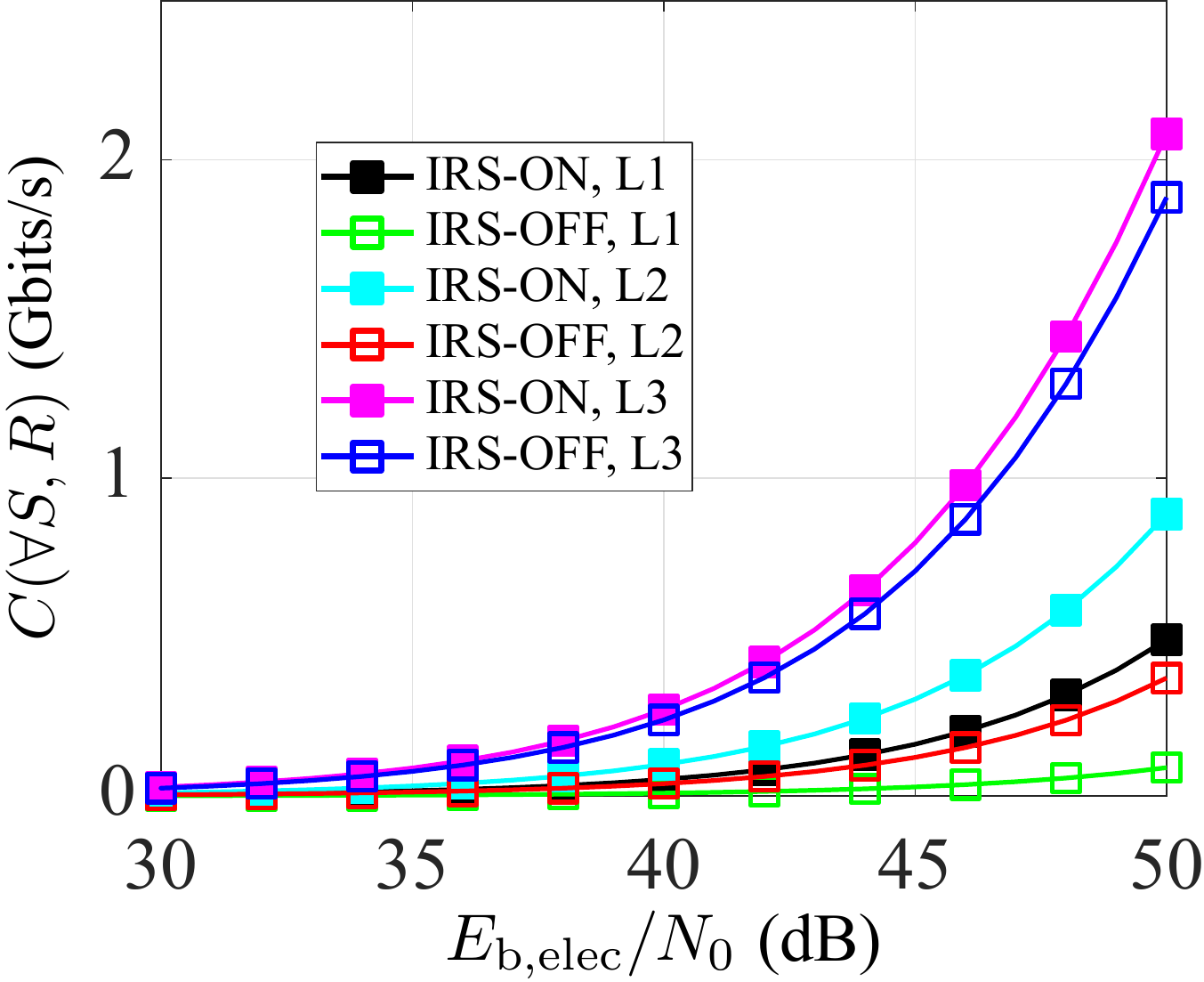}
		\caption{VL Band}
		\label{}
	\end{subfigure}\\
	\begin{subfigure}[t]{.7\columnwidth}
		\includegraphics[width=\columnwidth]{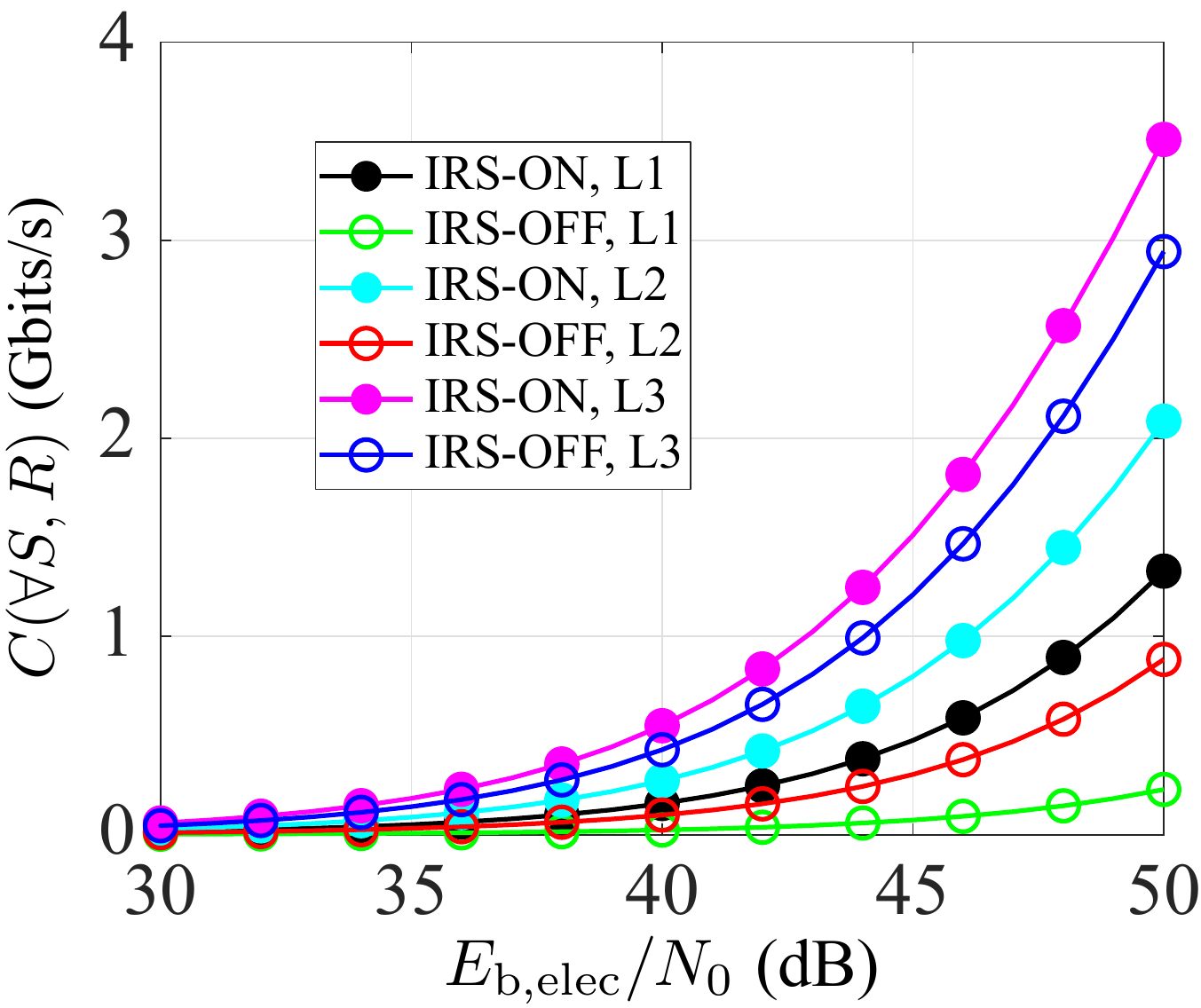}
		\caption{IR Band}
		\label{}
	\end{subfigure}
	\caption{Achievable capacity plots for (a) VL band and (b) IR band indoor LiFi for IRS-ON and IRS-OFF cases when the UE is located at L1, L2 and L3.}
	\label{fig:capacity}
\end{figure}\noindent Please note from \eqref{eq:capacity} and \eqref{eq:capacity_2} that typical \gls{LiFi} networks are not limited by channel bandwidth, but by the the electrical domain bandwidth of the front-end opto-electronic devices such as \glspl{LED} and \glspl{PD}. In this work, the optical channel bandwidth is considered as the main limiting factor for the system as it inherently emerges. On the contrary, the electrical channel bandwidth, caused by the font-end opto-electronic frequency response, is strictly dependent on the manufacturing technology behind the transmit \glspl{LED} and \glspl{PD}. Therefore, the channel impairments introduced by the electrical domain components could be avoided as technology advances. Moreover, in our simulations, the number of \gls{DFT} subcarriers and the effective optical channel bandwidth could be calculated by using the values in Table \ref{table:MCRT_param} as $N=1024$ and $B=5$ GHz.

In Figs. \ref{fig:capacity}, the achievable capacity plots for \gls{IRS}-ON and \gls{IRS}-OFF cases when the \gls{UE} is located at L1, L2 and L3. Accordingly, the \gls{VL} band plots are given for the $[30~ 50]$ dB \gls{SNR}-per-bit ($E_\text{b,elec}/N_0$) region in Fig. \ref{fig:capacity}(a). As can be seen from figure that the maximum achievable capacity increases exponentially within the low \gls{SNR} region. Please note from Figs. \ref{fig:CIR_VL} and \ref{fig:CIR_IR} that the channel magnitudes are in the $10^{-6}$ and $10^{-7}$ region, thus, the electrical domain path loss at the \gls{RX} becomes about $-120$ and $-140$ dB, respectively \cite{phd_Fath}. Therefore, the \gls{SNR} measured at the \gls{TX} side within the interval of $30-50$ dB will be classified as the low \gls{SNR} regime for this application. For \gls{UE} location L1, the achievable data rate becomes approximately $500$ and $90$ Mbits/sec when the \glspl{IRS} are ON and OFF, respectively. Similarly, for L2, the achievable rates of approximately $700$ and $370$ Mbit/sec for cases where the \glspl{IRS} are ON and OFF, respectively. Lastly, approximately $2.08$ and $1.88$ Gbit/sec data rates are achieved for the \gls{UE} location L3 when the \glspl{IRS} are active and inactive, respectively.

The \gls{IR} band achievable capacity plots for \gls{IRS}-ON and \gls{IRS}-OFF cases when the \gls{UE} is located at points L1, L2 and L3 are depicted in Fig. \ref{fig:capacity}(b). It is important to note that the overall capacity values are higher in the \gls{IR} band compared to the \gls{VL} band results due to the higher reflectivity profiles of the coating materials in the \gls{IR} spectra. For \gls{UE} location L1, the capacity rates of approximately $1.33$ Gbits/sec and $237$ Mbits/sec are achieved when \glspl{IRS} are ON and OFF, respectively. Moreover, the achievable rates become approximately $2.09$ Gbits/sec and $884$ Mbits/sec for \gls{IRS} active and inactive cases, when the \gls{UE} is located at L2. Finally, for \gls{UE} location L3, the achievable rate of $3.51$ and $2.94$ Gbits/sec are obtained for the \gls{IRS}-ON and \gls{IRS}-OFF cases, respectively. Please note that based on both \gls{VL} and \gls{IR} band plots, the maximum capacity values are achieved for each \gls{UE} location when the \glspl{IRS} are activated.
%\begin{align}
%    B_\text{dB}=10\log_{10} \left( \frac{(I_\text{max}+I_\text{min})^2}{4\sigma^2} + 1 \right)
%\end{align}
%\noindent where $\sigma = \sqrt{1/N}$
%
%\begin{table}[!t]
%\caption{The set of parameters used in the BER simulations.}{%
%		\begin{tabular}{|l|l|l|}
%			\hline
%			\textbf{Parameter} & \textbf{Description} & \textbf{Value} \\ \hline\hline
%			$N$ & Number of subcarriers & $1024$              \\ \hline
%			$N_{\text{CP}}$ & Length of the CP & $38$              \\ \hline
%			$M$ & Order of the QAM constellation & $4$             \\ \hline
%			$B_\text{dB}$ & DC bias value & $22.17$ dB              \\ \hline
%			$I_{\textrm{min}}$ & The lower limit for the $I_\textrm{f}$ & $100$ mA \cite{gwqsspa1.em,sfh4253}              \\ \hline
%			$I_{\textrm{max}}$ & The upper limit for the $I_\textrm{f}$ & $700$ mA \cite{gwqsspa1.em,sfh4253}             \\ \hline
%		\end{tabular}}{}
%	\label{table:BER_par}
%\end{table}
%####################################################################################################
\section{Challenges \& Research Directions } \label{sec:challenges}
The concept of \gls{IRS} provides opportunities for achieving unprecedented capabilities when it comes to signal and link manipulation. However, realising such capabilities is subject to overcoming the challenges associated with the realisation of this concept. %The achievable capacity and \gls{ISI} induced error performance trade-off in \gls{IRS}-aided \gls{LiFi} is carried to future work. 
In this section, we highlight some of the challenges and open research directions related to the integration of \gls{IRS} in \gls{OWC} systems and propose a road map for future research directions. 

\subsection{Modelling and Characterisation}
One of the challenges in the research and implementation of the concept of smart walls in \gls{OWC} systems is the development of realistic and accurate channel models. More specifically, there is a need to establish realistic models that take into account the type, capabilities, and limitations of different possible \gls{IRS} structures in order to capture the fundamental behaviour and performance limits of such systems. Moreover, since classical optical channel gain models might not be suitable, it is critical to understand the \gls{CIR} of different metasurfaces over the range of optical wavelengths. In particular, there is a need to quantify the efficiency and response time for achieving specific functionalities such as amplification factors, absorption capabilities, anomalous reflection, etc. Additionally, it is critical to understand how such systems perform in different mediums such as underwater and outdoor environments with high ambient noise.

\subsection{Inter-symbol Interference (ISI)}
It has been reported in the previous section that the deployment of the \glspl{IRS} in typical indoor \gls{LiFi} networks increases the achievable data rate values significantly both for \gls{VL} and \gls{IR} spectra. However, one of the biggest challenges for \gls{IRS}-aided \gls{LiFi} arises from the inflated channel time dispersion. The increase in the channel delay spread values could effectively reduce the system performance due to the increased \gls{ISI}. The \gls{ISI} could be avoided in multi-carrier transmission methods, especially in \gls{OFDM}, by the utilization of the \gls{CP}. Furthermore, with the activation of the \glspl{IRS},  the channel frequency selectivity will also increase which means that the \gls{SNR} in each subcarrier will significantly differ from each other. Consequently, in order to avoid further channel impairments which could emerge with the utilization of \glspl{IRS},  suitable \gls{CP} length selection as well as adaptive bit loading in \gls{OFDM} are required. A detailed investigation of the optical \gls{OFDM} systems and their error performances under the frequency selective channel that emerges in \gls{IRS}-aided \gls{LiFi} is necessary to obtain better insights on the performance of such systems.

\subsection{Channel Estimation}
The  availability  of accurate \gls{CSI} for all the channel paths created by massive numbers of tunable sub-wavelength reflecting elements proves to be a challenging task. Traditional \gls{CSI} estimation techniques might be unpractical for real-time estimations given the high dimensions of the channel vectors for each network user, particularly if higher order reflections are considered.    
There is a need to quantify the trade-off between distributed and centralized \gls{CSI} acquisition approaches and to come up with intelligent and cost-effective methodologies. Distributed \gls{CSI} acquisition approaches employ  local estimation at each \gls{IRS} array  which requires sensing and processing capabilities. Centralized \gls{CSI} acquisition approaches, on the other hand, employ a central control unit, typically at the \gls{AP}. The central unit is responsible for performing  the \gls{CSI} estimation for all network entities, which are then utilized for optimisation and reconfiguration decisions that are executed centrally. The obvious advantage of the centralized approach is that the \gls{IRS} is not required  to perform exhaustive sensing and processing, leading  to lower energy consumption and a simpler hardware design. The main challenge of this approach is the signalling overhead, which occurs between the central unit and the \gls{IRS}. 

\subsection{Real-time Operation}
Integrating \gls{IRS} arrays with a high number of reflecting elements and varying functionalities inevitably results in high operational complexity which  entails increased  computational, energy, and overhead cost. The \gls{IRS} elements need to be reconfigured in real-time to provide precise control over their optical functionalities, which is decided  based on the varying system conditions. Such an operation is not trivial and entails different trade-offs.  For example, there is a need to develop activation strategies that take into account different metrics, such as throughput and fairness,  in order to balance the desired performance enhancement and the associated energy consumption and time delay. Also, how often the reflecting element must be reconfigured must be determined i.e., whether this should be performed with each channel realisation or when a change occurs in the user location. Incorporating data-driven optimisation tools such as deep learning, reinforcement learning, and federated learning  could provide viable solutions to achieve time-efficient optimisation.

\backmatter

\bibliographystyle{unsrtnat}
\bibliography{ch01}%

\backmatter

\printindex

\end{document}